\documentclass[manuscript,screen]{acmart}

\usepackage[ruled,vlined]{algorithm2e}
\usepackage{subfigure, graphicx, epsfig}

\AtBeginDocument{%
  }

\setcopyright{acmcopyright}
\copyrightyear{2018}
\acmYear{2018}
\acmDOI{XXXXXXX.XXXXXXX}

\acmJournal{POMACS}
\acmVolume{37}
\acmNumber{4}
\acmArticle{111}
\acmMonth{8}

\usepackage{multirow}

\begin{document}

\title{Trading off Quality for Efficiency of Community Detection: An Inductive Method across Graphs}

\author{Meng Qin}
\email{mengqin\_az@foxmail.com}
\affiliation{%
  \institution{Department of CSE, HKUST}
  \country{Hong Kong SAR}}

\author{Chaorui Zhang}
\email{chaorui.zhang@gmail.com}
\affiliation{%
  \institution{Theory Lab, Huawei}
  \country{Hong Kong SAR}}

\author{Bo Bai}
\email{ee.bobbai@gmail.com}
\affiliation{%
 \institution{Theory Lab, Huawei}
 \country{Hong Kong SAR}}

\author{Gong Zhang}
\email{nicholas.zhang@huawei.com}
\affiliation{%
  \institution{Theory Lab, Huawei}
  \country{Hong Kong SAR}}

\author{Dit-Yan Yeung}
\email{dyyeung@cse.ust.hk}
\affiliation{%
  \institution{Department of CSE, HKUST}
  \country{Hong Kong SAR}}

\renewcommand{\shortauthors}{Meng Qin et al.}

\begin{abstract}
Many network applications can be formulated as NP-hard combinatorial optimization problems of community detection~(CD). Due to the NP-hardness, to balance the CD quality and efficiency remains a challenge. Most existing CD methods are \textit{transductive}, which are independently optimized only for the CD on a single graph. Some of these methods use advanced machine learning techniques to obtain high-quality CD results but usually have high complexity. Other approaches use fast heuristic approximation to ensure low runtime but may suffer from quality degradation. In contrast to these \textit{transductive} methods, we propose an alternative \textit{inductive} community detection (ICD) method across graphs of a system or scenario to alleviate the NP-hard challenge. ICD first conducts the \textit{offline} training of an adversarial dual GNN on historical graphs to capture key properties of the system. The trained model is then directly generalized to new unseen graphs for \textit{online} CD without additional optimization, where a better trade-off between quality and efficiency can be achieved. ICD can also capture the permutation invariant community labels in the \textit{offline} training and tackle the \textit{online} CD on new graphs with non-fixed number of nodes and communities. Experiments on a set of benchmarks demonstrate that ICD can achieve a significant trade-off between quality and efficiency over various baselines.
\end{abstract}




\maketitle

\section{Introduction}
For various complex systems, e.g., communication and social networks, graph is a generic model to describe entities and their relations using a set of nodes and edges. Community detection (CD), a.k.a. graph clustering \cite{schaeffer2007graph}, aims to partition the nodes of a graph into several groups (i.e., communities) with dense linkage distinct from other groups \cite{fortunato202220}. Since the extracted communities are believed to correspond to several real-world substructures of a system, e.g., cells in wireless networks \cite{dai2017optimal}, many network applications can be formulated as CD tasks \cite{mayer2018graph,qin2019towards,patil2021graph}.

Mathematically, CD can be described as some NP-hard combinatorial optimization problems, e.g., modularity maximization \cite{newman2006modularity} and normalized cut (NCut) minimization \cite{von2007tutorial}. Due to the NP-hardness, to balance the quality and efficiency remains a challenge but some real applications have both requirements of high quality and low runtime, e.g., accurate CD on a wireless cellular network with several thousand nodes in a few seconds \cite{dai2017optimal}.

\begin{figure}[t]
\centering
\includegraphics[width=0.75\columnwidth, trim=18 18 18 18,clip]{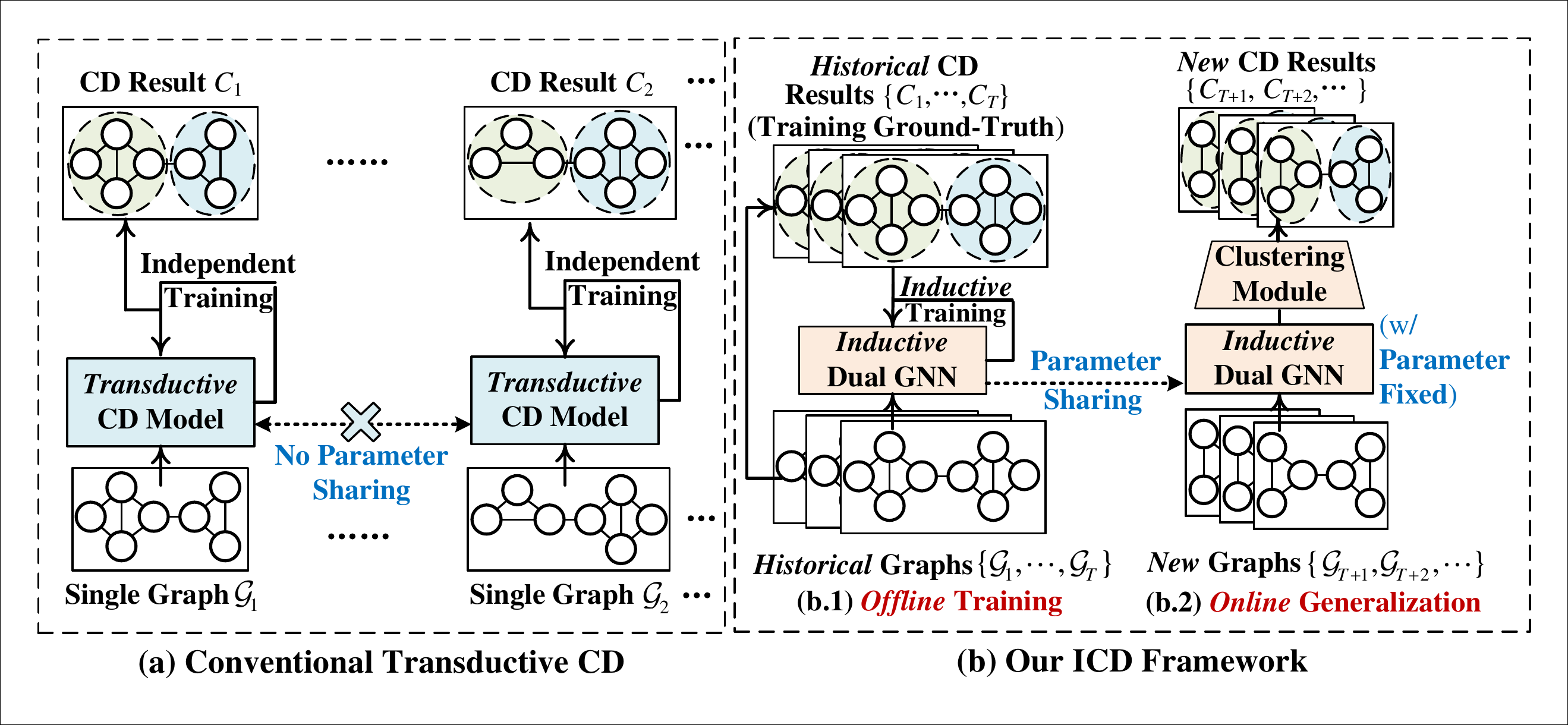}
\vspace{-0.2cm}
\caption{Overview of (a) conventional \textit{transductive} CD methods for each single graph and (b) our \textit{inductive} community detection (ICD) framework across graphs, including the (b.1) \textit{offline} training on historical graphs and (b.2) \textit{online} generalization to new graphs.}\label{Fig:Overview}
\vspace{-0.3cm}
\end{figure}

\textbf{Conventional \textit{Transductive} CD.}
As illustrated in Fig.~\ref{Fig:Overview} (a), most existing CD methods are \textit{transductive}, which independently optimize the CD model on each single graph and can only tackle CD on such a unique graph. Since model parameters are not shared across graphs, the model trained on one graph cannot be directly generalized to other graphs. Usually, one needs to optimize the model from scratch for each new graph. Conventional \textit{transductive} methods usually focus on either high quality or efficiency.

On the one hand, prior work has demonstrated the ability of advanced machine learning (ML) techniques to achieve high quality, e.g., high accuracy between the CD results and ground-truth.
Typical ML-based methods include non-negative matrix factorization \cite{wang2008clustering,wang2011community} and probabilistic graphical models \cite{karrer2011stochastic,zhang2014scalable}. Graph embedding emerges as a promising technique for CD in recent studies. These methods first learn low-dimensional node representations via random walk \cite{perozzi2014deepwalk,grover2016node2vec}, matrix factorization \cite{qiu2018network,liu2019general}, or deep learning \cite{wang2016structural,yang2016modularity} to capture high-order proximities and nonlinear characteristics of a graph. The CD result is derived by feeding the learned embedding into a downstream clustering module (e.g., $K$Means). However, aforementioned ML-based methods usually rely on iterative optimization algorithms (e.g., gradient descent for deep learning and expectation maximization for probabilistic models) to independently learn large-scale parameters for each graph with inevitable high complexity.

On the other hand, to reduce the overall runtime and satisfy real-time constraints of some applications (e.g., decomposing a wireless network with several thousand entities into cells in several seconds \citep{dai2017optimal}) 
is another major focus. Some \textit{transductive} methods use fast approximation for classic CD objectives, e.g., greedy modularity maximization \cite{blondel2008fast} and multilevel coarsening for NCut minimization \cite{dhillon2007weighted}. Several graph embedding approaches also adopt fast approximation to reduce the inference time, e.g., random projection for high-order proximities \cite{zhang2018billion} and randomized SVD to capture local structural information \cite{zhang2019prone}.
Despite their high efficiency (e.g., low runtime to derive CD results), they may suffer from quality declines due to the information loss of fast approximation.

\textbf{\textit{Inductive} CD across Graphs.}
In contrast to conventional \textit{transductive} methods, we try to \textit{achieve a better trade-off between quality and efficiency from an alternative \textit{inductive} perspective across multiple graphs of a system or scenario}. An \textit{inductive} community detection (ICD) method is proposed based on the fact that \textit{most real-world complex systems independently generate a set of graphs via common knowledge in terms of underlying distributions}, e.g., power-law distributions. This hypothesis is also widely adopted in the simulation of various systems \cite{wehrle2010modeling}. 
The multiple graphs can be (\romannumeral1) snapshots evolving over time or (\romannumeral2) independent graphs of a scenario without temporal dependency, e.g., ego-nets in social media. 
The \textit{inductiveness} of ICD implies that its model parameters are shared by all the associated graphs.

As illustrated in Fig.~\ref{Fig:Overview} (b), ICD includes two phases. (\romannumeral1) We first train a high-quality CD model on historical graphs of a system or scenario in an \textit{offline} way, aiming to fully capture the underlying distributions of the system or scenario regardless of time. (\romannumeral2) The trained model can be directly generalized to newly generated graphs of the same system or scenario for \textit{online} CD without additional optimization, which significantly reduces the runtime on these new graphs and is believed to have high-quality results. 
%For instance, we can conduct the \textit{offline} training of ICD on existing known ego-nets of a social network and generalize it to new unseen ego-nets. 
In real applications, we usually assume that one has enough time to train a high-quality model using historical data in an \textit{offline} way, which is also one-time-effort only.
Our main focus is to \textit{achieve a better trade-off between quality and efficiency of the \textit{online} CD on new graphs after deploying the trained model to a system with its model parameters fixed}.
%Moreover, ICD also has the potential to capture additional characteristics shared by historical graphs and achieve better quality on new graphs beyond conventional methods applied to one single snapshot.

\begin{table}\scriptsize
\centering
\caption{Summary of existing \textit{inductive} GNNs for unsupervised CD across graphs without attributes.}
\label{Tab:Sum-Ind}
\vspace{-0.3cm}
\begin{tabular}{l|p{1.4cm}|l|p{1.9cm}}
\hline
 & \textbf{Settings} & \textbf{Descriptions} & \textbf{Examples} \\ \hline
\multirow{2}{*}{\textbf{Feature Input}} & \textbf{Standard Settings} & \begin{tabular}[c]{@{}l@{}}Use a constant matrix/one-hot node degree representation as feature inputs.\\Not informative features to support high-quality CD.\end{tabular} & \textit{GraphSAGE}\cite{hamilton2017inductive}, \textit{GAT}\cite{velivckovic2017graph}, \textit{GIN}\cite{xu2018powerful} \\ \cline{2-4} 
 & \textbf{Feature Extraction} & \begin{tabular}[c]{@{}l@{}}Extract additional features via dimension reduction (e.g., PCA) on topology.\\Usually time-consuming.\end{tabular} & \textit{GAP}\cite{nazi2019gap}, \textit{ClusNet}\cite{wilder2019end}, \textit{LGNN} \cite{chen2020supervised} \\ \hline
\multirow{2}{*}{\begin{tabular}[c]{@{}l@{}}\textbf{How to derive}\\\textbf{CD results}\end{tabular}} & \textbf{Embedding} & \begin{tabular}[c]{@{}l@{}}Apply a downstream clustering module to the learned embedding (from GNN).\\ Can tackle \textit{inductive} CD across graphs with both fixed \& non-fixed K.\end{tabular} & \textit{GraphSAGE}\cite{hamilton2017inductive}, \textit{GAT}\cite{velivckovic2017graph}, \textit{GIN}\cite{xu2018powerful} \\ \cline{2-4} 
 & \textbf{End-to-End} (E2E) & \begin{tabular}[c]{@{}l@{}}Feed learned embeddings (from GNN) to an output layer to derive CD results.\\Can tackle \textit{inductive} CD with fixed $K$ but have to be optimized from scratch for graphs with non-fixed $K$.\end{tabular} & \textit{GAP}\cite{nazi2019gap}, \textit{ClusNet}\cite{wilder2019end}, \textit{LGNN}\cite{chen2020supervised} \\ \hline
\multirow{2}{*}{\textbf{Training Loss}} & \textbf{Unsupervised Loss} & \begin{tabular}[c]{@{}l@{}}Use unsupervised loss of existing graph embedding techniques (e.g., sikpgram-based loss).\\Usually lacks robustness for high-quality CD and cannot capture permutation invariant training labels.\end{tabular} & \textit{GraphSAGE}\cite{hamilton2017inductive}, \textit{GAT}\cite{velivckovic2017graph}, \textit{GIN}\cite{xu2018powerful} \\ \cline{2-4} 
 & \textbf{CD Objectives} & \begin{tabular}[c]{@{}l@{}}Use CD objectives (e.g., modularity maximization \& NCut minimization) for unsupervised training.\\Usually combines with the E2E framework and cannot capture permutation invariant training labels.\end{tabular} & \textit{GAP}\cite{nazi2019gap}, \textit{ClusNet}\cite{wilder2019end} \\ \hline
\end{tabular}
\vspace{-0.3cm}
\end{table}

ICD adopts a graph embedding scheme for CD across graphs using the \textit{inductive} nature of graph neural networks (GNN). Although existing \textit{inductive} GNNs (e.g., \textit{GraphSAGE} \cite{hamilton2017inductive}) can be generalized to new graphs for \textit{online} inference, they may suffer from the following limitations for CD as summarized in Table~\ref{Tab:Sum-Ind}.

First, in this study, we consider CD across graphs where topology is the only available information source, i.e., without attributes.
As some systems allow the addition and deletion of entities, we also assume the number of nodes $N$ can be non-fixed for each graph. Although most \textit{inductive} GNNs \cite{hamilton2017inductive,velivckovic2017graph} can be applied to graphs with non-fixed $N$, they are originally designed for attributed graphs with two input sources of topology and node features. In particular, \textit{their \textit{inductiveness} relies on the fixed dimensionality of feature inputs for all the graphs.} Our experiments indicate that some standard settings of \textit{inductive} GNNs for the case without attributes (e.g., using a constant matrix as feature inputs \cite{xu2018powerful}) cannot derive informative node features to support high-quality CD. Other GNN-based methods extract additional features via dimension reduction (e.g., PCA) to map the topology (e.g., adjacency matrix) with non-fixed $N$ to a fixed-dimensional feature space \cite{nazi2019gap}, which is usually time-consuming.

Second, graphs from a system or scenario can be assigned with different number of communities $K$. Some GNN-based methods adopt an end-to-end (E2E) framework to approximate some classic CD objectives (e.g., NCut minimization \cite{nazi2019gap}), with the output of GNN fed into a fully-connected output layer to directly derive CD results.
%CD results directly derived via a fully-connected output layer. 
However, they can only tackle the \textit{inductive} CD across graphs with fixed $K$, due to fixed dimensionality of the output layer. These E2E methods still need to be optimized from scratch for new graphs with non-fixed $K$ (i.e., using conventional \textit{transductive} settings), which is also time-consuming.

Third, although prior studies have validated the ability of \textit{inductive} GNNs to tackle \mbox{(semi-)supervised} tasks (e.g., node classification) on new nodes or graphs \cite{hamilton2017inductive,velivckovic2017graph}, few of them consider unsupervised node-level tasks (e.g., CD) across graphs. Our experiments indicate that some standard settings of \textit{inductive} GNNs for unsupervised tasks (e.g., using unsupervised training loss for GNN \cite{hamilton2017inductive}) lack robustness for the \textit{online} CD on new graphs.

\textbf{Present Work.}
ICD is a generic framework that can address the aforementioned limitations of \textit{inductive} GNNs with several original designs. \textbf{(\romannumeral1)} To enable ICD to tackle \textit{inductive} CD across graphs with non-fixed $N$, we develop an efficient feature extraction module for \textit{inductive} GNNs via graph coarsening, which can extract informative node features to support high-quality fast \textit{online} CD. \textbf{(\romannumeral2)} In contrast to E2E methods, ICD adopts an \textit{inductive} graph embedding scheme across graphs
with CD results derived via a downstream clustering module,
which can tackle the \textit{inductive} CD with non-fixed $K$.
\textbf{(\romannumeral3)} Note that CD is an unsupervised node-level task, where community labels are permutation invariant. For instance, label assignments $(l_1, l_2, l_3)=(1, 2, 2)$ and $(l_1, l_2, l_3)=(2, 1, 1)$ are the same in terms of CD with $l_i$ as the community label of node $v_i$.
Most existing GNN-based methods cannot directly utilize the permutation invariant training labels of CD.
In contrast, ICD can further incorporate such label information of historical graphs to the \textit{offline} training, which enhances the embedding optimization, by combining an adversarial dual GNN structure and a clustering regularization loss based on classic CD objectives, e.g., modularity maximization and NCut minimization.

We summarize our major contributions as follows.
\begin{itemize}
    \item In contrast to conventional \textit{transductive} CD methods, we propose a novel ICD method to achieve a better trade-off between quality and efficiency via an alternative \textit{inductive} graph embedding scheme across graphs.
    \item Compared with existing \textit{inductive} GNN based methods, ICD is a generic method that can tackle the \textit{inductive} CD across graphs with non-fixed $N$ and $K$ while capturing permutation invariant training labels based on several original designs, e.g., graph coarsening based feature extraction, adversarial dual GNN, and clustering regularization loss.
    \item We compare the quality and efficiency of ICD with $17$ baselines on $10$ datasets. Experiments demonstrate that ICD can achieve a better trade-off between the quality and efficiency of CD over various baselines.
\end{itemize}
Remainder of this paper is organized as follows. In Section~\ref{Sec:Rel}, we briefly review related work. The formal problem statements and preliminaries of this study are provided in Section~\ref{Sec:Prob-Pre}. Section~\ref{Sec:Meth} elaborates on the proposed ICD framework. Experiment settings and evaluation results are described in Section~\ref{Sec:Exp}. Finally, Section~\ref{Sec:Cons} concludes the paper.

\section{Related Work}\label{Sec:Rel}
In the past few decades, a series of methods have been proposed for CD (a.k.a. graph clustering). As reviewed in \cite{schaeffer2007graph,jin2021survey,dey2022community}, CD can be mathematically formulated as several NP-hard combinatorial optimization problems. Overviews of some typical objectives (e.g., NCut minimization and modularity maximization) can be found in \cite{von2007tutorial,chen2014community}.

%\subsection{Transductive CD Methods}
\textbf{\textit{Transductive} CD Methods.}
Most existing CD methods are \textit{transductive}, which usually focus on either high quality or efficiency. Some approaches try to obtain high-quality CD results via advanced ML techniques. Wang et al. \cite{wang2008clustering} used non-negative matrix factorization (NMF) with diffusion kernel similarity to tackle CD. In \cite{wang2011community}, CD on undirected, directed, and compound graphs was formulated as three NMF objectives. Karrer et al. \cite{karrer2011stochastic} proposed the degree-corrected stochastic blockmodel and discussed its relations to modularity maximization based on the generative probabilistic model.
%while Zhang et al. \cite{zhang2014scalable} introduced a belief propagation algorithm for modularity maximization.
Graph embedding, which learns distributed representations to capture key properties of graphs, has become a promising ML technique for various graph inference tasks including CD. Inspired by the skipgram-based word embedding, Perozzi et al. \cite{perozzi2014deepwalk} and Grover et al. \cite{grover2016node2vec} proposed \textit{DeepWalk} and \textit{node2vec} that learn embeddings based on truncated random walks on graphs.
%Qiu et al. \cite{qiu2018network} and Liu et al. \cite{liu2019general} introduced generic matrix factorization frameworks and theoretically analyzed their relations to various embedding methods.
Tian et al. \cite{tian2014learning} and Yang et al. \cite{yang2016modularity} explored the potential of deep learning to derive community-preserved embeddings by reconstructing features of NCut minimization and modularity maximization. In contrast, Wang et al. \cite{wang2017community} and Li et al. \cite{li2019learning} extracted community-preserved embeddings by incorporating modularity maximization and stochastic block model to NMF objectives. However, aforementioned \textit{transductive} ML-based methods usually rely on time-consuming iterative optimization algorithms (e.g., gradient descent)
%(e.g., expectation maximization for probabilistic models and gradient descent for deep learning) 
to independently learn large-scale parameters for each single graph despite their high quality.

Some other \textit{transductive} methods try to reduce their optimization or inference time via fast approximation to some objectives. Dhillon et al. \cite{dhillon2007weighted} developed a multilevel coarsening algorithm for graph-cut based CD using the equivalence between kernel $K$Means and some graph-cut objectives, e.g., NCut minimization. Clauset et al. \cite{clauset2004finding} introduced a hierarchical approach with a greedy strategy for modularity maximization, while Wang et al. \cite{wang2020community} formulated a relaxed version of modularity maximization as a low-cardinality semidefinite programming objective. In \cite{peixoto2014efficient}, a efficient algorithm based on the Markov chain Monte Carlo sampling was presented to inference the stochastic block model of CD.
Some graph embedding methods also try to reduce their inference time using fast approximation. In \cite{zhang2018billion}, Zhang et al. learned high-order proximity preserved embedding by Gaussian random projection. Dong et al. \cite{zhang2019prone} adopted sparse randomized SVD to capture local structural properties in the embedding inference. Yang et al. \cite{yang20209homogeneous} used BKSVD to initialize the learned embedding based on the personalized PageRank measure. However, aforementioned methods usually suffer from quality declines due to the information loss of fast approximation.

In summary, to balance the quality and efficiency of CD remains a challenge for aforementioned \textit{transductive} methods. In contrast, we consider an alternative \textit{inductive} graph embedding scheme across graphs to achieve a better trade-off between quality and efficiency using the \textit{inductive} nature of GNNs.

%\subsection{Inductive GNNs for CD}
\textbf{\textit{Inductive} GNNs for CD.}
In contrast to \textit{transductive} methods optimized for a single graph, the learned model parameters of \textit{inductive} GNNs can be generalized to new unseen nodes and graphs. Most GNNs are originally designed for attributed graphs and rely on the multi-layer aggregation of node features, a.k.a. message passing. Hamilton et al. \cite{hamilton2017inductive} proposed \textit{GraphSAGE}, a generic \textit{inductive} GNN framework based on the graph embedding scheme, with various feature aggregators and both supervised and unsupervised training losses. Veli{\v{c}}kovi{\'c} et al. \cite{velivckovic2017graph} introduced \textit{GAT} that can adaptively adjust the feature aggregation by applying self-attention to node features. You et al. \cite{you2021identity} developed a class of message passing GNNs by inductively considering node identities. Although aforementioned \textit{inductive} GNNs have demonstrated (semi-)supervised node-level tasks (e.g., node classification), few of them consider node-level unsupervised tasks (e.g., CD) across graphs. Our experiments demonstrate that some standard settings of GNN for \textit{inductive} CD across graphs without attributes (e.g., applying a sikpgram-based unsupervised training loss \cite{hamilton2017inductive} and using a constant matrix as input features of GNNs \cite{xu2018powerful}) lack robustness for high-quality CD.

Some other approaches use \textit{inductive} GNNs to approximate classic CD objectives in an E2E scheme, where the outputs of GNN are further fed into an output layer to directly derive the CD results. Nazi et al. \cite{nazi2019gap} and Wilder et al. \cite{wilder2019end} proposed \textit{GAP} and \textit{ClusNet}, which are E2E methods approximating NCut minimization and modularity maximization. However, their \textit{inductiveness} is designed only for CD with fixed number of communities $K$, due to the fixed dimensionality of their output layers.
One has to use conventional \textit{transductive} settings for graphs with non-fixed $K$ (i.e., optimizing them from scratch for each new graph), which is time-consuming. For \textit{inductive} CD with non-fixed number of nodes $N$, \textit{GAP} and \textit{ClusNet} use \textit{PCA} and \textit{node2vec} to map the original graph topology into another feature space \cite{nazi2019gap,wilder2019end} for input features of GNN, which also have high complexity.

Moreover, the aforementioned GNN-based methods cannot utilize the permutation invariant community labels in their \textit{offline} training. Although \textit{LGNN} \cite{chen2020supervised} can capture such label information, it relies on an E2E cross-entropy loss that finds the best mapping from the derived CD result to the training labels among $O(K!)$ cases. Hence, it can only tackle the \textit{inductive} CD with fixed $K$ and has high complexity to capture training labels, intractable to the case with large $K$. In contrast, our ICD framework can capture permutation invariant training labels to enhance the \textit{offline} embedding optimization with much lower complexity while supporting the fast \textit{online} CD with non-fixed $N$ and $K$.

\section{Problem Statements and Preliminaries}\label{Sec:Prob-Pre}
In this study, we consider CD on a set of graphs $S = \{ {{\mathcal{G}}_1}, \cdots, {{\mathcal{G}}_T}\}$ extracted from a common system or scenario.
Each graph ${{\mathcal{G}}_t} \in S$ can be represented as ${{\mathcal{G}}_t} = ({{\mathcal{V}}_t},{{\mathcal{E}}_t})$ with ${{\mathcal{V}}_t} = \{ v_1^t, \ldots ,v_{{N_t}}^t\}$ and ${{\mathcal{E}}_t} = \{ (v_i^t,v_j^t)\left| {v_i^t,v_j^t \in {{\mathcal{V}}_t}} \right.\}$ denoting the sets of nodes and edges. For each ${\mathcal{G}}_t$, topology is the only available information source without graph attributes. We use an adjacency matrix ${{\bf{A}}_{t}} \in \Re^{{N_t} \times {N_t}}$ to describe its topology with ${N_t}$ nodes, where ${({{\bf{A}}_t})_{ij}} = {({{\bf{A}}_t})_{ji}} = 1$ if $(v_i^t, v_j^t) \in {\mathcal{E}}_t$ and ${({{\bf{A}}_t})_{ij}} = {({{\bf{A}}_t})_{ji}} = 0$ otherwise. Since some systems allow the addition and deletion of entities, we assume different graphs in $S$ can have different node sets (i.e., $\exists {{\mathcal{G}}_t},{{\mathcal{G}}_s} \in S$ s.t.\ ${{\mathcal{V}}_t} \ne {{\mathcal{V}}_s}$) with non-fixed number of nodes $N_t$.

Graphs in $S$ can be (\romannumeral1) system snapshots evolving over time or (\romannumeral2) independent graphs of a scenario without temporal dependency and node correspondence, e.g., ego-nets in social media. To ensure the model can tackle both the cases, we assume that the node correspondence among $\{ {\mathcal{V}}_1,\cdots, {\mathcal{V}}_T\}$ is unavailable.
The proposed ICD framework follows the \textit{independent and identically distributed} (i.i.d.) hypothesis adopted in the simulation of various network systems \cite{wehrle2010modeling} that \textit{graphs in $S$ are independently generated via common underlying distributions of a system or scenario}.
%Moreover, we study the challenging case where the temporal dependency of graph topology in $S$ is unavailable, i.e., we do not consider the node index correspondence among the node sets $\{ {\mathcal{V}}_1,\cdots, {\mathcal{V}}_T\}$.

%\subsection{Combinatorial Optimization of CD}\label{Sec:NP-CD}
\textbf{Combinatorial Optimization of CD.}
Given a graph ${{\mathcal{G}}_t}$ and a pre-set number of communities $K_t$, CD aims to partition the node set ${{\mathcal{V}}_t}$ into $K_t$ subsets (i.e., communities) ${C_t} = \{ C_1^t, \cdots ,C_{{K_t}}^t\}$ so that (\romannumeral1) within each community the linkage is dense but (\romannumeral2) between different communities the linkage is relatively loose. ${C_t}$ also satisfies the disjoint constraint, i.e., $\forall r \ne s$ s.t. $C_r^t \cap C_s^t = \emptyset$. CD can be formulated as several NP-hard combinatorial optimization problems, e.g., NCut minimization \cite{von2007tutorial} and modularity maximization \cite{newman2006modularity}.

Given ${\mathcal{G}}_t$ and $K_t$, NCut minimization aims to get the CD result $C_t$ that minimizes the following NCut metric:
\begin{equation}\label{Eq:NCut-Min}
    \arg {\min}_{{C_t}}~{\mathop{\rm NCut}\nolimits} ({C_t}) = \frac{1}{2}\sum\nolimits_{r = 1}^{{K_t}} {[{\mathop{\rm cut}\nolimits}(C_r^t,\bar C_r^t)/{\mathop{\rm vol}\nolimits} (C_r^t)]}
\end{equation}
where $\bar C_r^t = {{\mathcal{V}}_t} - C_r^t$ is the complementary set of ${C_r^t}$; ${\mathop{\rm cut}\nolimits}(C_r^t,{\bar C_r^t}) = \sum\nolimits_{{v_i} \in C_r^t,{v_j} \in \bar C_r^t} {{{({{\bf{A}}_t})}_{ij}}}$ is defined as the cut between ${C_r^t}$ and ${{\bar C}_r^t}$; ${\mathop{\rm vol}\nolimits} (C_r^t) = \sum\nolimits_{{v_i} \in C_r^t, {v_j} \in \mathcal{V}_t} {{{({{\bf{A}}_t})}_{ij}}}$ is the volume of ${C_r^t}$. The objective (\ref{Eq:NCut-Min}) can be equivalently expressed in the following matrix form:
\begin{equation}\label{Eq:NCut-Min-Mat}
    \arg {\min}_{{{\bf{H}}_t}}~{\mathop{\rm tr}\nolimits} ({\bf{H}}_t^T{{\bf{L}}_t}{{\bf{H}}_t}) ~~{\rm{s}}{\rm{.t}}{\rm{. }}~{{\bf{H}}_t^T}{{\bf{H}}_t} = {{\bf{I}}_{{K_t}}},
\end{equation}
where ${{\bf{L}}_t} = {{\bf{I}}_{N_t}} - {{\bf{D}}_t^{ - {0.5}}}{{\bf{A}}_t}{{\bf{D}}_t^{ - 0.5}}$ is the normalized Laplacian matrix of ${\bf{A}}_t$; ${{\bf{D}}_t} = {\mathop{\rm diag}\nolimits} (d_1^t, \cdots ,d_{{N_t}}^t)$ is a diagonal matrix with $d_i^t = \sum\nolimits_j {{{({{\bf{A}}_t})}_{ij}}}$; ${{\bf{I}}_{N}}$ is an $N$-dimensional identity matrix. ${\bf{H}}_t \in {\Re ^{N_t \times K_t}}$ is the membership indicator, where ${({{\bf{H}}_t})_{ir}} = {[d_i^t \cdot {\mathop{\rm vol}\nolimits} {(C_r^t)^{ - 1}}]^{0.5}}$ if ${v_i^t} \in {C_r^t}$ and $({{\bf{H}}_t)_{ir}} = 0$ otherwise.

Modularity maximization is another classic NP-hard objective of CD. Given ${{\mathcal{G}}_t}$ and $K_t$, it aims to find a partition ${C_{t}}$ that maximizes the following modularity metric:
\begin{equation}\label{Eq:Mod-Max}
    \arg {\max} _{{C_t}}~{\rm{Mod}}({C_t}) = \frac{1}{{2e}}\sum\nolimits_{r = 1}^{{K_t}} {\sum\nolimits_{v_i^t,v_j^t \in C_r^t} {[{{({{\bf{A}}_t})}_{ij}} - d_i^td_j^t/(2e)]} },
\end{equation}
where $e = \sum\nolimits_i {d_i^t} /2$ is the number of edges. The objective (\ref{Eq:Mod-Max}) can also be rewritten in a matrix form:
\begin{equation}\label{Eq:Mod-Max-Mat}
    {\arg} {\min}{_{{\bf{H}}_t}} - {\rm{tr}}({\bf{H}}_t^T{{\bf{Q}}_t}{{\bf{H}}_t})~~{\rm{s}}.{\rm{t}}.~{\rm{tr}}({\bf{H}}_t^T{{\bf{H}}_t}) = {N_t},
\end{equation}
where ${{\bf{Q}}_t} \in {\Re ^{{N_t} \times {N_t}}}$ is the modularity matrix with ${({{\bf{Q}}_t})_{ij}} = {({{\bf{A}}_t})_{ij}} - d_i^td_j^t/(2e)$. ${{\bf{H}}_t} \in {\Re ^{{N_t} \times {K_t}}}$ is the membership indicator, where ${({{\bf{H}}_t})_{ir}} = 1$ if $v_i^t \in C_r^t$ and ${({{\bf{H}}_t})_{ir}} = 0$ otherwise.
%In particular, modularity measures the difference between the exact edge weights ${({\bf{A}}_{t})}_{ij}$ and the expected weights ${{{d^t_i}{d^t_j}} \mathord{\left/ {\vphantom {{{d^t_i}{d^t_j}} {(2w)}}} \right. \kern-\nulldelimiterspace} {(2w)}}$ of a randomly generated graph. Usually, larger difference (i.e., modularity) means more distinct community structure.

\textbf{Quality Evaluation Criteria.}
The NP-hardness of aforementioned objectives implies that there are so far no polynomial-time algorithms to obtain the optimal solutions. In real applications, we usually use relaxed algorithms (e.g., spectral clustering \cite{von2007tutorial} for NCut minimization) to derive feasible CD results and do not expect the optimality.
When ground-truth of real applications (e.g., cells in wireless cellular network decomposition) is available, which may not be optimal for a CD objective but is highly related to application requirements, one can evaluate CD quality by measuring the correspondence between CD results and ground-truth, where better correspondence implies higher quality. Moreover, the derived objective values of NCut minimization and modularity maximization can also be unsupervised quality metrics, where smaller NCut and larger modularity implies higher quality.

%\subsection{Transductive and Inductive Graph Embedding}\label{Sec:Ind-Emb}
\textbf{\textit{Transductive} and \textit{Inductive} Graph Embedding.}
For a graph ${{\mathcal{G}}_t}$, conventional \textit{transductive} graph embedding learns a function $f:\{ v_i^t\}  \mapsto \{ {\bf{u}}_i^t \in {\Re ^{1 \times k}}\}$ that maps each node $v_i^t$ to a $k$-dimensional vector ${\bf{u}}_i^t$, where $f$ is optimized only for ${\mathcal{G}}_t$ without parameter sharing across other graphs. In particular, the learned embedding space is expected to preserve the key properties of ${{\mathcal{G}}_t}$, where a node pair $(v_i^t,v_j^t)$ with similar properties (e.g., in the same community) should have similar representations $({\bf{u}}_i^t,{\bf{u}}_j^t)$.
The derived embeddings $\{ {\bf{u}}_i^t\}$ can be used as the input of several downstream tasks on ${{\mathcal{G}}_t}$ including CD. For each new graph $\mathcal{G}_{t'}$, \textit{transductive} methods should optimize $f$ from scratch.

To alleviate the NP-hard challenge of CD, we consider an \textit{inductive} graph embedding scheme, where parameters of $f$ are shared by all the graphs in $S$. We divide $S$ into a training set $\Gamma  \subset S$ and a test set $\Gamma' = S - \Gamma$, which represent the sets of historical known and newly generated graphs. We first train $f$ on $\Gamma$ in an \textit{offline} way to fully capture properties (i.e., underlying distributions) of the system or scenario regardless of time. After the \textit{offline} training, we directly obtain the embedding $\{ {\bf{u}}_i^{t'} \}$ from $f$ for each new graph ${{\mathcal{G}}_{t'}} \in \Gamma'$ with model parameters fixed, which significantly save the inference time. The \textit{online} CD on ${\mathcal{G}}_{t'}$ is the downstream task, where \textit{we assume the number of communities $K_{t'}$ is given}. Finally, we apply a clustering algorithm (e.g., $K$Means) to $\{ {\bf{u}}_i^{t'} \}$ for the fast high-quality \textit{online} CD.
%Our strategy of \textit{offline} training (regardless of time cost) and \textit{online} fast generalization is also adopted in the control network systems \citep{chen2018auto}.

\textbf{Sources of Training Labels.} Although graph embedding and CD are unsupervised, we assume the \textit{permutation invariant} CD result ${C_t}$ (i.e., label information) of each graph $\mathcal{G}_t \in \Gamma$ is available in the \textit{offline} training, which can be used to enhance the \textit{offline} embedding optimization.
%Note that community labels in ${C_t}$ are permutation invariant, e.g., label sequences $(0, 1, 1)$ and $(1, 0, 0)$ are the same for CD.
Note that we do not need to ensure the training labels $\{C_t | {\mathcal{G}}_t \in \Gamma \}$ are optimal for a specific CD objective. Consistent with our evaluation criteria of CD quality, $\{C_t | {\mathcal{G}}_t \in \Gamma \}$ can be from (\romannumeral1) ground-truth of real applications (e.g., cells in a wireless cellular network) or (\romannumeral2) results of a strong (but usually time-consuming) baseline with good objective values (e.g., small NCut or large modularity). Namely, we can also use the CD results of a strong baseline to regularize the \textit{offline} embedding optimization.

\section{Methodology}\label{Sec:Meth}
\begin{figure}[t]
\centering
\includegraphics[width=0.65\columnwidth, trim=19 18 18 20,clip]{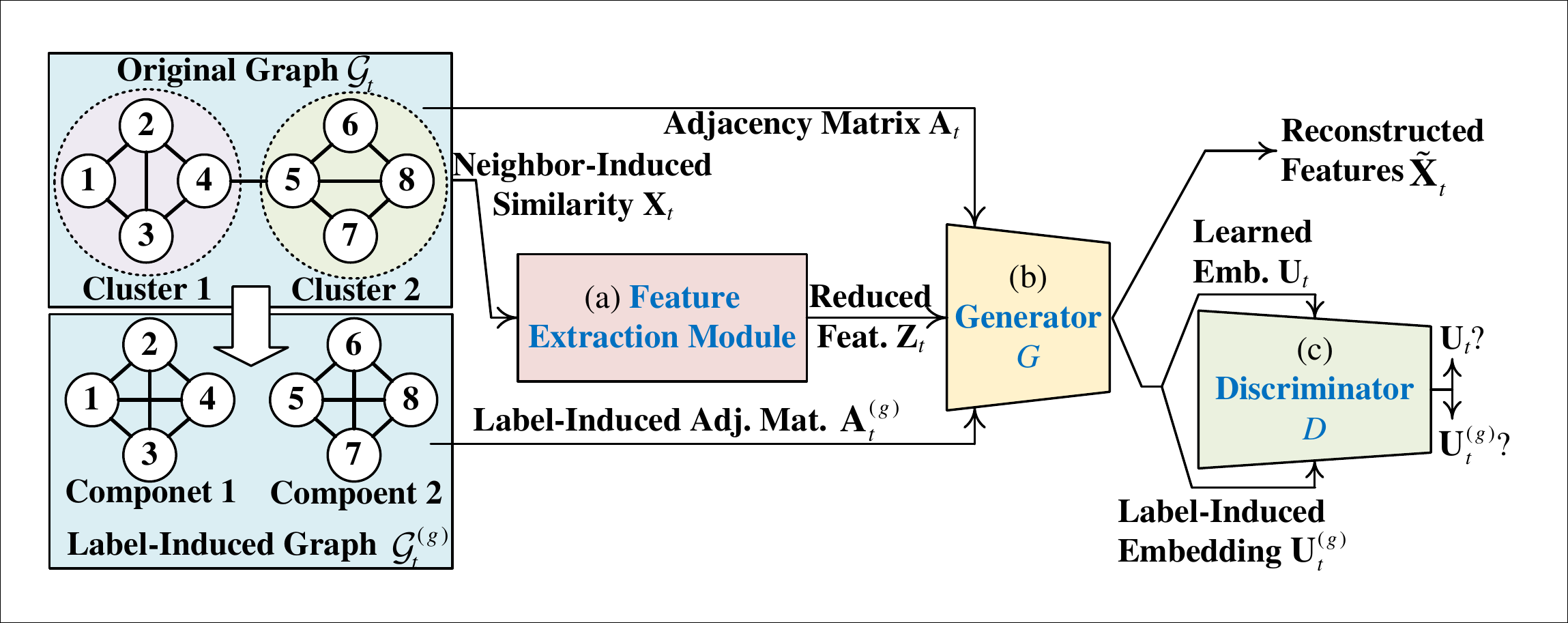}
%\vspace{-0.2cm}
\caption{Sketch of the \textit{offline} training of our adversarial dual GNN structure.
%The feature extraction module merges a graph with $8$ nodes into $2$ supernodes via a greedy 
}
\label{Fig:Example}
\vspace{-0.3cm}
\end{figure}

For a better trade-off between quality and efficiency of CD, we propose a novel ICD method with an overview shown in Fig.~\ref{Fig:Overview} (b) including the (\romannumeral1) \textit{offline} training of an adversarial dual GNN model on historical known graphs and (\romannumeral2) \textit{online} generalization to newly generated graphs via an \textit{inductive} graph embedding scheme. Fig.~\ref{Fig:Example} gives a sketch of the adversarial dual GNN with a running example for the \textit{offline} training.
In the rest of this section, we elaborate on the (\romannumeral1) model architecture as well as the (\romannumeral2) \textit{offline} training and \textit{online} generalization of ICD.

\subsection{Model Architecture}\label{Sec:Method-Model}
Inspired by adversarial auto-encoder (AAE) \cite{makhzani2015adversarial}, we introduce a novel adversarial dual GNN structure. As shown in Fig.~\ref{Fig:Example}, it includes (a) a feature extraction module, (b) a generator $G$, and (c) a discriminator $D$.
%$G$ consists of a \textit{feature encoder} and an \textit{auxiliary label-induced encoder} which are both multi-layer GNNs with shared parameters, while $D$ is an fully-connected network.

To the best of our knowledge, the adversarial dual GNN is our original design, which is different from existing AAE-based graph embedding methods \cite{pan2018adversarially} that regularize the learned embedding using a prior distribution. In contrast, our dual GNN incorporates the permutation invariant label information of historical graphs $\{ {\mathcal{G}}_t \in \Gamma \}$ to the \textit{offline} training via a novel adversarial process between (\romannumeral1) original training graphs $\{ {\mathcal{G}}_t \}$ and (\romannumeral2) corresponding \textit{auxiliary label-induced graphs} $\{ {\mathcal{G}}_t^{(g)} \}$.

\textbf{(a) Feature Extraction Module.}
Most GNNs are originally designed for attributed graphs where each node $v_i^t$ has a feature vector input described by the $i$-th row of a feature matrix ${\bf{X}}_t$. In this study, we adopt GNN as a basic building block of ICD but consider CD without attributes. Instead of using the classic settings of GNN for graphs without attributes (e.g., use a constant matrix as node features \cite{xu2018powerful}), we extract additional structural features ${\bf{X}}_t$ based on the neighbor-induced similarity encoded in classic CD objectives of modularity maximization (\ref{Eq:Mod-Max-Mat}) and NCut minimization (\ref{Eq:NCut-Min-Mat}), resulting in two variants of ICD.

For modularity maximization (\ref{Eq:Mod-Max-Mat}), the modularity matrix ${{\bf{Q}}_t}$ encodes the neighbor similarity of each graph $\mathcal{G}_t$, so we let ${{\bf{X}}_t} = {{\bf{Q}}_t}$. For NCut minimization (\ref{Eq:NCut-Min-Mat}), the Laplacian matrix ${\bf{L}}_t$ gives primary characteristics of graph structure, where ${{\bf{M}}_t} = {\bf{D}}_t^{ - 0.5}{{\bf{A}}_t}{\bf{D}}_t^{ - 0.5}$ is the key component regarding neighbor similarity, so we let ${{\bf{X}}_t} = {{\bf{M}}_t}$. Both ${\bf{Q}}_t$ and ${\bf{M}}_t$ are the reweighting of ${\bf{A}}_t$, where nodes $(v_i^t, v_j^t)$ with similar neighbor-induced features $({({{\bf{X}}_t)}_{i,:}}, {({{\bf{X}}_t)}_{j,:}})$ are more likely to be partitioned into the same community.
Prior work \cite{tian2014learning,yang2016modularity}, which uses ${{\bf{Q}}_t}$ or ${{\bf{M}}_t}$ to learn graph embeddings based on auto-encoders, has validated our motivation to use ${{\bf{Q}}_t}$ and ${{\bf{M}}_t}$ as informative structural features.
Our experiments also demonstrate that the extraction of ${\bf{X}}_t$ is essential to supporting the high-quality \textit{online} CD.

However, the feature dimensionality of ${\bf{X}}_t \in \{ {\bf{Q}}_t, {\bf{M}}_t \}$ is the number of nodes ${N_t}$, which may not be fixed over $t$. Since the \textit{inductiveness} of GNNs relies on the fixed feature dimensionality of ${\bf{X}}_t$, we introduce an efficient feature extraction module via the heavy-edge matching (HEM) graph coarsening \cite{hendrickson1995multi}, which enable ICD to tackle CD across graphs with non-fixed $N_t$. It maps ${{\bf{X}}_t} \in \Re^{{N_t} \times {N_t}}$ to another feature matrix ${\bf{Z}}_t \in \Re^{{N_t} \times {L}}$ with fixed feature dimensionality $L$. Algorithm~\ref{Alg:Feat} summarizes the feature extraction procedure.

\begin{figure}[t]
\centering
\includegraphics[width=0.7\columnwidth, trim=18 18 18 18,clip]{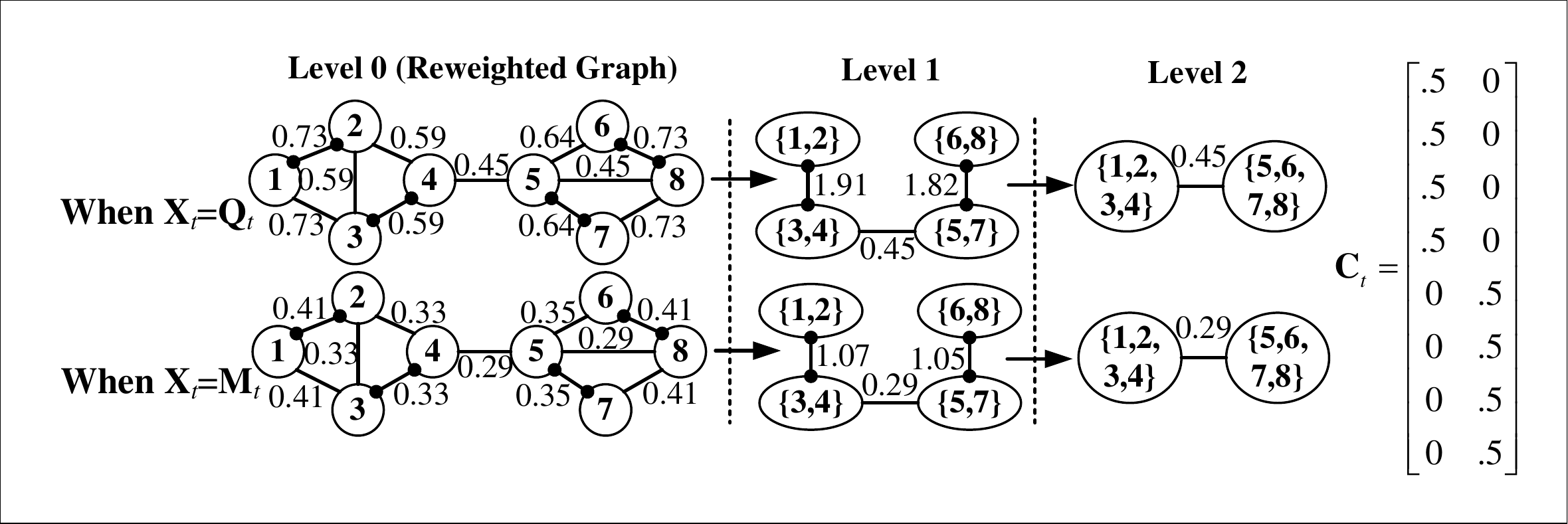}
\vspace{-0.3cm}
\caption{Running examples of the feature extraction that merges a graph with $8$ nodes into $2$ supernodes.}
\label{Fig:Feat-Example}
\vspace{-0.2cm}
\end{figure}

When ${N_t}>L$, we first extract reweighted edges ${\mathcal{E}}_t^{w} = \{ w(v_i^t,v_j^t)|w(v_i^t,v_j^t) = {({{\bf{X}}_t})_{ij}}, {({{\bf{A}}_t})_{ij}} = 1\}$ and then apply HEM to ${\mathcal{E}}_t^{w}$. For simplicity, we summarize the HEM procedure in Appendix~\ref{Sec:HEM}. Fig.~\ref{Fig:Feat-Example} gives running examples of HEM w.r.t. the input graph in Fig.~\ref{Fig:Example}. Concretely, HEM merges the original graph ${{\mathcal{G}}_t}$ with ${N_t}$ nodes into a supergraph ${\mathcal{G}}_t^*$ with $L$ supernodes via a greedy multi-level strategy of continuously merging the node pair with largest weight (in the original graph or induced supergraph) into a supernode. For instance, in Fig.~\ref{Fig:Feat-Example}, it merges a graph with $8$ nodes into $2$ supernodes $v_1^*=\{ 1, 2, 3, 4\}$ and $v_2^*=\{ 5, 6, 7, 8 \}$ via the coarsening with two levels. HEM finally outputs a \textit{coarsening matrix} ${{\bf{C}}_t} \in \Re^{{N_t} \times L}$, where ${({{\bf{C}}_t})_{ij}} = |v_j^{t*}{|^{ - 0.5}}$ if node $v_i^t$ is merged into supernode $v_j^{t*}$ and $({\bf{C}}_t)_{ij} = 0$ otherwise. We then let ${{\bf{Z}}_t} = {{\bf{X}}_t}{{\bf{C}}_t}$ be the reduced features. Since ${{\bf{C}}_t}$ is a sparse matrix,
%describing the merging membership, 
one can obtain ${{\bf{Z}}_t}$ by setting its $j$-th column to ${({{\bf{Z}}_t})_{:,j}} = \sum\nolimits_{{v_i} \in v_j^{t*}} {{{({{\bf{C}}_t})}_{ij}}{{({{\bf{X}}_t})}_{:,i}}}$.
When ${N_t} \le L$, we set ${{\bf{Z}}_t} = [{{\bf{X}}_t},{{\bf{0}}_{{N_t} \times (L - {N_t})}}]$ via a padding strategy. 

To the best of our knowledge, using HEM to extract feature input ${\bf{Z}}_t$ for \textit{inductive} GNNs is our original design. It is more efficient than the dimension reduction (e.g., PCA of ${{\bf{A}}_t}$) used in existing GNN-based approaches \cite{nazi2019gap,wilder2019end} since HEM is widely adopted in some fast \textit{transductive} methods \cite{dhillon2007weighted} with low complexity. The derived ${\bf{Z}}_t$ is also more informative than the node features in some classic settings of inductive GNNs (e.g., let ${\bf{Z}}_t$ be a constant matrix \cite{xu2018powerful}). The extracted ${\bf{Z}}_t$ is further fed into the generator $G$.

\begin{algorithm}[t]\small
\caption{Feature Extraction of ICD}
\label{Alg:Feat}
\LinesNumbered
\KwIn{input graph ${\mathcal{G}}_t$; neighbor-induced features ${{\bf{X}}_t}$; number of nodes ${N_t}$; reduced dimensionality $L$}
\KwOut{reduced features ${{\bf{Z}}_t} \in \Re^{N_t \times L}$}
\If{${N_t} > {L}$}
    {
        get supernode membership ${\mathcal{V}}_t^*$ \& coarsening matrix ${{\bf{C}}_t}$ via HEM (see Algorithm~\ref{Alg:HEM} in Appendix~\ref{Sec:HEM})\\
        \For{{\bf{each}} supernode $v_j^* \in {\mathcal{V}}_t^*$}
        {
            Initialize $j$-th column of ${\bf{Z}}_t$:
            ${({{\bf{Z}}_t})_{:,j}} \leftarrow {{\bf{0}}_{{N_t} \times 1}}$\\
            \For{{\bf{each}} node $v_i \in v_j^*$}
            {
                ${({{\bf{Z}}_t})_{:,j}} \leftarrow {({{\bf{Z}}_t})_{:,j}} + {({{\bf{C}}_t})_{ij}}{({{\bf{X}}_t})_{:,i}}$
            }
        }
    }
    \ElseIf{${N_t} \le {L}$}
    {
        Pad ${\bf{X}}_t$ with zeros: ${{\bf{Z}}_t} \leftarrow [{{\bf{X}}_t},{{\bf{0}}_{{N_t} \times (L - {N_t})}}]$
    }
\end{algorithm}

\begin{figure}[t]
\centering
\includegraphics[width=0.75\columnwidth, trim=19 18 18 15,clip]{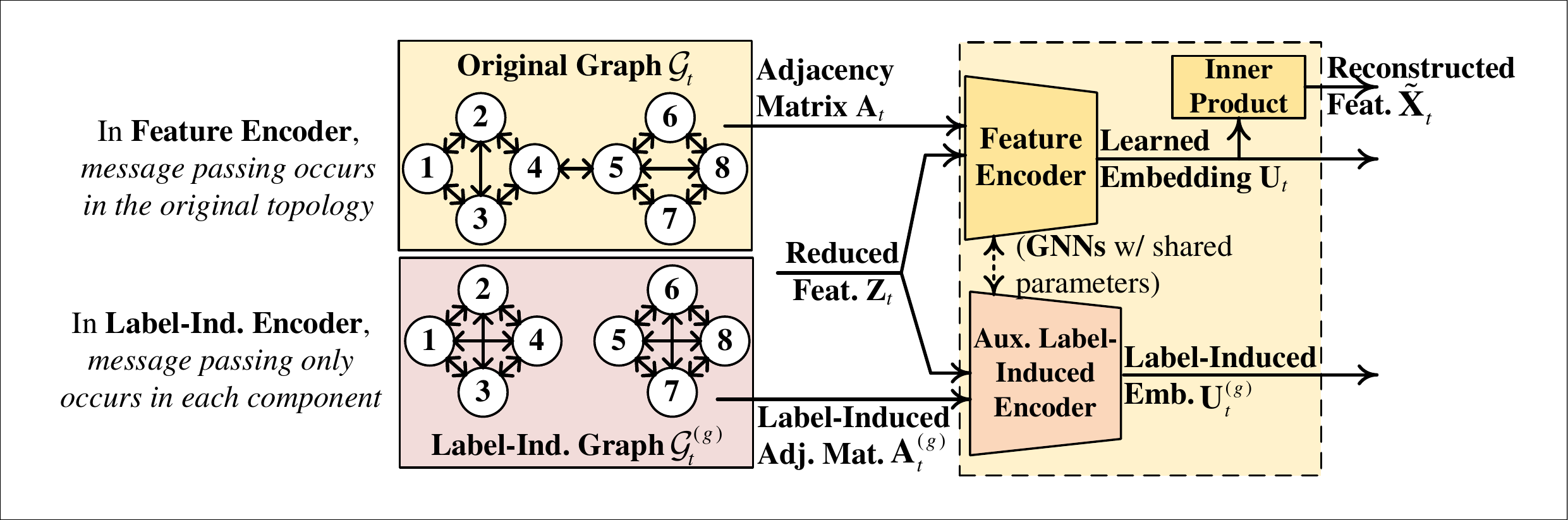}
\vspace{-0.2cm}
\caption{Details of the generator $G$ w.r.t. the running example in Fig.~\ref{Fig:Example}.}
\label{Fig:Gen-Example}
\vspace{-0.3cm}
\end{figure}

\textbf{(b) Generator $G$.}
As highlighted in Fig.~\ref{Fig:Gen-Example}, $G$ consists of a \textit{feature encoder} and an \textit{auxiliary label-induced encoder} which are multi-layer GNNs with shared parameters ${\delta _G}$. Both the encoders have the topology and feature inputs described by an adjacency matrix and a feature matrix. Concretely, the \textit{feature encoder} takes adjacency matrix ${{\bf{A}}_t}$ (w.r.t. original graph $\mathcal{G}_t$) and reduced features ${{\bf{Z}}_t}$ as inputs, and derives graph embedding ${\bf{U}}_t$. The \textit{label-induced encoder} takes adjacency matrix ${\bf{A}}_t^{(g)}$ (w.r.t. an \textit{auxiliary label-induced graph} $\mathcal{G}_t^{(g)}$) and ${{\bf{Z}}_t}$ as inputs, and outputs \textit{label-induced embedding} ${\bf{U}}_t^{(g)}$. For simplicity, we denote the \textit{feature} and \textit{label-induced encoders} as ${{\bf{U}}_t} = G({{\bf{A}}_t},{{\bf{Z}}_t};{\delta _G})$ and ${{\bf{U}}_t^{(g)}} = G({\bf{A}}_t^{(g)},{{\bf{Z}}_t};{\delta _G})$.

In the \textit{offline} training, we assume that the topology and CD result (used as training `ground-truth') of each historical graph ${{\mathcal{G}}_t} \in \Gamma$ are available, which are described by an adjacency matrix ${{\bf{A}}_t} \in \Re^{{N_t} \times {N_t}}$ and an \textit{indicator matrix} ${{\bf{R}}_t} \in \Re^{{N_t} \times {K_t}}$. We define $({\bf{R}}_t)_{ir}=1$ if node ${v_i^t}$ is in community $C_r^t$ by the training `ground-truth' and $({\bf{R}}_t)_{ir}=0$ otherwise. 
Due to the disjoint constraint of CD, only one entry in each row of ${{\bf{R}}_t}$ is $1$ with other entries in the same row set to $0$. In addition to the original graph ${\mathcal{G}}_t$, an \textit{auxiliary label-induced graph} ${\mathcal{G}}_t^{(g)}$, with ${\bf{A}}_t^{(g)} = {{\bf{R}}_t}{\bf{R}}_t^{\rm T}$ as the adjacency matrix describing its topology, is introduced to encode the structure of `ground-truth' based on the following \textbf{Fact 1}.

\textbf{(Fact 1)} \textit{For a graph ${{\mathcal{G}}_t}$ with ${K_t}$ communities, its \textit{auxiliary label-induced graph} ${\mathcal{G}}_t^{(g)}$ has ${K_t}$ fully connected components with each component corresponding to one unique community of ${{\mathcal{G}}_t}$. Namely, there is an edge between nodes $v_i^t$ and $v_j^t$ in ${\mathcal{G}}_t^{(g)}$ only when they are in the same community.}

The proof of \textbf{Fact 1} is given in Appendix~\ref{Sec:Proof}. For the example in Fig.~\ref{Fig:Gen-Example} with $8$ nodes $\{1, 2, \cdots, 8\}$, ${\mathcal{G}}_t^{(g)}$ has $2$ fully connected components $\{1, 2, 3, 4\}$ and $\{5, 6, 7, 8\}$ w.r.t. the $2$ communities in the given CD `ground-truth' of ${\mathcal{G}}_t$.

Let ${\bf{F}}_t^{(l-1)}$ and ${\bf{F}}_t^{(l)}$ be the input and output of the $l$-th GNN layer of the \textit{feature} or \textit{label-induced encoder} in $G$ with ${\bf{F}}_t^{(0)} = {{\bf{Z}}_t}$. The $l$-th GNN layer is defined as
\begin{equation}\label{Eq:GCN}
    {\bf{F}}_t^{(l)} = \tanh ({\bf{\hat D}}_t^{ - 0.5}{{{\bf{\hat A}}}_t}{\bf{\hat D}}_t^{ - 0.5}{\bf{F}}_t^{(l - 1)}{\bf{W}}_G^{(l - 1)}),
\end{equation}
where ${{{\bf{\hat A}}}_t} = {{\bf{A'}}_t} + {{\bf{I}}_{{N_t}}}$ (${{{\bf{A'}}}_t} \in \{ {{\bf{A}}_t},{\bf{A}}_t^{(g)}\}$) is the adjacency matrix with self-edges; ${{{\bf{\hat D}}}_t} = {\mathop{\rm diag}\nolimits} (\hat d_1^t, \cdots ,\hat d_{{N_t}}^t)$ is the diagonal degree matrix; ${\bf{W}}_G^{(l - 1)}$ is the trainable parameter shared by the two encoders; $({\bf{F}}_t^{(l)})_{i,:}$ is the latent feature of node $v_i^t$.
In particular, $({\bf{F}}_t^{(l)})_{i,:}$ is the nonlinear aggregation of features of $\{ v_i^t\}  \cup n(v_i^t)$ from the previous layer, where $n(v_i^t)$ is the neighbor set of $v_i^t$. It is also known as the \textit{message passing} of GNN, where neighbors of $v_i^t$ propagate their features to $v_i^t$ for aggregation. By \textbf{Fact 1}, ${\bf{A}}_t^{(g)}$ ensures that \textit{\textit{message passing} of the \textit{label-induced encoder} only occurs in each connected component w.r.t. each community in the `ground-truth'} (e.g., $2$ components as highlighted in Fig.~\ref{Fig:Gen-Example}), while \textit{\textit{message passing} of the \textit{feature encoder} occurs in the original topology of ${\mathcal{G}}_t$}.
%Hence, ${{\bf{U}}_t^{(g)}} = G({\bf{A}}_t^{(g)},{{\bf{X}}_t})$ preserves more informative label-induced properties than ${{\bf{U}}_t} = G({{\bf{A}}_t},{{\bf{X}}_t})$. 
Here we use GCN \cite{kipf2016semi} as an example building block of the two encoders. One can easily extend ICD to include other advanced \textit{inductive} GNNs, e.g., GAT \cite{velivckovic2017graph} and GIN \cite{xu2018powerful}.

\begin{comment}
Furthermore, we also apply the $l_2$-normalization to each column of ${{\bf{F}}_t^{(l)}}$ before feeding it to the next GNN layer:
\begin{equation}
    {({\bf{F}}_t^{(l)})_{:,r}} = {({\bf{F}}_t^{(l)})_{:,r}} \cdot {[\sum\nolimits_i {({\bf{F}}_t^{(l)})_{ir}^2} ]^{ - 0.5}}.
\end{equation}
\end{comment}

The last layers of the \textit{feature} and \textit{label-induced encoders} output the learned graph embedding ${{\bf{U}}_t} \in \Re^{{N_t} \times k}$ and \textit{auxiliary label-induced embedding} ${{\bf{U}}_t^{(g)}} \in \Re^{{N_t} \times k}$, where \textit{${{\bf{U}}_t^{(g)}}$ preserves more informative label-induced properties than ${{\bf{U}}_t}$}. We further use the non-linear inner product of ${{\bf{U}}_t}$ to reconstruct the neighbor-induced features ${{\bf{X}}_t}$:
\begin{equation}\label{Eq:Decoder}
    {{{\bf{\tilde X}}}_t} = \tanh ({{\bf{U}}_t}{\bf{U}}_t^{\mathop{\rm T}\nolimits} ).
\end{equation}
The derived embeddings $\{ {\bf{U}}_t, {\bf{U}}_t^{(g)} \}$ are then in turns fed into the discriminator $D$ to support the adversarial process between $G$ and $D$, while ${\bf{\tilde X}}$ is further used to regularized ${\bf{U}}_t$ in the \textit{offline} optimization.

%\subsubsection{Discriminator $D$}\label{Sec:Meth-D}
\textbf{(c) Discriminator $D$.}
$D$ is an auxiliary classifier to distinguish ${\bf{U}}_t^{(g)}$ from ${\bf{U}}_t$, while $G$ tries to generate plausible embedding ${\bf{U}}_t$ to fool $D$. Such an adversarial process helps $G$ to generate embedding ${\bf{U}}_t$ close to ${\bf{U}}_t^{(g)}$ which encodes the structures of training `ground-truth'. 

For simplicity, we denote $D$ as ${{\bf{y}}_t} = D({{\bf{S}}_t};{\delta _D})$, where ${{\bf{S}}_t} \in \{ {\bf{U}}_t , {\bf{U}}_t^{(g)}\}$ and $\delta_{D}$ are embedding input and set of model parameters; ${{\bf{y}}_t} \in \Re^{N_t}$ is a column vector with $({\bf{y}}_t)_i$ as the probability that $({{\bf{S}}_t})_{i,:} = ({\bf{U}}_t^{(g)})_{i,:}$ rather than $({{\bf{S}}_t})_{i,:} = ({\bf{U}}_t)_{i,:}$. Let ${\bf{P}}_t^{(l-1)}$ and ${\bf{P}}_t^{(l)}$ be input and output of the $l$-th layer in $D$ with ${\bf{P}}_t^{(0)} = {{\bf{S}}_t}$. The $l$-th layer of $D$ is defined as
\begin{equation}
    {\bf{P}}_t^{(l)} = {\mathop{\rm ReLU}\nolimits} ({\bf{P}}_t^{(l - 1)}{\bf{W}}_D^{(l - 1)} + {\bf{b}}_D^{(l - 1)}),
\end{equation}
where $\{{\bf{W}}_D^{(l-1)},{\bf{b}}_D^{(l-1)}\}$ are trainable model parameters. 
%Suppose there are $L_D$ layers in $D$. 
In particular, we use $\rm{sigmoid}$ as activation function of the last layer instead of $\rm{ReLU}$.
%i.e., ${y_t} = {\mathop{\rm sigmoid}\nolimits} ({\bf{P}}_t^{(L_D)})$.

\subsection{Offline Training and Online Generalization}\label{Sec:Meth-Opt}
In this subsection, we elaborate on the \textit{offline} training and \textit{online} generalization procedures of ICD across a set of graphs $S = \{\mathcal{G}_1, \cdots, \mathcal{G}_T \}$. For simplicity, we divide $S$ into a training set $\Gamma_{\rm{T}}$, a validation set $\Gamma_{\rm{V}}$, and a test set $\Gamma'$, where $\Gamma = \Gamma_{\rm{T}}\cup \Gamma_{\rm{V}}$ and $\Gamma'$ represent the sets of historical known and newly generated graphs, respectively. 
%Namely, graphs in $\Gamma$ should occur before those in $\Gamma'$.
%Since we assume the CD results ${C_t} = \{ C^t_{1}, \cdots, C^t_{K_t}\}$ (used as training `ground-truth' with two possible sources as defined in Section~\ref{Sec:Prob-Pre}) are available for graphs in $\Gamma$, we correlate each graph ${{\mathcal{G}}_t} \in \Gamma$ with a $2$-tuple $( {{\bf{A}}_{t}}, {C_t})$.

The \textit{offline} training of ICD is based on three objectives of (\romannumeral1) adversarial learning (AL), (\romannumeral2) feature reconstruction (FR), and (\romannumeral3) clustering regularization (CR).

The AL objective tries to incorporate the permutation invariant label information of historical graphs to the \textit{offline} embedding optimization via an adversarial process between $D$ and $G$. On the one hand, $D$ tries to distinguish ${\bf{U}}_t^{(g)}$ from ${\bf{U}}_t$. The objective of $D$ w.r.t. a graph ${\mathcal{G}}_t$ is
\begin{equation}\label{Eq:D-loss}
    \arg {\min}_{{\delta _D}}~{L_D}({{\mathcal{G}}_t}) =  - [\sum\nolimits_i {\log (1 - D{{({{\bf{U}}_t})}_i})}  + \sum\nolimits_i {\log D{{({\bf{U}}_t^{(g)})}_i}} ]/{N_t}
\end{equation}
On the other hand, $G$ tries to fool $D$ by minimizing the following loss w.r.t. a graph $\mathcal{G}_t$:
%\begin{equation}\label{Eq:AL-loss}
%    {L_{\rm AL}}({{\mathcal{G}}_t}) =  - \log D({{\bf{U}}_t};{\delta _G}).
%\end{equation}
\begin{equation}\label{Eq:AL-loss}
    {L_{{\rm{AL}}}}({\mathcal{G}_t}) =  - [ {\sum\nolimits_i {\log D({{\bf{U}}_t})_i} }]/{N_t}.
\end{equation}
Such an adversarial process directs $G$ to output the embedding ${{\bf{U}}_t}$ close to ${\bf{U}}_t^{(g)}$, enabling $G$ to capture the permutation invariant training labels of historical graphs.

The FR loss further forces $G$ to derive embedding ${\bf{U}}_t$ that preserves key properties of the structural features ${\bf{X}}_t$ by minimizing the reconstruction error between ${{\bf{X}}_t}$ and ${\bf{\tilde X}}_t$:
\begin{equation}\label{Eq:FR-loss}
    {L_{\rm{FR}}}({{\mathcal{G}}_t}) = ||{{{\bf{\tilde X}}}_t} - {{\bf{X}}_t}||_F^2 = ||\tanh ({\bf{U}}_t {\bf{U}}_t^{\rm{T}}) - {{\bf{X}}_t} ||_F^2.
\end{equation}

In addition to the AL loss, ICD can also capture the permutation invariant training labels via the regularization of some classic CD objectives (e.g., modularity maximization and NCut minimization) corresponding to the two variants (with ${\bf{X}}_t = {\bf{Q}}_t$ and ${\bf{X}}_t = {\bf{M}}_t$) as described in Section~\ref{Sec:Method-Model}. We introduce the CR objective w.r.t. a graph ${\mathcal{G}}_t$ that minimizes the following loss:
\begin{equation}\label{Eq:CR-loss}
    {L_{\rm{CR}}}({{\mathcal{G}}_t}) =  - {\mathop{\rm tr}\nolimits} ({\bf{H}}_t^{\rm{T}}{{{\bf{\tilde X}}}_t}{{\bf{H}}_t}),
\end{equation}
where ${{\bf{H}}_t}$ is the \textit{membership indicator} encoding the training `ground-truth' with same definitions as (\ref{Eq:Mod-Max-Mat}) and (\ref{Eq:NCut-Min-Mat}). Since the constraints on ${{\bf{H}}_t}$ (i.e., ${\mathop{\rm tr}\nolimits} ({\bf{H}}_t^T{{\bf{H}}_t}) = {N_t}$ and ${\bf{H}}_t^T{{\bf{H}}_t} = {{\bf{I}}_{{K_t}}}$) are always satisfied for the given `ground-truth', we do not need to consider discrete constraints on ${{\bf{H}}_t}$. To the best of our knowledge, we are the first to use the CR objective to incorporate permutation invariant label information (encoded by ${\bf{H}}_t$ of training data) to the unsupervised embedding learning, whereas ${\bf{H}}_t$ is the parameter to be optimized or output in existing CD methods \cite{nazi2019gap,wilder2019end}.

Finally, we derive the objective of $G$ w.r.t. each ${{\mathcal{G}}_t} \in \Gamma_{\rm{T}}$ by combining (\ref{Eq:AL-loss}), (\ref{Eq:FR-loss}), and (\ref{Eq:CR-loss}):
\begin{equation}\label{Eq:G-loss}
    \mathop {\arg \min }{_{\delta _G}} {L_G}({{\mathcal{G}}_t}) = {L_{{\mathop{\rm AL}\nolimits} }}({{\mathcal{G}}_t}) + \alpha {L_{{\mathop{\rm FR}\nolimits} }}({{\mathcal{G}}_t}) + \beta {L_{\rm CR}}({{\mathcal{G}}_t}),
\end{equation}
where $\{ \alpha ,\beta \}$ are parameters to balance $L_{\rm FR}$ and $L_{\rm CR}$.

In the joint \textit{offline} optimization of $D$ and $G$, we use the Xavier method \citep{glorot2010understanding} 
to initialize model parameters $\{ \delta_{D}, \delta_G \}$. The Adam optimizer is applied to iteratively update $\delta_D$ and $\delta_G$ based on gradients of (\ref{Eq:D-loss}) and (\ref{Eq:G-loss}). In each epoch, we randomly sample a certain number $p$ of historical graphs from $\Gamma_{\rm{T}}$ to update $\{ \delta_D, \delta_G\}$. For a selected training sample in each epoch, we can repeatedly updating model parameters $m \ge 1$ times. Finally, we save the model parameters $\{ \delta_D^* , \delta_G^*\}$ with the best average CD quality on $\Gamma_{\rm{V}}$ within a pre-set number of epochs $n$, based on the quality evaluation criteria defined in Section~\ref{Sec:Prob-Pre}.

After the \textit{offline} training, we generalize ICD to new graphs for fast \textit{online} CD. Concretely, for each new graph ${\mathcal{G}}_{t'} \in \Gamma'$, we derive embedding ${\bf{U}}_{t'}$ by passing $\{ {\bf{A}}_{t'}, {\bf{X}}_{t'}\}$ forward the \textit{feature encoder} of $G$ with its parameters $\delta_G^*$ fixed. Since we assume \textit{the number of communities $K_{t'}$ is given}, we apply $K$Means to ${\bf{U}}_{t'}$, which outputs the CD result w.r.t. the given $K_{t'}$. Hence, the runtime of the \textit{online} CD on each graph ${\mathcal{G}}_{t'} \in \Gamma'$ includes (\romannumeral1) the feature extraction of ${\bf{Z}}_{t'}$, (\romannumeral2) one feedforward propagation through the \textit{feature encoder}, and (\romannumeral3) downstream clustering. 
%ICD also has the potential to estimate the proper $K_{t'}$ value by applying an auxiliary model selection method, e.g., XMenas \citep{pelleg2000x}, to ${\bf{U}}_{t'}$.

In summary, we conclude the \textit{offline} training and \textit{online} generalization procedures of ICD in Algorithm~\ref{Alg:Train} and \ref{Alg:Gen}.

\begin{algorithm}[t]\small
\caption{\textit{Offline} Training of ICD}
\label{Alg:Train}
\LinesNumbered
\KwIn{training set ${\Gamma_{\rm{T}}}$; validation set ${\Gamma_{\rm{V}}}$; feature dimensionality $L$; hyper-parameters $\{ \alpha, \beta, p, m\}$; learning rates $\{ {\eta_{D}}, {\eta_{G}}\}$; number of epochs $n$}
\KwOut{optimized model parameters $\{ {\delta^*_{G}}, {\delta^*_{D}} \}$}
%initialize ${{\bar q}^*}$ that saves best average quality metric\\
initialize model parameters $\{ {\delta_{G}}, {\delta_{D}} \}$ via Xavier method\\
\For{$epoch$ {\bf{from}} $1$ {{\bf{to}}} $n$}
{
    \For{$sample\_count$ {\bf{from}} $1$ {\bf{to}} $p$}
    {
        randomly sample ${\mathcal{G}}_t \in \Gamma_{\rm{T}}$\\
        get structural features ${{\bf{X}}_t}$ \& indicator ${\bf{H}}_t$ w.r.t.${\mathcal{G}}_t$\\
        get reduced features ${{\bf{Z}}_t}$ from $\{ {{\bf{A}}_t}, {{\bf{X}}_t} \}$ via Algorithm~\ref{Alg:Feat}\\
        \For{$update\_count$ {\bf{from}} $1$ {\bf{to}} $m$}
        {
            update ${\delta _D}$ via ${\delta _D} \leftarrow {\mathop{\rm Opt}\nolimits} ({\eta _D},{\delta _D},\partial {L_D}/\partial {\delta _D})$ with $\delta_G$ fixed\\
            update ${\delta _G}$ via ${\delta _G} \leftarrow {\mathop{\rm Opt}\nolimits} ({\eta _G},{\delta _G},\partial {L_G}/\partial {\delta _G})$ with $\delta_D$ fixed\\
        }
    }
    %$\bar q \leftarrow 0${\footnotesize{//Initialize current epoch's average quality metric}}\\
    \For{{\bf{each}} ${\mathcal{G}}_{t'} \in \Gamma_{\rm{V}}$}
    {
        get structural features ${{\bf{X}}_{t'}}$ w.r.t. $\mathcal{G}_t$ \\
        get reduced features ${{\bf{Z}}_{t'}}$ from $\{{{\bf{A}}_{t'}},{{\bf{X}}_{t'}}\}$ via Algorithm~\ref{Alg:Feat}\\
        get graph embedding ${{\bf{U}}_{t'}}$ from \textit{feature encoder}\\
        apply $K$Means to ${{\bf{U}}_{t'}}$ to get CD result ${\tilde C_{t'}}$\\
        evaluate CD quality $q$ w.r.t. ${\tilde C_{t'}}$\\
        %$\bar q \leftarrow \bar q + q$ //Update $\bar q$\\
        update average quality $\bar q$ on $\Gamma_{\rm{V}}$
    }
    %$\bar q \leftarrow {{\bar q} \mathord{\left/ {\vphantom {{\bar q} {|{P_V}|}}} \right. \kern-\nulldelimiterspace} {|{\Gamma_V}|}}$//Compute average quality metric\\
    \If{\bf{have} \bf{better} $\bar q$}
    {
      Update best model param.: $\{ \delta _D^*,\delta _G^*\} \leftarrow \{ {\delta _D},{\delta _G}\}$
    }
}
\end{algorithm}

\begin{algorithm}[t]\small
\caption{\textit{Online} Generalization of ICD}
\label{Alg:Gen}
\LinesNumbered
\KwIn{new graph ${{\mathcal{G}}_{t'}} \in {\Gamma'}$, feature dimensionality $L$, number of communities $K_{t'}$}
\KwOut{\textit{online} CD result ${\tilde C_{t'}}$ w.r.t. ${{\mathcal{G}}_{t'}}$}
get structural features ${{\bf{X}}_{t'}}$ w.r.t. $\mathcal{G}_{t'}$\\
get reduced features ${{\bf{Z}}_{t'}}$ from $\{ {{\bf{A}}_{t'}}, {{\bf{X}}_{t'}}\}$ via Algorithm~\ref{Alg:Feat}\\
get graph embedding ${{\bf{U}}_{t'}}$ from \textit{feature encoder}\\
apply $K$Means to ${{\bf{U}}_{t'}}$ to get CD result $\tilde C_{t'}$
\end{algorithm}

For each graph ${{\mathcal{G}}_t}$, the complexity to get reweighted edges $\mathcal{E}_t^w$ is $O(|\mathcal{E}_t|)$. When $N_t>L$, there are in total $(N_t-L)$ merging operations for graph coarsening. In each level of graph coarsening, we continuously select the edge with largest weight to merge all the possible node pair, where the number of edges in a new level is much less than that in the previous level, i.e., $|\mathcal{E}_t^{(k)}|>|\mathcal{E}_t^{(k+1)}|$. The total cost of selecting edges with largest weights is no more than $O(|\mathcal{E}_t|+|\mathcal{E}_t^{(0)}|log|\mathcal{E}_t^{(0)}|+|\mathcal{E}_t^{(1)}|log|\mathcal{E}_t^{(1)}|+\cdots) = O(|\mathcal{E}_t|log|\mathcal{E}_t|)$. Hence, the complexity of graph coarsening (i.e., Algorithm~\ref{Alg:HEM}) is no more than $O((N_t-L)+|\mathcal{E}_t|log|\mathcal{E}_t|)$. Moreover, the complexity to derive ${{\bf{Z}}_t}$ via ${\bf{X}}_t$ and ${\bf{C}}_t$ is no more than $O({N_t}L{n^*})$, with $n^*$ as the maximum number of nodes merged into a supernode. 
In summary, the total complexity to derive the reduced feature matrix ${\bf{Z}}_t$ based on ${{\bf{X}}_t}$ is no more than $O((N_t-L)+|\mathcal{E}_t|log|\mathcal{E}_t|+{N_t}L{n^*})=O(|\mathcal{E}_t|log|\mathcal{E}_t|+{N_t}L{n^*}) \approx O(|\mathcal{E}_t|log|\mathcal{E}_t| + {N_t})$ with $L \ll N_t$ and $n^* \ll N_t$.

To extract ${\bf{M}}_t$ and $\bf{Q}_t$ from ${\bf{A}}_t$, we first derive node degrees ${\bf{d}}_t = [d_1^t, \cdots ,d_{{N_t}}^t]$ with cost of $O(|\mathcal{E}_t|)$ by utilizing the sparsity of ${\bf{A}}_t$. Since ${\bf{M}}_t$ is also a sparse matrix, the complexity to derive ${\bf{M}}_t$ is no more than $O(|\mathcal{E}_t|)$. Although the complexity to derive ${\bf{Q}}_t$ is $O(N_t^2)$, we use the efficient matrix multiplication ${{\bf{Q}}_t} = {{\bf{A}}_t} - {\bf{d}}_t^{\rm{T}}{{\bf{d}}_t}/(2w)$, which can be easily paralleled via GPUs with significant speedup rate.

Let $L^{(l-1)}$ be the feature dimensionality of the $l$-th GNN layer of encoder in $G$, with $L^{(0)}=L$. We used the efficient sparse-dense matrix multiplication to implement the graph convolutional operation described in (\ref{Eq:GCN}). The complexity of one feedforward propagation through the \textit{feature encoder} is no more than $O(|{{\mathcal{E}}_t}|{L^{(0)}}{L^{(1)}} + |{{\mathcal{E}}_t}|{L^{(1)}}{L^{(2)}} +  \cdots ) = O(|{{\mathcal{E}}_t}|LL')$, with $L' = {L^{(1)}}$ and ${L^{(l)}} < {L^{(l-1)}}$. Hence, the complexity of one feedforward propagation is linear in $|{\mathcal{E}}_t|$ w.r.t. $\mathcal{G}_t$. One can further speed up the feedforward propagation using GPUs.

Let $k$ be the dimensionality of graph embedding. For a graph ${\mathcal{G}}_t$ with $K_t$ communities, the complexity of downstream $K$Means clustering is no more than $O({N_t}{K_t}kI)$ for $I$ iterations. The runtime of $K$Means can also be significantly reduced via parallel implementations.

\section{Experiments}\label{Sec:Exp}
In this section, we elaborate on our experiments. Concretely, we first introduce the experiment setup (e.g., datasets, baselines, and metrics) in Section \ref{Sec:Setup}. Evaluation results of CD quality and efficiency are analyzed in Section~\ref{Sec:Exp-Eva}. Based on the evaluation results, we quantitatively compare the trade-off between quality and efficiency via a novel \textit{trade-off score} in Section~\ref{Sec:Exp-TOS}. Further experiments of ablution study and convergence analysis are given in Section~\ref{Sec:Abl} and \ref{Sec:Conv}.
%\textcolor{red}{In addition, we also consider extended applications of (\romannumeral1) estimating the number of communities $K$ of CD when $K$ is not given and (\romannumeral2) transferring ICD from one training scenario to other test scenarios in Appendix.}

\subsection{Experiment Setup}\label{Sec:Setup}

\begin{table}\scriptsize
\centering
\caption{Statistic detials of all the datasets with $T$, $N$, $|\mathcal{E}|$, $K$ as the number of graphs, nodes, edges, and communities.}
\label{Tab:Data}
\vspace{-0.3cm}
\begin{tabular}{l|l|p{0.5cm}p{0.5cm}p{0.6cm}|p{0.6cm}p{0.6cm}p{0.7cm}|p{0.5cm}p{0.5cm}p{0.6cm}|l}
\hline
 & \textit{T} & Min~\textit{N} & Max~\textit{N} & Avg~\textit{N} & Min~$|\mathcal{E}|$ & Max~$|\mathcal{E}|$ & Avg~$|\mathcal{E}|$ & Min~\textit{K} & Max~\textit{K} & Avg~\textit{K} & \textbf{Descriptions} \\ \hline
\textit{GN-}0.4 & \multirow{2}{*}{2,000} & \multirow{2}{*}{5,000} & \multirow{2}{*}{5,000} & \multirow{2}{*}{5,000} & 48,335 & 49,698 & 48,999 & \multirow{2}{*}{250} & \multirow{2}{*}{250} & \multirow{2}{*}{250} & \multirow{2}{*}{Fixed \textit{N}, Fixed \textit{K}, w/ ground-truth} \\
\textit{GN-}0.3 &  &  &  &  & 48,488 & 49,981 & 49,246 &  &  &  & \\ \hline
\textit{L(f,0.3)} & \multirow{2}{*}{2,000} & \multirow{2}{*}{5,000} & \multirow{2}{*}{5,000} & \multirow{2}{*}{5,000} & 22,539 & 25,707 & 23,925 & 57 & 104 & 79 & \multirow{2}{*}{Fixed \textit{N}, Non-fixed \textit{K}, w/ ground-truth}\\
\textit{L}(\textit{f},0.6) &  &  &  &  & 22,585 & 25,639 & 23,927 & 58 & 107 & 79 & \\ \hline
\textit{L}(\textit{n},0.3) & \multirow{2}{*}{2,000} & 5,000 & 5,999 & 5,503 & 22,999 & 29,670 & 26,311 & 59 & 117 & 87 & \multirow{2}{*}{Non-fixed \textit{N}, Non-fixed \textit{K}, w/ ground-truth}\\
\textit{L}(\textit{n},0.6) &  & 5,000 & 6,000 & 5,493 & 22,680 & 29,931 & 26,289 & 61 & 118 & 87 & \\ \hline
\textit{Taxi} & 3,000 & 1,279 & 1,279 & 1,279 & 38,554 & 40,171 & 39,291 & 7 & 12 & 9 & Fixed \textit{N}, Non-fixed \textit{K}, w/o ground-truth\\ \hline
\textit{Reddit} & 1,870 & 501 & 3,648 & 946 & 525 & 4,780 & 1,133 & 14 & 121 & 32 & Non-fixed \textit{N}, Non-fixed \textit{K}, w/o ground-truth\\ \hline
\textit{Enron} & 410 & 502 & 3,261 & 1,048 & 554 & 5,097 & 1,376 & 10 & 31 & 17 & Non-fixed \textit{N}, Non-fixed \textit{K}, w/o ground-truth\\ \hline
\textit{AS} & 733 & 103 & 6,474 & 4,183 & 487 & 26,467 & 16,324 & 7 & 39 & 29 & Non-fixed \textit{N}, Non-fixed \textit{K}, w/o ground-truth\\ \hline
\end{tabular}
\vspace{-0.3cm}
\end{table}

\textbf{Datasets.} We evaluate ICD on $2$ synthetic benchmarks (with $6$ settings) and $4$ real datasets. Statistics of the datasets are depicted in Table~\ref{Tab:Data}, where $T$, $N$, $|{\mathcal{E}}|$, and $K$ are the number of graphs, nodes, edges, and communities.

\textit{GN-Net} \cite{girvan2002community} is a widely-used synthetic benchmark for CD with the topology and CD ground-truth of each graph generated via the stochastic block model \cite{abbe2017community}. We fixed the number of nodes to $N=5,000$ and evenly partitioned the nodes into $K = 250$ communities. For two uniformly selected nodes ${v_i}$ and ${v_j}$, if they are in the same community (i.e., ${v_i}, {v_j} \in C_r$), the probability to generate an edge $({v_i}, {v_j})$ is $p_{\rm{in}}$. Otherwise the probability to generate an edge $({v_i}, {v_j})$ is ${{(1 - {p_{{\rm{in}}}})} \mathord{\left/ {\vphantom {{(1 - {p_{{\rm{in}}}})} {(K - 1)}}} \right. \kern-\nulldelimiterspace} {(K - 1)}}$. We set ${p_{in}} \in \{0.4, 0.3\}$ to generate $2$ datasets, where we independently generated $2,000$ graphs for each dataset. \textit{With the decrease of $p_{\rm{in}}$, the community structures become increasingly difficult to identify.}  
For simplicity, we denote the \textit{GN-Net} with a specific setting of ${p_{\rm{in}}}$ as \textit{GN-}${p_{\rm{in}}}$. 
%In each graph, we also randomly shuffled the node indices to ensure that there is no index correspondence between any two graphs in the dataset (as we defined in Section~\ref{Sec:Prob-Pre}).

\textit{LFR-Net} \cite{lancichinetti2008benchmark} is a more challenging synthetic benchmark for CD that can simulate the power-law properties of real-world network systems. It uses $(d,{d_{\max }},{c_{\min }},{c_{\max }},\mu )$ to independently generate each graph, where $d$ and ${d_{\max}}$ are the average and maximum node degree; ${c_{\min }}$ and ${c_{\max }}$ are the minimum and maximum community size; $\mu$ is the mixing ratio between the external degree and total degree of each node $v_i$ w.r.t. the community $v_i$ belongs to. We fixed $(d,{d_{\max }},{c_{\min }},{c_{\max }}) = (10,100,10,200)$ and set $\mu  \in \{ 0.3,0.6\}$. \textit{With the increase of $\mu$, the community structures are increasingly difficult to identify.} We consider both cases with fixed and non-fixed $N$, where we respectively fixed $N=5,000$ and randomly set $N \in [5000, 6000]$ to independently generate $2,000$ graphs for each case.
%Similar to \textit{GN-Net}, we also independently generated each graph in a dataset and ensure that there is no index correspondence between any two graphs via the random shuffle of node indices. 
For simplicity, we use \textit{L}$(c, \mu)$ ($c \in \{f, n\}$) to denote the dataset with a setting of $\mu$, where $f$ and $n$ represent the case with fixed and non-fixed $N$.
%Moreover, $K$ is non-fixed in \textit{LFR-Net}.

\textit{Taxi}\footnote{https://www.microsoft.com/en-us/research/publication/t-drive-driving-directions-based-on-taxi-trajectories/} \cite{piorkowski2009parsimonious} is a real GPS dataset including trajectories of $1,279$ taxis in Beijing, which forms a mobile ad hoc network system. We independently sampled $3,000$ graphs from the dataset.
In each graph, we treated each taxi as a node and constructed the topology based on the top-$50$ neighbors with the closest distance for each node.
%Since \textit{Taxi} does not provide ground-truth regarding $K$ for CD, we used \textit{Louvain} algorithm \cite{blondel2008fast} to estimate the $K$ value for each graph in this dataset. Hence, it is with fixed $N$ and non-fixed $K$.

\textit{Reddit}\footnote{https://github.com/weihua916/powerful-gnns/blob/master/dataset.zip} \cite{yanardag2015deep} is a social network dataset from Reddit. We extracted $1,870$ graphs with $N \ge 500$ from the dataset, where each graph corresponds to an online discussion thread. The comment interactions between users in each thread are described as the topology of each graph.
%Different graphs are with different number of users (i.e., nodes), varying from $501$ to $3,648$. 
%Since \textit{Reddit} does not provide ground-truth regarding $K$, we also used \textit{Louvain} to estimate $K$ for each graph. Hence, \textit{Reddit} is with non-fixed $N$ and $K$. Note that different snapshots have different node sets (i.e., with non-fixed $N$) and node indices in each snapshot start from 0. It indicates that there is no correspondence between node indices of any two snapshots in the dataset.

\textit{Enron}\footnote{http://networkrepository.com/ia-enron-email-dynamic.php} \cite{rossi2015network} is a real interaction network extracted from the email system of Enron company, which contains email records from 1980-01-01 to 2004-02-04. We extracted 410 graphs with $N \ge 500$ from the records 
%in chronological order, 
where we treated each user as a unique node and constructed the topology based on email interactions.
%Note that different graphs have different sets of users (i.e., nodes). 
%Moreover, we also used \textit{Louvain} algorithm to estimate $K$ for each graph, since \textit{Enron} does not provide ground-truth regarding $K$. Therefore, it is with non-fixed $N$ and $K$. We also ensure that there is no node index correspondence between any two graphs by permuting node indices of each graph from 0.

\textit{AS}\footnote{http://snap.stanford.edu/data/as-733.html} \cite{leskovec2005graphs} is a dataset from the border gateway protocol (BGP) logs of a communication network, which describe the topology between a set of autonomous systems. It contains 733 daily instances from 1997-08-11 to 2000-01-02. Each BGP router is abstracted as a unique node with the communication between routers extracted as graph topology. 
%As the dataset allows addition and deletion of routers, different graphs are with different set of nodes. 
%Since \textit{AS} does not provide the ground-truth $K$, we also applied \textit{Louvain} to determine $K$ for each snapshot. Therefore, it is with non-fixed $N$ and $K$, where there is no index correspondence between any two graphs.

In summary, the $2$ datasets of \textit{GN-Net} are with fixed $N$ and $K$. The \textit{LFR-Net} covers cases of (\romannumeral1) fixed $N$ but non-fixed $K$ and (\romannumeral2) non-fixed $N$ and $K$. \textit{Taxi} is with fixed $N$ but non-fixed $K$, while \textit{Reddit}, \textit{Enron}, and \textit{AS} are with non-fixed $N$ and $K$. 
\textit{GN-Net} and \textit{LFR-Net} are synthetic benchmarks that can simultaneously generate graph topology and CD ground-truth of each graph, while \textit{Taxi}, \textit{Reddit}, \textit{Enron}, and \textit{AS} are real datasets without ground-truth.
%As defined in Section~\ref{Sec:Prob-Pre}, we assume \textit{$K$ is given by the downstream CD}. Most baselines in our experiments also need pre-set $K$. To ensure the fairness of comparison, where \textit{each graph is assigned with a common $K$ for all the methods to be evaluated}, we adopted a widely-used strategy \cite{qin2018adaptive} to estimate $K$ for each graph of the four real datasets using an auxiliary method (e.g., \textit{Louvain} \cite{blondel2008fast}).

\begin{table}[t]\scriptsize
\centering
\caption{Summary of all the methods to be evaluated.}
\label{Tab:Meth}
\vspace{-0.3cm}
\begin{tabular}{l|p{0.25cm}p{0.02cm}p{0.45cm}p{0.45cm}p{0.3cm}p{0.15cm}p{0.3cm}p{0.25cm}p{0.3cm}p{0.5cm}p{0.25cm}p{0.45cm}|p{0.5cm}p{0.2cm}p{0.2cm}p{0.2cm}p{0.6cm}|l}
\hline
 & \textit{SNMF} & \textit{SC} & \textit{GraClus} & \textit{MCSBM} & \textit{Locale} & \textit{N2V} & \textit{MNMF} & \textit{DNR} & \textit{NECS} & \textit{RandNE} & \textit{NRP} & \textit{ProNE} & \textit{GSAGE} & \textit{GAT} & \textit{GIN} & \textit{GAP} & \textit{ClusNet} & \textbf{ICD} \\ \hline
Trans & $\surd$ & $\surd$ & $\surd$ & $\surd$ & $\surd$ &  & $\surd$ & $\surd$ & $\surd$ & $\surd$ & $\surd$ & $\surd$ & &  &  & $\Delta$ & $\Delta$ &  \\ \hline
Ind &  &  &  &  &  &  &  &  &  &  &  &  & $\surd$ & $\surd$ & $\surd$ & $\Delta$ & $\Delta$ & $\surd$ \\ \hline
E2E & $\surd$ & $\surd$ & $\surd$ & $\surd$ & $\surd$ &  &  &  &  &  &  &  &  &  &  & $\surd$ & $\surd$ &  \\ \hline
Emb &  &  &  &  &  & $\surd$ & $\surd$ & $\surd$ & $\surd$ & $\surd$ & $\surd$ & $\surd$ & $\surd$ & $\surd$ & $\surd$ &  &  & $\surd$ \\ \hline
\end{tabular}
\vspace{-0.3cm}
\end{table}

\textbf{Baselines.} We compared ICD over $17$ baselines which can be divided into $2$ categories of \textit{transductive} and \textit{inductive} approaches. Table~\ref{Tab:Meth} gives an overview of these approaches.

\textit{SNMF} \cite{wang2011community}, \textit{spectral clustering} (\textit{SC}) \cite{von2007tutorial}, \textit{GraClus} \cite{dhillon2007weighted}, \textit{MC-SBM} \cite{peixoto2014efficient} and \textit{Locale} \cite{wang2020community} are \textit{transductive} E2E CD methods, while \textit{node2vec} (\textit{N2V}) \cite{grover2016node2vec}, \textit{M-NMF} \cite{wang2017community}, \textit{DNR} \cite{yang2016modularity}, \textit{NECS} \cite{li2019learning}, \textit{RandNE} \cite{zhang2018billion}, \textit{NRP} \cite{yang20209homogeneous}, and \textit{ProNE} \cite{zhang2019prone} are \textit{transductive} graph embedding approaches that can suppport CD. In particular, \textit{GraClus}, \textit{MC-SBM}, \textit{Locale}, \textit{RandNE}, \textit{NRP}, and \textit{ProNE} are fast methods with heuristic approximation, while the rest \textit{transductive} baselines focus on how to derive high-quality CD results or embeddings. Moreover, \textit{M-NMF}, \textit{DNR}, \textit{NECS} are typical community-preserved embedding approaches.

\textit{GraphSAGE} (\textit{GSAGE}) \cite{hamilton2017inductive}, \textit{GAT} \cite{velivckovic2017graph}, and \textit{GIN} \cite{xu2018powerful} are typical \textit{inductive} GNNs following the graph embedding framework, while \textit{GAP} \cite{nazi2019gap} and \textit{ClusNet} \cite{wilder2019end} are E2E baselines using \textit{inductive} GNNs.
%Since the CD results are directly derived by an output layer for \textit{GAP} and \textit{ClusNet}, their inductiveness is only available for the case with fixed $K$. One has to use the \textit{transducitve} settings (i.e., training them from scratch for inference) for graphs with non-fixed $K$, due to the fixed dimensionality of the output layer.

Both variants of ICD were also evaluated. For simplicity, we use ICD-M and ICD-C to denote the variants adopting objectives of modularity maximization and NCut minimization, respectively.

\textbf{Evaluation Metrics.} In our evaluation, we \textit{consider CD with given $K$}. By the quality evaluation criteria in Section~\ref{Sec:Prob-Pre}, we adopted quality metrics of \textbf{modularity} (\ref{Eq:Mod-Max}) and \textbf{NCut} (\ref{Eq:NCut-Min}) for all the datasets. For datasets with ground-truth (i.e., \textit{GN-Net} and \textit{LFR-Net}), we also used metrics of normalized mutual information (\textbf{NMI}) and accuracy (\textbf{AC}) to measure the correspondence between CD results and ground-truth. For a given graph ${{\mathcal{G}}_t}$ with ${K_t}$ communities, we can use ${H_t}=\{ {H^t_{1}}, \cdots, {H^t_{K_t}}\}$ and ${C_t}=\{ {C^t_{1}}, \cdots, {C^t_{K_t}}\}$ to denote the ground-truth and CD result with $H_r^t \subseteq {{\mathcal{V}}_t}$ and $C_r^t \subseteq {{\mathcal{V}}_t}$ as node member sets w.r.t. the $r$-th community. Given ${H_t}$ and ${C_t}$, \textbf{NMI} is defined as
\begin{equation}\label{Eq:NMI}
    {\mathop{\rm NMI}\nolimits} ({H_t},{C_t}) = \frac{{ - 2\sum\limits_{H_r^t \in {H_t}} {\sum\limits_{C_s^t \in {C_t}} {\frac{{{n_{r,s}}}}{n}\log \frac{{n \times {n_{r,s}}}}{{{n_r} \times {N_s}}}} } }}{{\sum\limits_{H_r^t \in {H_t}} {\frac{{{n_r}}}{n}\log \frac{{{n_r}}}{n} + \sum\limits_{C_s^t \in {C_t}} {\frac{{{n_s}}}{n}\log \frac{{{n_s}}}{n}} } }},
\end{equation}
where ${n_r} = |H_r^t|$, ${n_s} = |C_s^t|$, ${n_{r,s}} = |H_r^t \cap C_s^t|$, and $n = N_t$. Let ${L_t} = ( l_1^t, \cdots ,l_{{N_t}}^t)$ and ${R_t} = ( r_1^t, \cdots ,r_{{N_t}}^t)$ be label sequences of CD result and ground-truth w.r.t. node sequence $( v_1^t, \cdots ,v_{{N}}^t)$, where $l^t_i$ and $r^t_i$ are community labels of node ${v^t_i}$. \textbf{AC} is defined as
\begin{equation}\label{Eq:AC}
    {\mathop{\rm AC}\nolimits} ({R_t},{L_t}) = \frac{{\sum\nolimits_{i = 1}^{{N_t}} {\delta (r_i^t,{\mathop{\rm map}\nolimits} (l_i^t))} }}{{{N_t}}},
\end{equation}
where ${\mathop{\rm map}\nolimits} ( \cdot )$ is the Kuhn-Munkres mapping that gives the best map from ${L_t}$ to ${R_t}$; $\delta (a,b)$ is the Kronecker delta function with $\delta (a,b) = 1$ if $a=b$ and $\delta (a,b) = 0$ otherwise.

Usually, \textit{larger \textbf{modularity}, \textbf{NMI} and \textbf{AC}, as well as smaller \textbf{NCut} imply better quality}. Moreover, the total \textbf{runtime} of a method to derive its final CD result was used as the efficiency metric with \textit{lower \textbf{runtime} indicating higher efficiency}.

\textbf{Evaluation Settings.} All the \textit{transductive} baselines are independently optimized to tackle CD on each single graph, while ICD ties to alleviate the NP-hard challenge via an \textit{inductive} framework across graphs. To illustrate the superiority of ICD beyond \textit{transductive} methods, for each dataset, we used the first $80\%$ graphs as training set $\Gamma_{\rm{T}}$ with the remaining $10\%$ and $10\%$ deployed as validation set ${\Gamma_{\rm{V}}}$ and test set $\Gamma'$.
%We conducted the \textit{offline} training of ICD on $\Gamma_{\rm{T}}$ and generalized it to $\Gamma'$ for \textit{online} CD.
In the \textit{offline} training, we saved model parameters of ICD with the best average quality on $\Gamma_{\rm{V}}$, where \textbf{NMI} and \textbf{modularity} are used as the validation quality metrics for datasets with and without ground-truth. Since we focus on the trade-off between quality and efficiency on new test graphs, the evaluation was conducted on $\Gamma'$, where we had to train \textit{transductive} baselines from scratch for each graph in $\Gamma'$.

For \textit{inductive} embedding baselines (i.e., \textit{GSAGE}, \textit{GAT}, and \textit{GIN}), we use the same training and test settings with ICD, i.e., \textit{offline} training on $\Gamma_{\rm{T}}$ and \textit{online} generalization to $\Gamma'$. Since we consider CD on graphs without attributes, we use a constant matrix ${{\bf{1}}_{{N_t} \times L}}$, where each entry is set to $1$, as the input features of these GNN-based methods, which is a widely-used setting of \textit{inductive} GNNs for the case without attributes \cite{xu2018powerful}. Since \textit{GAT} \cite{velivckovic2017graph} and \textit{GIN} \cite{xu2018powerful} do not give the recommended unsupervised training loss for node-level embedding, we used the unsupervised loss of \textit{GSAGE} \cite{hamilton2017inductive}
%which is also a widely-used unsupervised loss for \textit{inductive} GNNs, 
to train \textit{GSAGE}, \textit{GAT}, and \textit{GIN} on $\Gamma_{\rm{T}}$ of each dataset. Note that most existing GNN-based embedding methods cannot directly integrate the permutation invariant training labels even using a supervised loss, e.g., cross-entropy.

For the E2E GNN-based methods (i.e., \textit{GAP} and \textit{ClusNet}), we could only use the same \textit{inductive} settings as ICD on datasets with fixed $K$ (i.e., \textit{GN-Net}) but had to train them from scratch on $\Gamma'$ of other datasets with non-fixed $K$, due to the fixed dimensionality of their E2E output layers. As recommended in \cite{nazi2019gap} and \cite{wilder2019end}, we used \textit{PCA} and \textit{node2vec} to map adjacency matrices $\{ {{\bf{A}}_{t}} \}$ to a feature space with fixed dimensionality for the GNN input features of \textit{GAP} and \textit{ClusNet}. Moreover, we used their own unsupervised training losses defined in \cite{nazi2019gap} and \cite{wilder2019end}.
%which approximate objectives of NCut minimization and modularity maximization, respectively.

For all the embedding methods, we used $K$Means as the downstream clustering algorithm. We ran $K$Means $10$ times to avoid its limitation of getting locally optimal solutions, which was included in the total \textbf{runtime}. 
%To ensure the fairness of comparison, 
The embedding dimensionality of all the embedding methods was set to be the same on each dataset. For other E2E baselines, we got CD results directly from their outputs. 

For datasets with ground-truth (i.e., \textit{GN-Net} and \textit{LFR-Net}), we directly used their CD ground-truth to derive the \textit{membership indicator} ${{\bf{H}}_{t}}$ for ICD, i.e., the first source of training labels defined in Section~\ref{Sec:Prob-Pre}. Since \textit{Taxi}, \textit{Reddit}, \textit{Enron}, and \textit{AS} do not provide ground-truth of each training graph, we used CD results of the baseline with best average \textbf{modularity} on $\Gamma_{\rm{V}}$ to derive ${{\bf{H}}_{t}}$ for the \textit{offline} training of ICD, i.e., the second source of training labels in Section~\ref{Sec:Prob-Pre}.
Note that we do not need to derive ${{\bf{H}}_{t}}$ in the evaluation phase when generalizing the trained model to test set $\Gamma'$.

Some other details of experiment setups (e.g., paramter settings and layer configurations) are given in Appendix~\ref{Sec:App-Setup}.

\subsection{Quantitative Evaluation}\label{Sec:Exp-Eva}

\begin{figure}[t]
\centering
 \begin{minipage}{\linewidth}
 \subfigure[\textit{GN}-0.3]{
  \frame{
  \includegraphics[width=\textwidth,trim=20 18 38 30,clip]{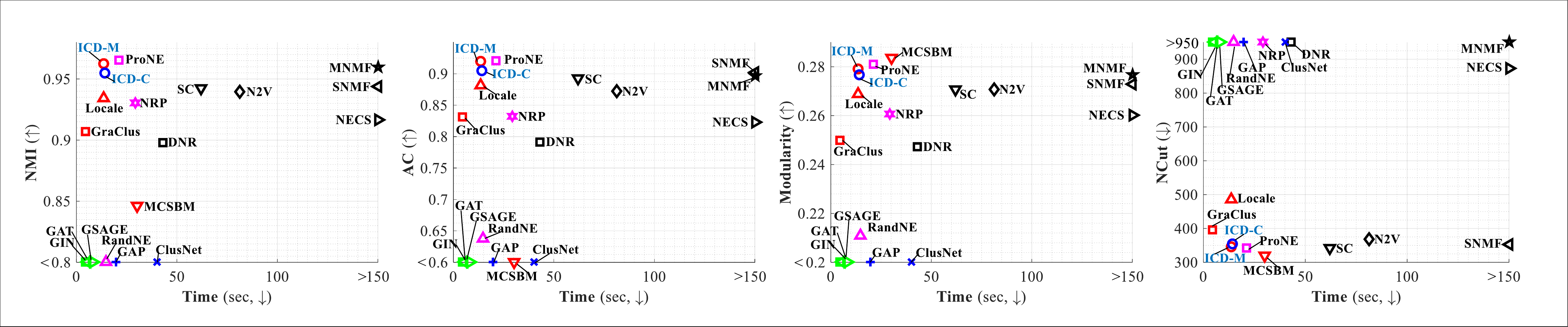}}
  }
 \end{minipage}
 \begin{minipage}{\linewidth}
 \subfigure[\textit{GN}-0.4]{
  \frame{
  \includegraphics[width=\textwidth,trim=20 18 38 30,clip]{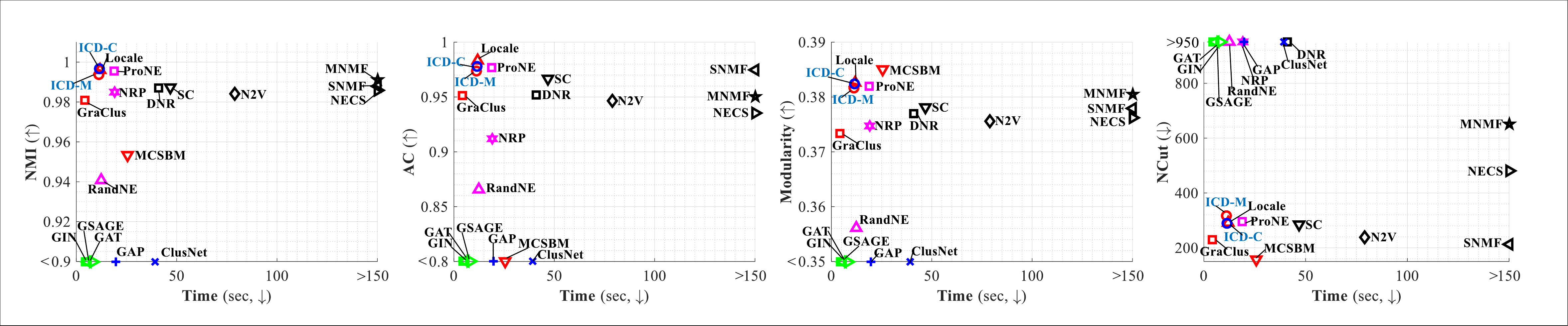}}
  }
 \end{minipage}
 \vspace{-0.3cm}
 \caption{Evaluation results on \textit{GN-Net} w.r.t. quality metrics of \textbf{NMI}$\uparrow$, \textbf{AC}$\uparrow$, \textbf{modularity}$\uparrow$, \& \textbf{NCut}$\downarrow$ ($y$-axis) and efficiency metric of \textbf{runtime}$\downarrow$ ($x$-axis).
 }\label{Fig:GN-Vis}
 \vspace{-0.5cm}
\end{figure}

\begin{figure}[t]
\centering
 \begin{minipage}{\linewidth}
 \subfigure[\textit{L}(\textit{f},0.3)]{
  \frame{
  \includegraphics[width=\textwidth,trim=20 18 38 30,clip]{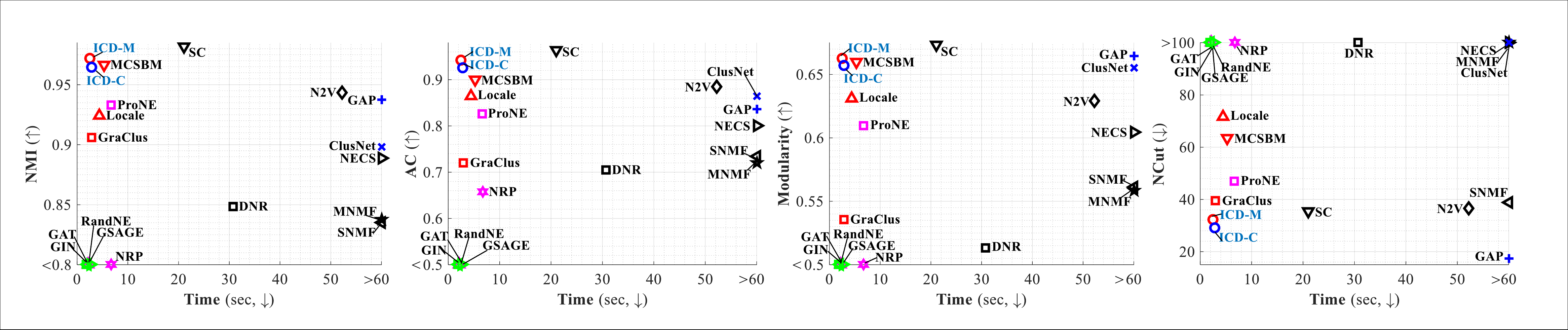}}
  }
 \end{minipage}
 \begin{minipage}{\linewidth}
 \subfigure[\textit{L}(\textit{f},0.6)]{
  \frame{
  \includegraphics[width=\textwidth,trim=20 18 38 30,clip]{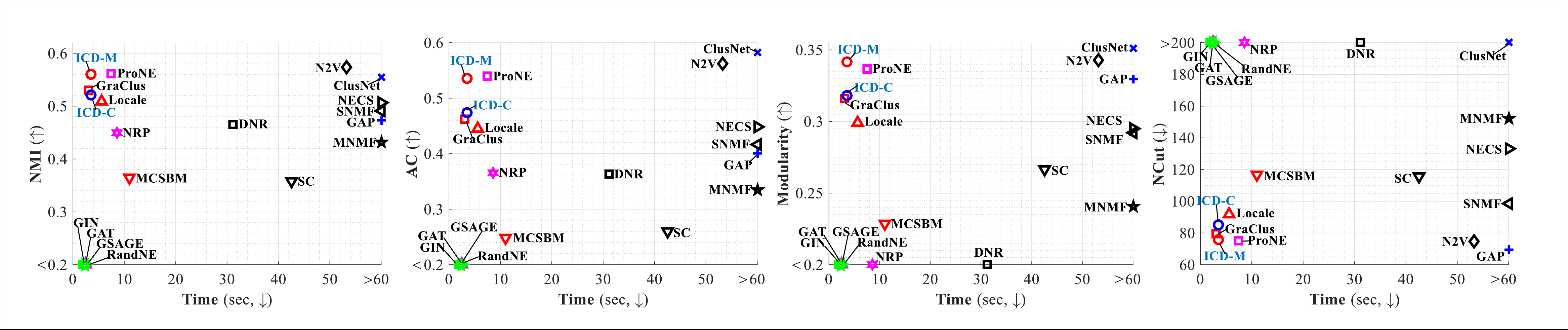}}
  }
 \end{minipage}
  \begin{minipage}{\linewidth}
 \subfigure[\textit{L}(\textit{n},0.3)]{
  \frame{
  \includegraphics[width=\textwidth,trim=20 18 38 30,clip]{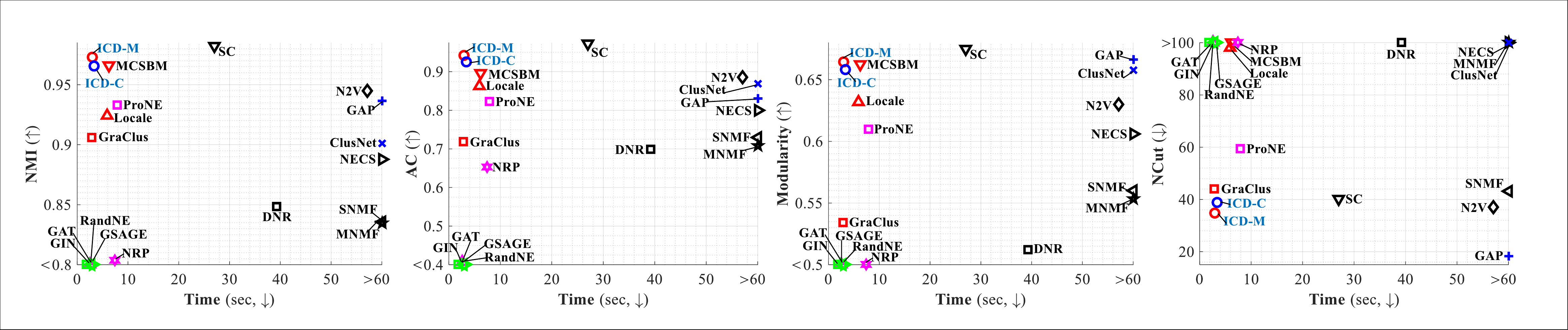}}
  }
 \end{minipage}
 \begin{minipage}{\linewidth}
 \subfigure[\textit{L}(\textit{n},0.6)]{
  \frame{
  \includegraphics[width=\textwidth,trim=20 18 38 30,clip]{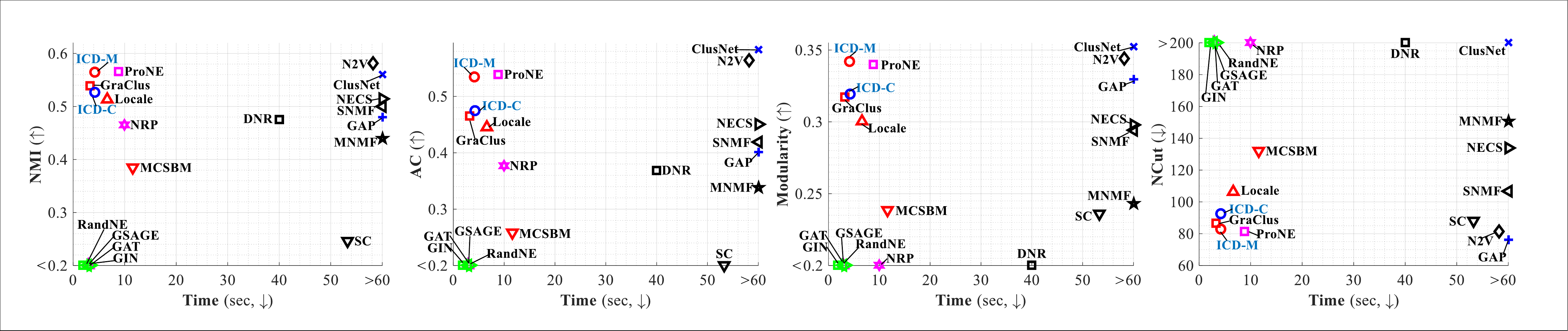}}
  }
 \end{minipage}
 \vspace{-0.3cm}
 \caption{Evaluation results on \textit{LFR-Net} w.r.t. quality metrics of \textbf{NMI}$\uparrow$, \textbf{AC}$\uparrow$, \textbf{modularity}$\uparrow$, \& \textbf{NCut}$\downarrow$ ($y$-axis) and efficiency metric of \textbf{runtime}$\downarrow$ ($x$-axis).
 }\label{Fig:LFR-Vis}
 \vspace{-0.5cm}
\end{figure}

\begin{figure}[t]
\centering
 \begin{minipage}{0.49\linewidth}
 \subfigure[\textit{Taxi}]{
  \frame{
  \includegraphics[width=\textwidth,trim=20 18 38 30,clip]{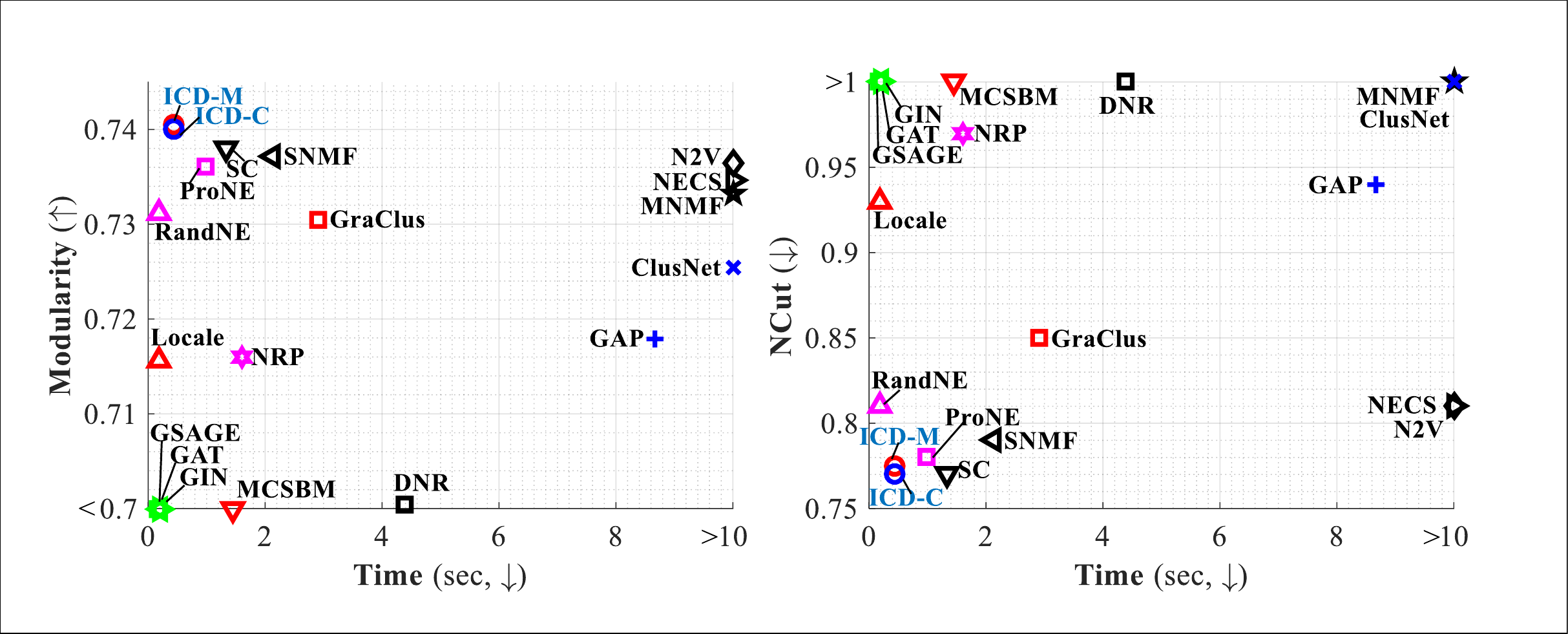}}
  }
 \end{minipage}
 \begin{minipage}{0.49\linewidth}
 \subfigure[\textit{Reddit}]{
  \frame{
  \includegraphics[width=\textwidth,trim=20 18 38 30,clip]{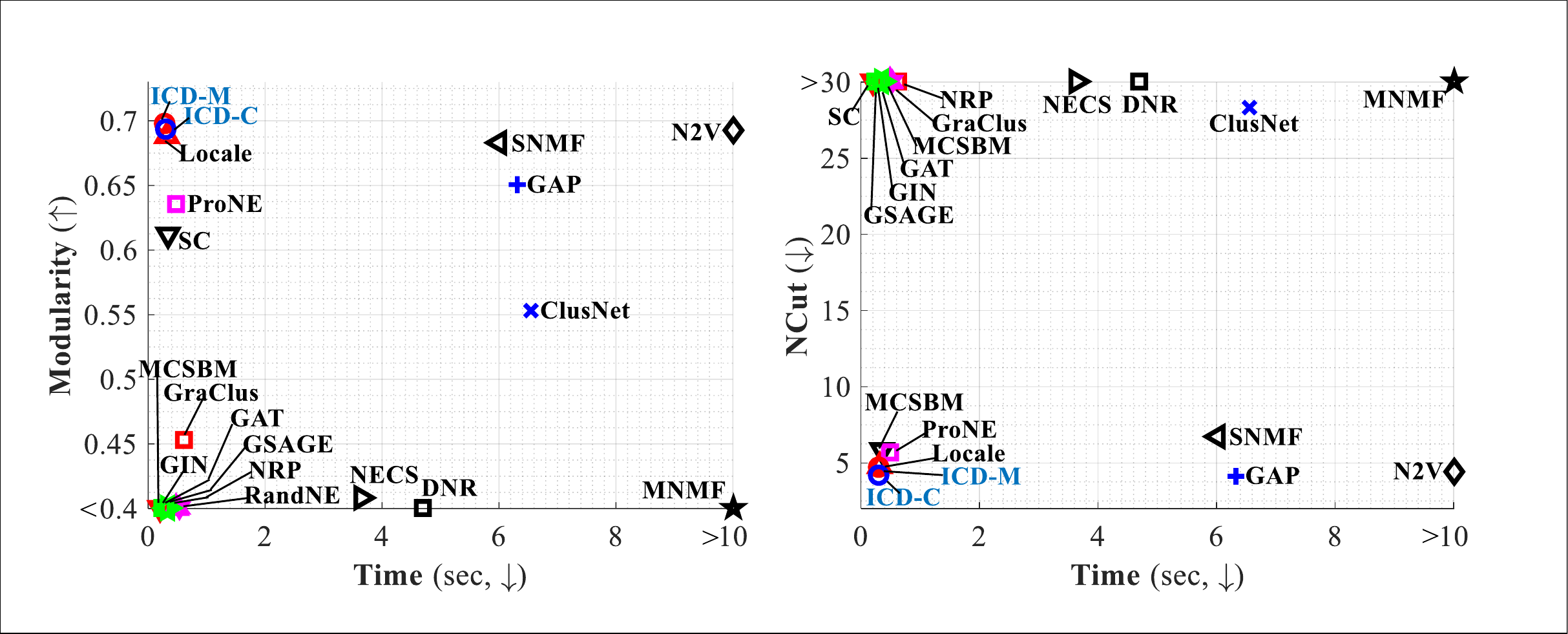}}
  }
 \end{minipage}
  \begin{minipage}{0.49\linewidth}
 \subfigure[\textit{AS}]{
  \frame{
  \includegraphics[width=\textwidth,trim=20 18 38 30,clip]{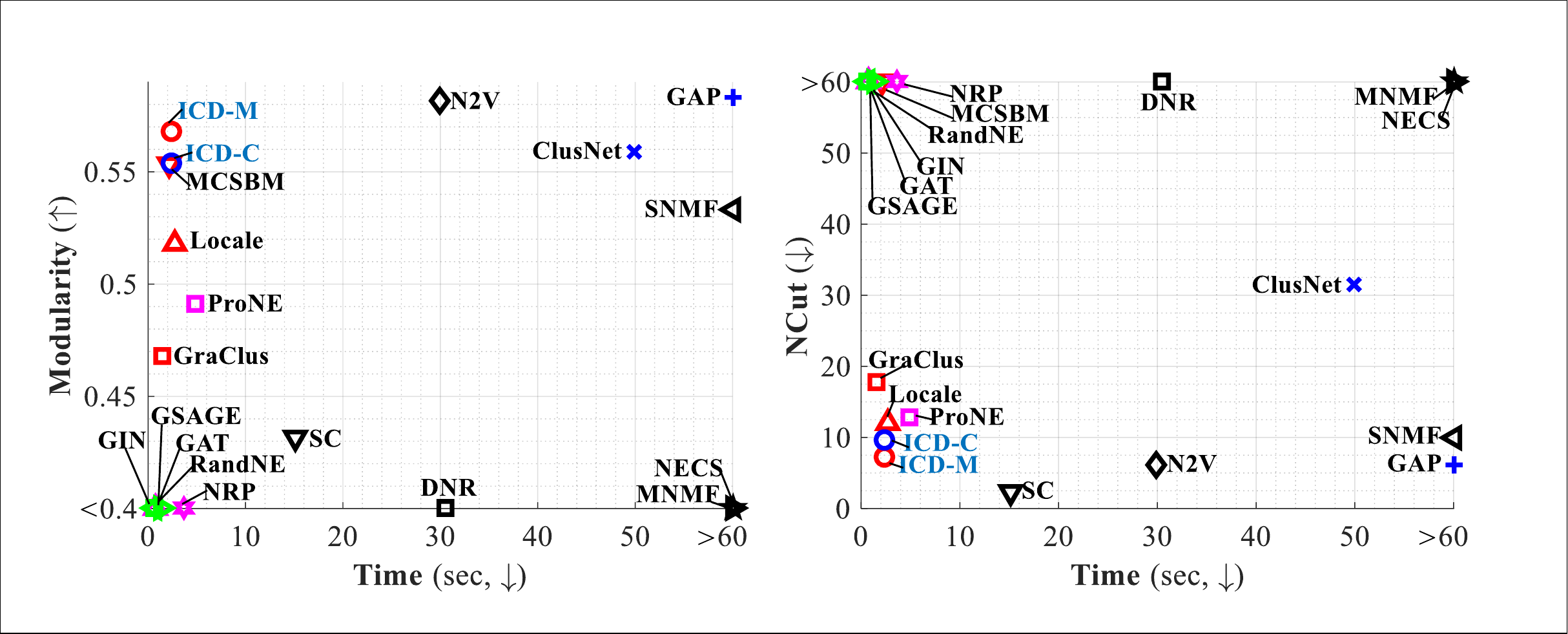}}
  }
 \end{minipage}
 \begin{minipage}{0.49\linewidth}
 \subfigure[\textit{Enron}]{
  \frame{
  \includegraphics[width=\textwidth,trim=20 18 38 30,clip]{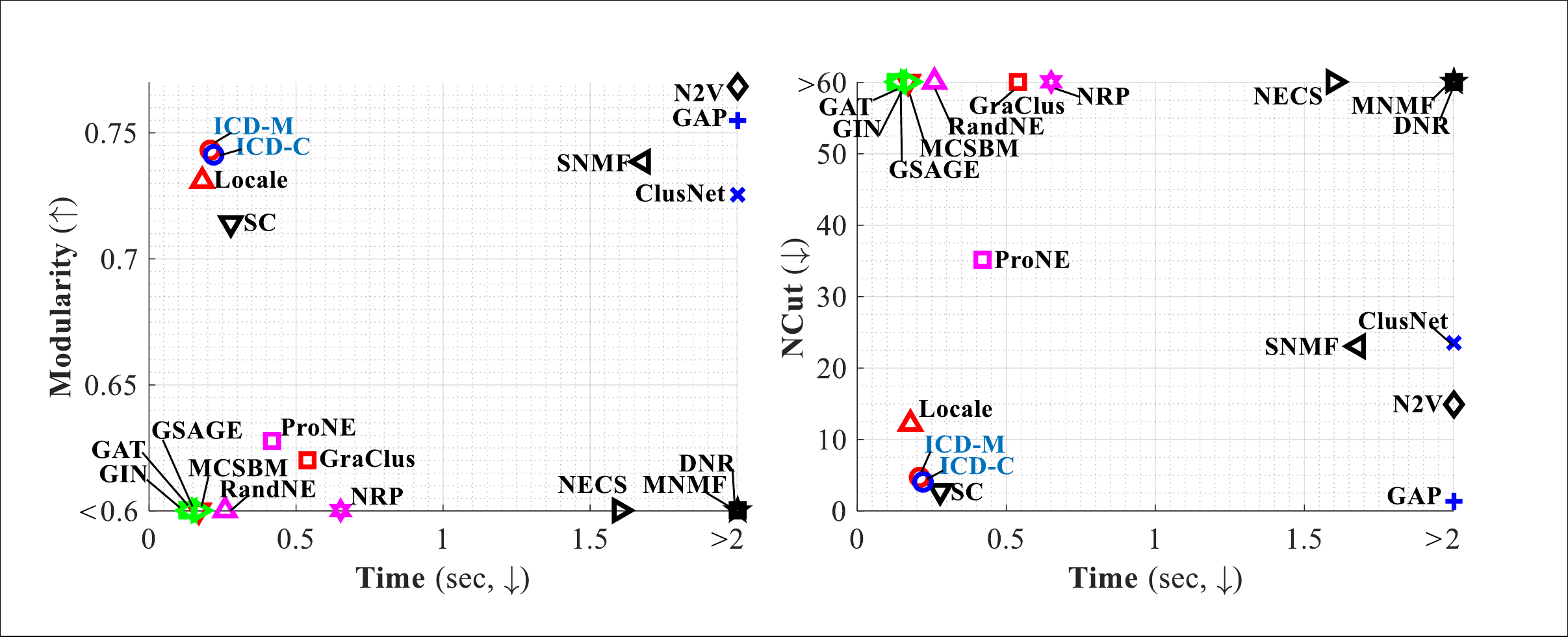}}
  }
 \end{minipage}
 \vspace{-0.3cm}
 \caption{Evaluation results on datasets without ground-truth w.r.t. quality metrics of \textbf{modularity}$\uparrow$, \& \textbf{NCut}$\downarrow$ ($y$-axis) and efficiency metric of \textbf{runtime}$\downarrow$ ($x$-axis).
 }\label{Fig:Data-wognd-Vis}
 \vspace{-0.3cm}
\end{figure}

In our evaluation, both the mean and standard deviation of all the quality and efficiency metrics on $\Gamma'$ of each dataset were recorded. Evaluation results regarding the trade-off between each quality metric (e.g., average \textbf{NMI}, \textbf{AC}, \textbf{modularity}, and \textbf{NCut}) and the unique efficiency metric of average \textbf{runtime} are visualized in Fig.~\ref{Fig:GN-Vis}, \ref{Fig:LFR-Vis}, and \ref{Fig:Data-wognd-Vis}. Concretely, we map each method to a data point in a 2D space with its $x$-axis and $y$-axis corresponding to the unique efficiency metric (i.e., \textbf{runtime}) and a quality metric. Number records of the mean and standard deviation w.r.t. each metric are given in the Appendix~\ref{Sec:App-Eva-Res}.
According to Fig.~\ref{Fig:GN-Vis}, \ref{Fig:LFR-Vis}, and \ref{Fig:Data-wognd-Vis}, we have the following observations.

For all the datasets, ICD-M and ICD-C are in the top groups with best quality. In some cases, they even have the best or second-best quality metrics. 
%Note that ICD is an \textit{inductive} framework without additional optimization on the test set $\Gamma'$, while conventional \textit{transductive} baselines can fully explore properties of each test graph in $\Gamma'$. 
%It indicates the potential of ICD to capture additional characteristics shared by multiple historical graphs beyond conventional \textit{transductive} methods only applied to a single graph. 
Moreover, the runtime of ICD-M and ICD-C is also competitive to the \textit{transductive} baselines with fast approximation (e.g., \textit{GraClus} and \textit{RandNE}) and other \textit{inductive} GNNs with fast \textit{online} generalization (e.g., \textit{GSAGE}). 
From the view of alleviating the NP-hard challenge, \textit{ICD can achieve a significant trade-off between quality and efficiency of \textit{online} CD}.
We further quantitatively evaluate the trade-off achieved by all the methods based on the visualization results of Fig.~\ref{Fig:GN-Vis}, \ref{Fig:LFR-Vis}, and \ref{Fig:Data-wognd-Vis} in Section~\ref{Sec:Exp-TOS}.

Although the \textit{inductive} GNN based embedding baselines (i.e., \textit{GSAGE}, \textit{GAT}, and \textit{GIN}) have slightly faster runtime than ICD, they suffer from poor quality on all the datasets. The results validate that the extracted features ${\bf{Z}}_t$ and hybrid training loss of ICD, which incorporates the permutation invariant training labels, are essential to ensure high CD quality.
On datasets with fixed $K$ (i.e., \textit{GN-Net}), we used the same \textit{inductive} settings for the E2E GNN-based methods (i.e., \textit{GAP} and \textit{ClusNet}). However, the \textit{online} CD of \textit{GAP} and \textit{ClusNet} on \textit{GN-Net} is still with poor quality, indicating that the graph embedding scheme in ICD is more robust than \textit{inductive} E2E baselines. Moreover, we had to use the time-consuming \textit{transductive} settings for \textit{GAP} and \textit{ClusNet} on datasets with non-fixed $K$, training them from scratch for each test graph. In contrast, ICD can still tackle the fast \textit{online} CD with non-fixed $K$ via its graph embedding scheme.

%For the total runtime of ICD on all the datasets, the downstream clustering (e.g., $K$Means with $10$ independent runs) is a major bottleneck but it is the key component to tackle the \textit{online} CD with non-fixed $K$. In our future work, we intend to further reduce the runtime by replacing the downstream clustering with a generic E2E module that can directly derive the CD result w.r.t. a non-fixed $K$.

In most cases, ICD-M and ICD-C have similar CD quality. ICD-M has better quality than ICD-C on datasets where community structures are difficult to identify (e.g., \textit{GN-}0.3, $L$($f$, 0.6) and $L$($n$, 0.6)). A possible reason is that the modularity matrix ${\bf{Q}}_t$ is more informative than the normalized adjacency matrix ${\bf{M}}_t$ to be structural features ${\bf{X}}_t$ of ICD.
%Moreover, the neighbor-induced feature ${\bf{X}}_t$ is also the key component to ensure the strong robustness of ICD, especially when community structures of a given graph are indistinct.

\subsection{Trade-Off Analysis}\label{Sec:Exp-TOS}

\begin{figure}[t]
\centering
\includegraphics[width=0.5\columnwidth, trim=25 19 18 18,clip]{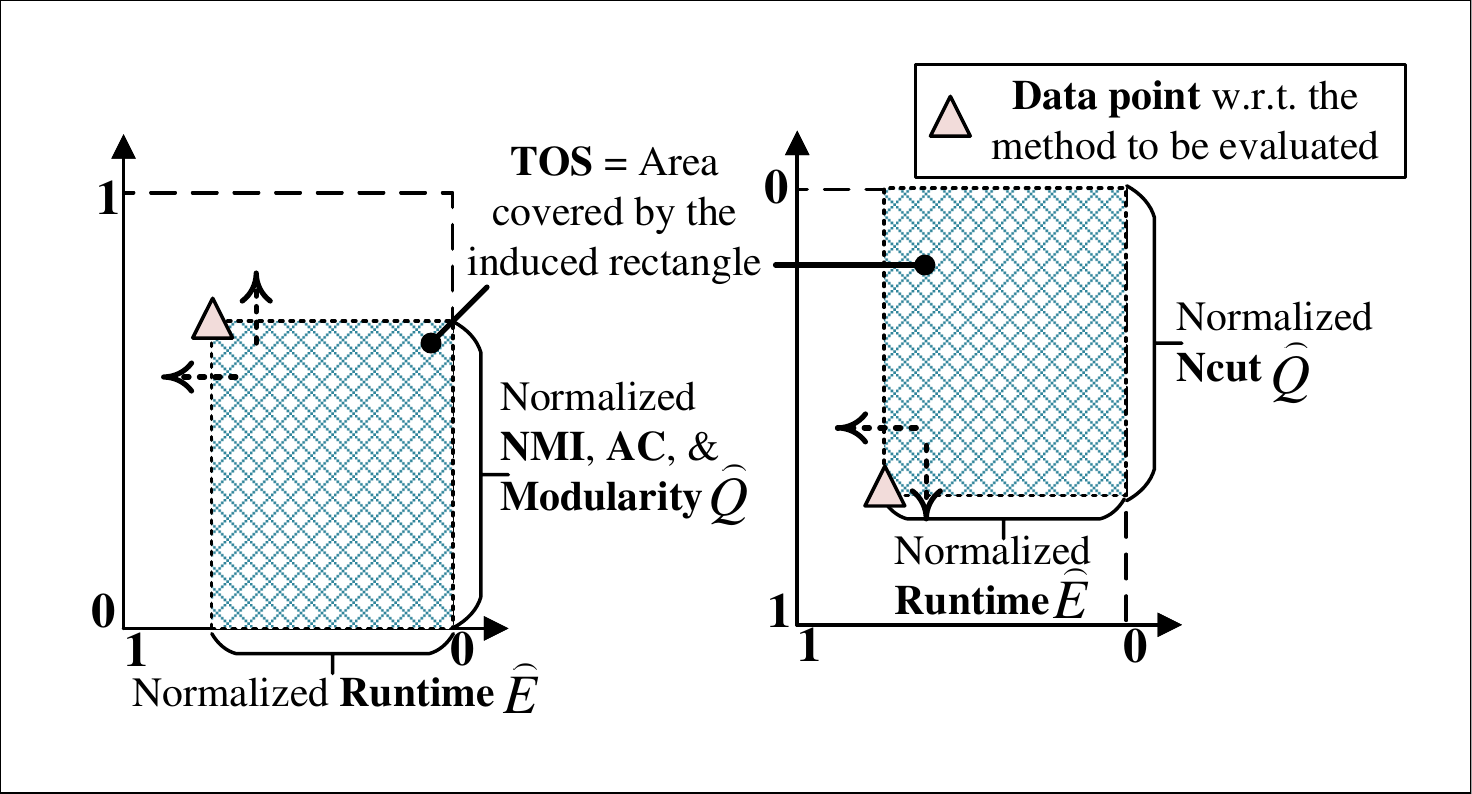}
\vspace{-0.3cm}
\caption{Intuitions of the \textit{trade-off score} (TOS) based on the visualization results of CD quality (i.e., $y$-axis) and efficiency (i.e., $x$-axis).
%with larger TOS indicating a better trade-off achieved by a method.
}
\label{Fig:TOS-Int}
\vspace{-0.5cm}
\end{figure}

In this study, we focus on the trade-off between quality and efficiency of CD. However, to evaluate the trade-off between two aspects that conflict with each other using a comprehensive metric remains an open issue. We propose a novel \textit{trade-off score} (TOS) to quantitatively evaluate the trade-off achieved by a method, based on the normalized area covered by an induced rectangle of a data point in the efficiency-quality visualization results like Fig.~\ref{Fig:GN-Vis}, \ref{Fig:LFR-Vis}, and \ref{Fig:Data-wognd-Vis}. Intuitions of TOS are shown in Fig.~\ref{Fig:TOS-Int}. Our goal is to let \textit{larger TOS indicate a better trade-off between quality and efficiency}.

In our evaluation, we used \textbf{NMI}, \textbf{AC}, \textbf{modularity}, and \textbf{NCut} as quality metrics, while we adopted the \textbf{runtime} of a method as the unique efficiency metric. Let $Q$ and $E$ be values of the quality and efficiency metrics achieved by a method. Given the tuple $(E, Q)$, each method can be mapped to a point in a 2D space as shown in Fig.~\ref{Fig:GN-Vis}, \ref{Fig:LFR-Vis}, and \ref{Fig:Data-wognd-Vis}.

Since there is magnitude difference between $E$ and $Q$, we first normalize them into the range of $[0, 1]$. When using \textbf{NMI} or \textbf{AC} as the quality metric $Q$, we derive the normalized metric via $\mathord{\buildrel{\lower3pt\hbox{$\scriptscriptstyle\frown$}} \over Q}  = Q/1$, since $Q \in [0, 1]$ and larger $Q$ indicates better quality. When using \textbf{modularity} as $Q$, we have $Q \in [-1, 1]$, where larger $Q$ implies better quality. We can normalize $Q$ via $\mathord{\buildrel{\lower3pt\hbox{$\scriptscriptstyle\frown$}} \over Q}  = (Q + 1)/2$.
When using \textbf{NCut} as $Q$, we have $Q \in [0, +\infty)$ with smaller $Q$ indicating better quality, so we normalize $Q$ via $\mathord{\buildrel{\lower3pt\hbox{$\scriptscriptstyle\frown$}}  \over Q}  = ({Q_m} - Q)/{Q_m}$, with $Q_m$ as the maximum NCut among all the methods. For the efficiency metric $E$ (i.e., \textbf{runtime}), smaller $E$ implies higher efficiency with $E \in [0, +\infty)$, so we normalize $E$ via $ \mathord{\buildrel{\lower3pt\hbox{$\scriptscriptstyle\frown$}} \over E}  = ({E_m} - E)/{E_m}$, where $E_m$ is the maximum runtime among all the methods.

As highlighted in Fig.~\ref{Fig:TOS-Int}, when using \textbf{NMI}, \textbf{AC}, and \textbf{modularity} as quality metrics, a data point close to the top left corner in a visualization result indicates the corresponding method can achieve a better trade-off between quality and efficiency. For \textbf{NCut}, a data point close to the bottom left corner implies a better trade-off. Based on this intuition, we define TOS w.r.t. the given $(Q, E)$ as
\begin{equation}
    {\mathop{\rm TOS}\nolimits} (Q,E) = \mathord{\buildrel{\lower3pt\hbox{$\scriptscriptstyle\frown$}} \over Q}  \times \mathord{\buildrel{\lower3pt\hbox{$\scriptscriptstyle\frown$}} \over E},
\end{equation}
which is the area covered by the induced rectangle illustrated in Fig.~\ref{Fig:TOS-Int} with \textit{larger TOS indicating a better trade-off between quality and efficiency}.

\begin{table}[t]\scriptsize
\centering
\caption{TOS between quality metrics of \textbf{NMI} and \textbf{AC} as well as efficiency metric of \textbf{runtime}.}
\vspace{-0.3cm}
\label{Tab:TOS-NMI-AC}
\begin{tabular}{l|llllll|llllll}
\hline
\multirow{2}{*}{} & \multicolumn{6}{c|}{TOS between \textbf{NMI} \& \textbf{runtime}} & \multicolumn{6}{c}{TOS between \textbf{AC} \& \textbf{runtime}} \\ \cline{2-13} 
 & \textit{GN}-0.3 & \textit{GN}-0.4 & \textit{L}(\textit{f},0.3) & \textit{L}(\textit{f},0.6) & \textit{L}(\textit{n},0.3) & \textit{L}(\textit{n},0.6) & \textit{GN}-0.3 & \textit{GN}-0.4 & \textit{L}(\textit{f},0.3) & \textit{L}(\textit{f},0.6) & \textit{L}(\textit{n},0.3) & \textit{L}(\textit{n},0.6) \\ \hline
SNMF & 0.76 & 0.83 & 0.69 & 0.41 & 0.66 & 0.40 & 0.73 & 0.82 & 0.61 & 0.35 & 0.58 & 0.33 \\
SC & 0.91 & 0.96 & 0.95 & 0.34 & 0.95 & 0.23 & 0.86 & 0.94 & 0.94 & 0.25 & 0.94 & 0.12 \\
GraClus & 0.90 & 0.98 & 0.90 & 0.53 & 0.90 & 0.54 & 0.83 & 0.95 & 0.72 & 0.46 & 0.72 & 0.46 \\
MC-SBM & 0.83 & 0.94 & 0.96 & 0.36 & 0.96 & 0.38 & 0.35 & 0.66 & 0.89 & 0.25 & 0.89 & 0.26 \\
Locale & 0.93 & 0.99 & 0.92 & 0.50 & 0.92 & 0.51 & 0.87 & 0.98 & 0.86 & 0.44 & 0.86 & 0.44 \\
N2V & 0.89 & 0.93 & 0.88 & 0.54 & 0.87 & 0.54 & 0.83 & 0.90 & 0.82 & 0.53 & 0.82 & 0.53 \\
M-NMF & 0.69 & 0.68 & 0.76 & 0.39 & 0.73 & 0.39 & 0.64 & 0.66 & 0.65 & 0.30 & 0.62 & 0.30 \\
DNR & 0.87 & 0.96 & 0.81 & 0.45 & 0.80 & 0.45 & 0.77 & 0.93 & 0.68 & 0.35 & 0.66 & 0.35 \\
NECS & 0.00 & 0.00 & 0.00 & 0.00 & 0.00 & 0.00 & 0.00 & 0.00 & 0.00 & 0.00 & 0.00 & 0.00 \\
RandNE & 0.78 & 0.93 & 0.57 & 0.19 & 0.58 & 0.19 & 0.63 & 0.86 & 0.41 & 0.13 & 0.41 & 0.13 \\
NRP & 0.91 & 0.97 & 0.79 & 0.45 & 0.80 & 0.46 & 0.82 & 0.90 & 0.65 & 0.36 & 0.65 & 0.37 \\
ProNE & 0.95 & 0.98 & 0.92 & 0.56 & 0.92 & 0.56 & 0.91 & 0.96 & 0.82 & 0.53 & 0.81 & 0.53 \\ \hline
GSAGE & 0.43 & 0.46 & 0.22 & 0.16 & 0.25 & 0.17 & 0.07 & 0.08 & 0.08 & 0.06 & 0.08 & 0.06 \\
GAT & 0.44 & 0.45 & 0.23 & 0.15 & 0.26 & 0.16 & 0.08 & 0.09 & 0.08 & 0.06 & 0.09 & 0.06 \\
GIN & 0.43 & 0.43 & 0.18 & 0.13 & 0.18 & 0.13 & 0.07 & 0.07 & 0.07 & 0.06 & 0.06 & 0.05 \\
GAP & 0.55 & 0.64 & 0.85 & 0.43 & 0.84 & 0.44 & 0.20 & 0.29 & 0.76 & 0.37 & 0.75 & 0.37 \\
ClusNet & 0.13 & 0.17 & 0.81 & 0.50 & 0.79 & 0.51 & 0.03 & 0.03 & 0.78 & 0.53 & 0.77 & 0.53 \\ \hline
\textbf{ICD-M} & \textbf{0.95} & \textbf{0.99} & \textbf{0.97} & \textbf{0.56} & \textbf{0.97} & \textbf{0.56} & \textbf{0.91} & \underline{0.97} & \textbf{0.94} & \textbf{0.53} & \textbf{0.94} & \textbf{0.53} \\
Ranking & 1 & 1 & 1 & 1 & 1 & 1 & 1 & 2 & 1 & 1 & 1 & 1 \\ \hline
\textbf{ICD-C} & \textbf{0.95} & \textbf{0.99} & \underline{0.96} & \underline{0.52} & \underline{0.96} & 0.52 & \underline{0.90} & \underline{0.97} & \underline{0.92} & \underline{0.47} & \underline{0.92} & \underline{0.47} \\
Ranking & 1 & 1 & 2 & 4 & 2 & 3 & 2 & 2 & 2 & 2 & 2 & 2 \\ \hline
\end{tabular}
\vspace{-0.2cm}
\end{table}

\begin{table}[t]\scriptsize
\centering
\caption{TOS between quality metrics of \textbf{modularity} and \textbf{NCut} as well as efficiency metric of \textbf{runtime}.}
\vspace{-0.3cm}
\label{Tab:TOS-Mod-NCut}
\begin{tabular}{l|p{0.2cm}p{0.2cm}p{0.25cm}p{0.25cm}p{0.25cm}p{0.25cm}p{0.22cm}p{0.27cm}p{0.2cm}p{0.37cm}|p{0.2cm}p{0.2cm}p{0.25cm}p{0.25cm}p{0.25cm}p{0.25cm}p{0.22cm}p{0.27cm}p{0.2cm}p{0.37cm}}
\hline
\multirow{2}{*}{} & \multicolumn{10}{c|}{TOS between \textbf{modularity} \& \textbf{runtime}} & \multicolumn{10}{c}{TOS between \textbf{NCut} \& \textbf{runtime}} \\ \cline{2-21} 
 & \textit{GN}.3 & \textit{GN}.4 & \textit{L}(\textit{f},.3) & \textit{L}(\textit{f},.6) & \textit{L}(\textit{n},.3) & \textit{L}(\textit{n},.6) & \textit{Taxi} & \textit{Reddit} & \textit{AS} & \textit{Enron} & \textit{GN}.3 & \textit{GN}.4 & \textit{L}(\textit{f},.3) & \textit{L}(\textit{f},.6) & \textit{L}(\textit{n},.3) & \textit{L}(\textit{n},.6) & \textit{Taxi} & \textit{Reddit} & \textit{AS} & \textit{Enron} \\ \hline
SNMF & 0.51 & 0.58 & 0.64 & 0.54 & 0.62 & 0.51 & 0.83 & 0.64 & 0.44 & 0.57 & 0.80 & 0.84 & 0.82 & 0.83 & 0.79 & 0.79 & 0.95 & 0.76 & 0.58 & 0.65 \\
SC & 0.61 & 0.67 & 0.81 & 0.60 & 0.81 & 0.58 & 0.84 & 0.79 & 0.68 & 0.81 & 0.96 & 0.97 & 0.97 & 0.95 & 0.96 & 0.94 & 0.97 & 0.99 & 0.94 & 0.94 \\
GraClus & 0.62 & 0.68 & 0.76 & 0.66 & 0.76 & 0.66 & 0.81 & 0.71 & 0.73 & 0.72 & 0.99 & 1.00 & 0.99 & 0.99 & 1.00 & 0.99 & 0.94 & 0.65 & 0.99 & 0.84 \\
MC-SBM & 0.63 & 0.68 & 0.82 & 0.61 & 0.82 & 0.61 & 0.69 & 0.51 & 0.77 & 0.55 & 0.98 & 0.98 & 0.99 & 0.98 & 0.99 & 0.99 & 0.93 & 0.75 & 0.97 & 0.66 \\
Locale & 0.63 & 0.69 & 0.81 & 0.65 & 0.81 & 0.65 & 0.85 & 0.83 & 0.75 & 0.83 & 0.99 & 0.99 & 0.99 & 0.99 & 0.99 & 0.99 & 1.00 & 0.99 & 0.99 & 0.96 \\
N2V & 0.60 & 0.65 & 0.76 & 0.63 & 0.75 & 0.63 & 0.50 & 0.00 & 0.70 & 0.10 & 0.95 & 0.95 & 0.93 & 0.93 & 0.92 & 0.94 & 0.58 & 0.00 & 0.89 & 0.11 \\
M-NMF & 0.46 & 0.48 & 0.70 & 0.56 & 0.68 & 0.55 & 0.42 & 0.27 & 0.22 & 0.05 & 0.71 & 0.69 & 0.89 & 0.89 & 0.86 & 0.89 & 0.45 & 0.40 & 0.34 & 0.07 \\
DNR & 0.61 & 0.67 & 0.73 & 0.56 & 0.72 & 0.56 & 0.77 & 0.43 & 0.50 & 0.15 & 0.95 & 0.97 & 0.94 & 0.87 & 0.93 & 0.89 & 0.91 & 0.18 & 0.86 & 0.17 \\
NECS & 0.00 & 0.00 & 0.00 & 0.00 & 0.00 & 0.00 & 0.00 & 0.60 & 0.00 & 0.53 & 0.00 & 0.00 & 0.00 & 0.00 & 0.00 & 0.00 & 0.00 & 0.82 & 0.00 & 0.66 \\
RandNE & 0.60 & 0.67 & 0.59 & 0.51 & 0.59 & 0.52 & 0.86 & 0.52 & 0.53 & 0.57 & 0.96 & 0.98 & 0.93 & 0.64 & 0.94 & 0.74 & 1.00 & 0.19 & 0.44 & 0.46 \\
NRP & 0.62 & 0.68 & 0.66 & 0.56 & 0.66 & 0.57 & 0.83 & 0.53 & 0.52 & 0.44 & 0.96 & 0.98 & 0.98 & 0.91 & 0.98 & 0.93 & 0.97 & 0.48 & 0.97 & 0.64 \\
ProNE & 0.63 & 0.68 & 0.80 & 0.66 & 0.80 & 0.66 & 0.85 & 0.80 & 0.73 & 0.74 & 0.99 & 0.99 & 0.99 & 0.99 & 0.99 & 0.99 & 0.98 & 0.98 & 0.98 & 0.91 \\ \hline
GSAGE & 0.50 & 0.50 & 0.50 & 0.50 & 0.50 & 0.50 & 0.61 & 0.53 & 0.53 & 0.53 & 0.07 & 0.00 & 0.00 & 0.00 & 0.00 & 0.00 & 0.99 & 0.00 & 0.11 & 0.27 \\
GAT & 0.50 & 0.50 & 0.50 & 0.50 & 0.51 & 0.50 & 0.50 & 0.54 & 0.52 & 0.51 & 0.11 & 0.54 & 0.47 & 0.15 & 0.32 & 0.38 & 0.00 & 0.21 & 0.55 & 0.64 \\
GIN & 0.50 & 0.50 & 0.50 & 0.50 & 0.52 & 0.52 & 0.54 & 0.52 & 0.53 & 0.54 & 0.00 & 0.29 & 0.29 & 0.25 & 0.07 & 0.26 & 0.98 & 0.06 & 0.00 & 0.00 \\
GAP & 0.52 & 0.54 & 0.76 & 0.61 & 0.75 & 0.61 & 0.70 & 0.62 & 0.58 & 0.29 & 0.91 & 0.95 & 0.91 & 0.92 & 0.90 & 0.92 & 0.81 & 0.75 & 0.73 & 0.33 \\
ClusNet & 0.49 & 0.49 & 0.75 & 0.61 & 0.73 & 0.61 & 0.51 & 0.58 & 0.63 & 0.00 & 0.79 & 0.84 & 0.89 & 0.88 & 0.88 & 0.88 & 0.59 & 0.74 & 0.81 & 0.00 \\ \hline
\textbf{ICD-M} & \textbf{0.63} & \textbf{0.69} & \textbf{0.83} & \textbf{0.67} & \textbf{0.83} & \textbf{0.67} & \textbf{0.86} & \textbf{0.84} & \textbf{0.78} & \textbf{0.83} & \textbf{0.99} & \underline{0.99} & \textbf{1.00} & \textbf{0.99} & \textbf{1.00} & \textbf{0.99} & \underline{0.99} & \textbf{0.99} & \textbf{0.99} & \textbf{0.96} \\
Ranking & 1 & 1 & 1 & 1 & 1 & 1 & 1 & 1 & 1 & 1 & 1 & 2 & 1 & 1 & 1 & 1 & 2 & 1 & 1 & 1 \\ \hline
\textbf{ICD-C} & \textbf{0.63} & \textbf{0.69} & \textbf{0.83} & \underline{0.66} & \textbf{0.83} & \underline{0.66} & \textbf{0.86} & \textbf{0.84} & \underline{0.77} & \textbf{0.83} & \textbf{0.99} & \underline{0.99} & \textbf{1.00} & \textbf{0.99} & \underline{0.99} & \textbf{0.99} & \underline{0.99} & \textbf{0.99} & \textbf{0.99} & \underline{0.95} \\
Ranking & 1 & 1 & 1 & 2 & 1 & 2 & 1 & 1 & 2 & 1 & 1 & 2 & 1 & 1 & 2 & 1 & 2 & 1 & 1 & 2 \\ \hline
\end{tabular}
\vspace{-0.2cm}
\end{table}

The TOS values w.r.t. the trade-off between four quality metrics (i.e., \textbf{NMI}, \textbf{AC}, \textbf{modularity}, and \textbf{NCut}) and efficiency metric of \textbf{runtime} are depicted in Table~\ref{Tab:TOS-NMI-AC} and \ref{Tab:TOS-Mod-NCut}. As described in Section~\ref{Sec:Setup}, \textit{we used \textbf{NMI} and \textbf{modularity} as validation quality metrics for datasets with and without ground-truth}, respectively. For the TOS w.r.t. \textbf{NMI} and \textbf{modularity} on datasets with and without ground-truth, ICD-M and ICD-C have the best or second-best TOS in most cases.
%indicating that ICD can achieve the best trade-off when using \textbf{NMI} and \textbf{modularity} as validation quality metrics.
For the rest cases (i.e., \textbf{modularity} on datasets with ground-truth as well as \textbf{AC} and \textbf{NCut} on all the datasets), 
ICD-M and ICD-C also have the top-$3$ TOS values. In summary, ICD can achieve a significant trade-off between quality and efficiency of the \textit{online} CD over various baselines.

\subsection{Ablation Study}\label{Sec:Abl}

\begin{table}[t]\scriptsize
\centering
\caption{Ablation study on \textit{L}(\textit{n},0.6), \textit{L}(\textit{f},0.6), and \textit{GN-}0.3.}
\label{Tab:Abl}
\vspace{-0.3cm}
\begin{tabular}{p{0.7cm}|p{0.7cm}p{0.7cm}p{0.7cm}p{1.1cm}|p{0.7cm}p{0.7cm}p{0.7cm}p{1.1cm}|p{0.7cm}p{0.7cm}p{0.7cm}p{1.1cm}}
\hline
\multirow{2}{*}{} & \multicolumn{4}{c|}{\textit{L}(\textit{n},0.6)} & \multicolumn{4}{c|}{\textit{L}(\textit{f},0.6)} & \multicolumn{4}{c}{\textit{GN-}0.3} \\ \cline{2-13} 
 & \textbf{NMI}$\uparrow$(\%) & \textbf{AC}$\uparrow$(\%) & \textbf{Mod}$\uparrow$(\%) & \textbf{Ncut}$\downarrow$ & \textbf{NMI}$\uparrow$(\%) & \textbf{AC}$\uparrow$(\%) & \textbf{Mod}$\uparrow$(\%) & \textbf{Ncut}$\downarrow$ & \textbf{NMI}$\uparrow$(\%) & \textbf{AC}$\uparrow$(\%) & \textbf{Mod}$\uparrow$(\%) & \textbf{Ncut}$\downarrow$ \\ \hline
\textbf{ICD-M} & \textbf{56.48}(2.08) & \textbf{53.42}(2.04) & \textbf{34.22}(0.47) & \textbf{82.76}(8.97) & \textbf{56.07}(2.33) & \textbf{53.57}(2.21) & \textbf{34.14}(0.53) & \textbf{75.92}(8.29) & \textbf{96.26}(0.32) & \textbf{91.96}(0.73) & \textbf{27.92}(0.22) & \textbf{345.82}(19.15) \\
w/oAL & 52.13(2.52) & 49.30(2.41) & 33.35(0.58) & 84.05(8.63) & 51.73(2.86) & 49.45(2.59) & 33.27(0.63) & 77.05(7.33) & 93.31(0.57) & 87.30(1.02) & 26.98(0.30) & 427.44(97.97) \\
w/oFR & 11.19(2.04) & 6.28(0.45) & 10.08(1.77) & 1.01e3(2.37e2) & 11.89(1.80) & 6.57(0.34) & 9.14(1.38) & 6.90e2(1.59e2) & 47.29(0.69) & 16.87(0.57) & 7.27(0.27) & 1.95e4(2.84e3) \\
w/oCR & 49.97(2.79) & 46.70(2.73) & 32.70(0.68) & 86.08(8.71) & 49.50(3.06) & 46.83(2.84) & 32.61(0.72) & 78.93(7.19) & 93.75(0.45) & 87.85(0.83) & 27.11(0.26) & 404.90(85.24) \\
w/oFeat & 16.39(1.59) & 6.14(0.26) & 1.27(0.47) & 3.01e4(4.76e3) & 14.13(1.65) & 5.71(0.13) & 0.24(0.15) & 2.62e4(4.49e3) & 44.20(0.18) & 7.34(0.12) & 0.03(0.03) & 1.68e5(1.28e4) \\
w/oGNN & 1.14(0.25) & 3.92(0.18) & 0.01(0.02) & 1.08e4(2.86e3) & 1.03(0.20) & 4.22(0.15) & 0.01(0.01) & 1.70e3(4.99e2) & 30.08(0.69) & 12.99(0.50) & 4.53(0.19) & 6.11e3(646.84) \\ \hline
\textbf{ICD-C} & \textbf{52.71}(2.52) & \textbf{47.47}(2.59) & \textbf{31.95}(0.62) & \textbf{92.40}(10.05) & \textbf{52.11}(2.79) & \textbf{47.38}(2.68) & \textbf{31.81}(0.65) & \textbf{85.00}(9.02) & \textbf{95.50}(0.43) & \textbf{90.47}(0.85) & \textbf{27.68}(0.25) & \textbf{354.95}(29.97) \\
w/oAL & 35.12(3.01) & 26.15(2.61) & 26.70(0.74) & 113.10(10.23) & 34.83(3.29) & 26.64(2.79) & 26.88(0.80) & 1.01e2(8.61) & 95.00(0.41) & 89.75(0.85) & 27.51(0.25) & 363.44(42.56) \\
w/oFR & 15.46(1.68) & 5.46(0.18) & 0.45(0.11) & 2.08e4(4.79e3) & 15.60(1.75) & 8.25(0.55) & 5.73(0.62) & 1.02e3(8.22e2) & 72.47(0.93) & 56.06(1.46) & 15.06(0.43) & 1.34e3(827.27) \\
w/oCR & 35.16(2.92) & 26.22(2.50) & 26.78(0.71) & 111.62(10.39) & 34.84(3.25) & 26.57(2.65) & 26.90(0.76) & 1.02e2(8.70) & 95.03(0.41) & 89.86(0.77) & 27.52(0.24) & 371.18(83.18) \\
w/oFeat & 13.88(1.39) & 5.57(0.18) & 0.28(0.15) & 2.88e4(5.25e3) & 15.39(1.81) & 5.79(0.14) & 0.76(0.19) & 1.03e4(4.42e3) & 44.91(0.17) & 7.73(0.12) & 0.03(0.03) & 1.65e5(1.32e4) \\
w/oGNN & 8.45(0.45) & 4.78(0.22) & 0.32(0.32) & 5.12e4(7.94e3) & 8.75(0.55) & 5.17(0.15) & 0.86(0.23) & 4.01e4(7.24e3) & 56.25(1.28) & 33.66(1.20) & 11.80(0.32) & 996.69(439.41) \\ \hline
\end{tabular}
\vspace{-0.3cm}
\end{table}

For ICD-M and ICD-C, we also validated the effectiveness of (\romannumeral1) AL loss, (\romannumeral2) FR loss, (\romannumeral3) CR loss, (\romannumeral4) feature input ${{\bf{Z}}_t}$, and (\romannumeral5) GNN by respectively excluding the corresponding components from the original model. In case (\romannumeral4), we used a constant matrix ${\bf{1}}_{N_t \times L}$ with all entries set to $1$ to replace ${\bf{Z}}_t$ (i.e., a standard setting of GNNs for graphs without attributes \cite{xu2018powerful}), while we use a fully-connected network that only takes ${\bf{Z}}_t$ as input to replace GNNs in $G$ with the same layer configurations. We conducted ablation studies on \textit{L}(\textit{f},0.6), \textit{L}(\textit{n},0.6), and \textit{GN}-0.3 whose community structures are more difficult to identify than other datasets. In each case, the mean value $\mu$ and standard deviation $\sigma$ of all the quality metrics (on test set $\Gamma'$) were recorded. Results in the format of $\mu (\sigma)$ are shown in Table~\ref{Tab:Abl}.

According to Table~\ref{Tab:Abl}, the FR loss, feature input ${\bf{Z}}_t$, and GNN are key components to ensure the high quality of ICD because there are significant quality declines in cases without these three components. The AL and CR losses are components to further enhance the CD quality, as they can incorporate the permutation invariant training labels to the \textit{offline} optimization of ICD. In summary, all the components are essential to ensure the high quality of ICD.

\subsection{Convergence Analysis}\label{Sec:Conv}

\begin{figure}[t]
\centering
\begin{minipage}{0.45\linewidth}
 \subfigure[ICD-M, NMI]{
   \frame{
   \includegraphics[width=\textwidth, trim=20 32 38 5, clip]{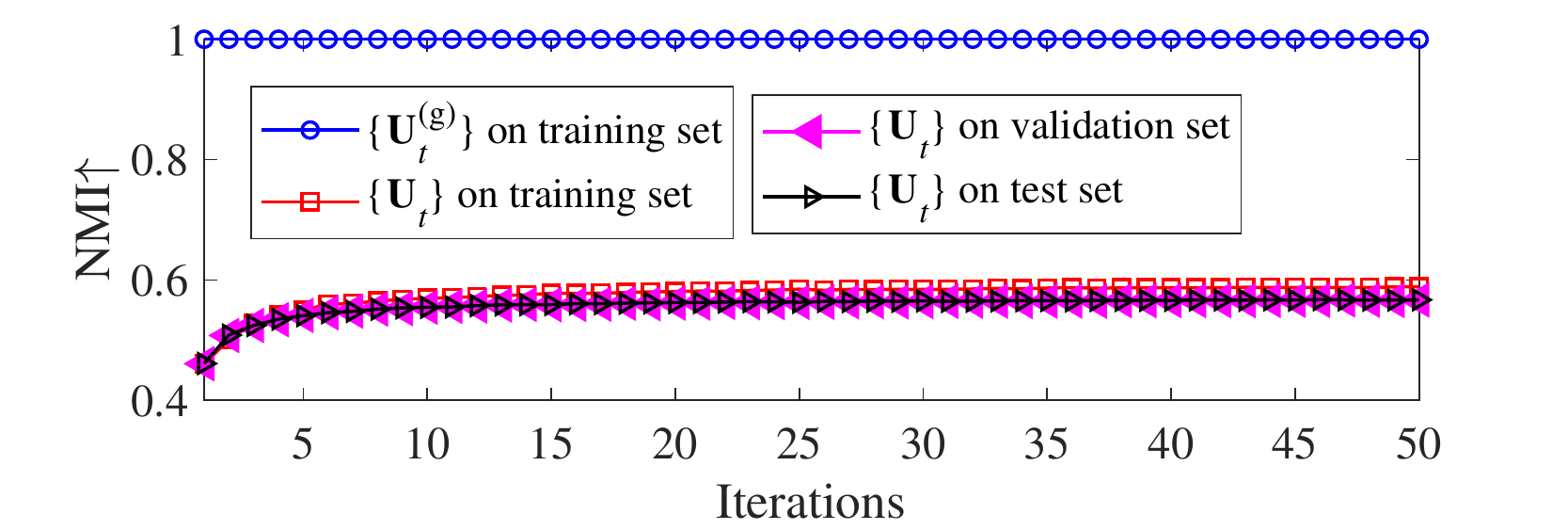}} %.pdf .eps
   }
\end{minipage}
\begin{minipage}{0.45\linewidth}
 \subfigure[ICD-C, NMI]{
   \frame{
   \includegraphics[width=\textwidth, trim=20 32 38 5, clip]{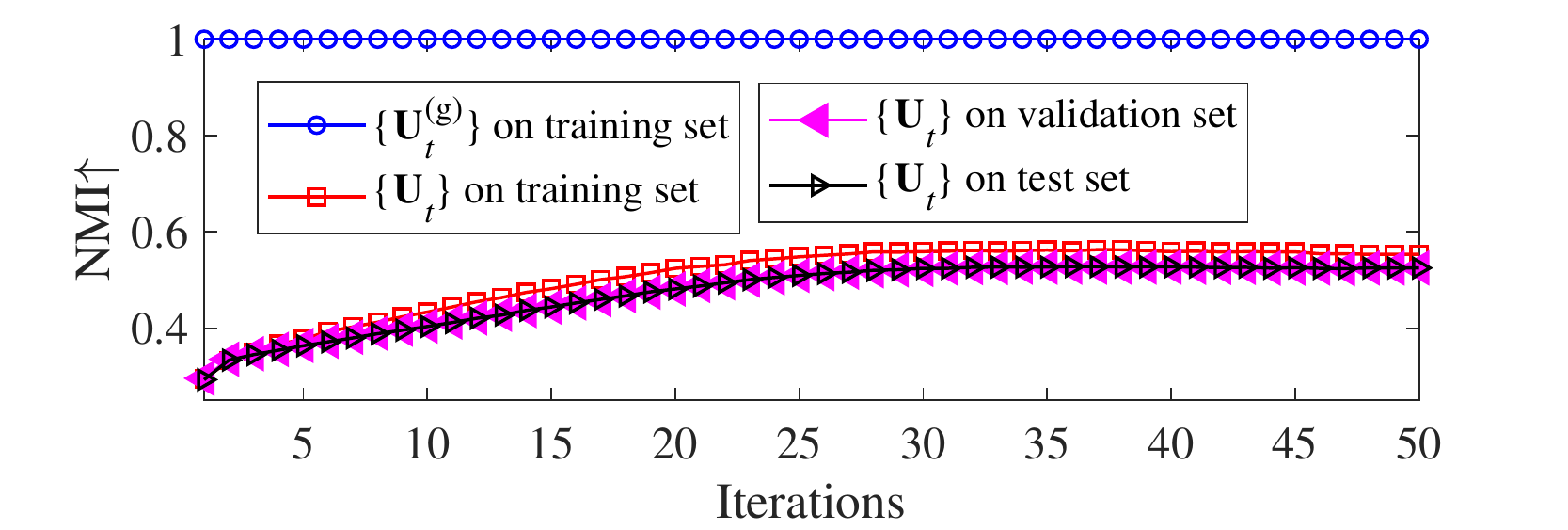}} %.pdf .eps
   }
\end{minipage}
\begin{minipage}{0.45\linewidth}
 \subfigure[ICD-M, AC]{
   \frame{
   \includegraphics[width=\textwidth, trim=20 32 38 5, clip]{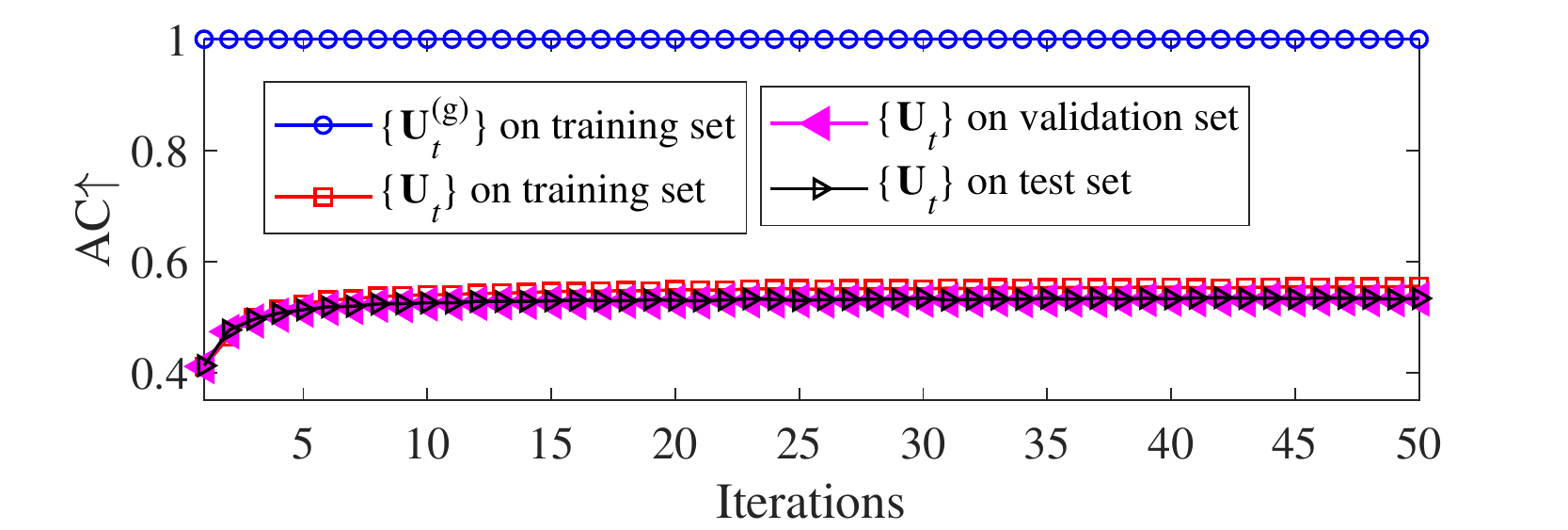}} %.pdf .eps
   }
\end{minipage}
\begin{minipage}{0.45\linewidth}
 \subfigure[ICD-C, AC]{
   \frame{
   \includegraphics[width=\textwidth, trim=20 32 38 5, clip]{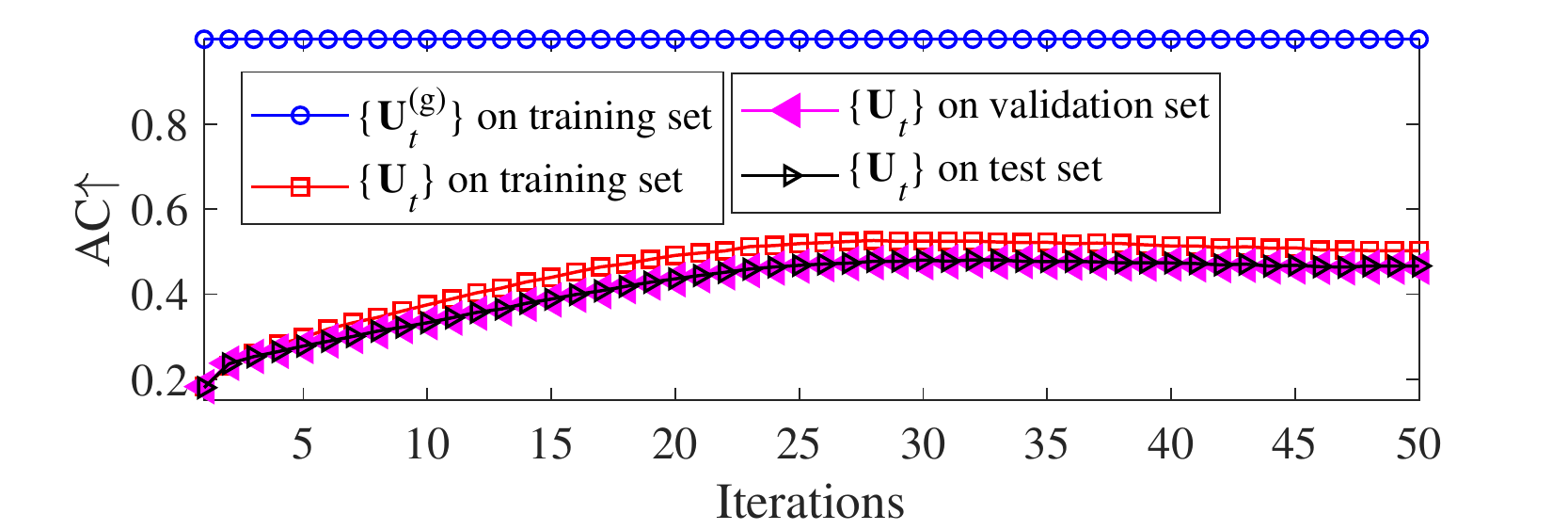}} %.pdf .eps
   }
\end{minipage}
\begin{minipage}{0.45\linewidth}
 \subfigure[ICD-M, modularity]{
   \frame{
   \includegraphics[width=\textwidth, trim=20 32 38 5, clip]{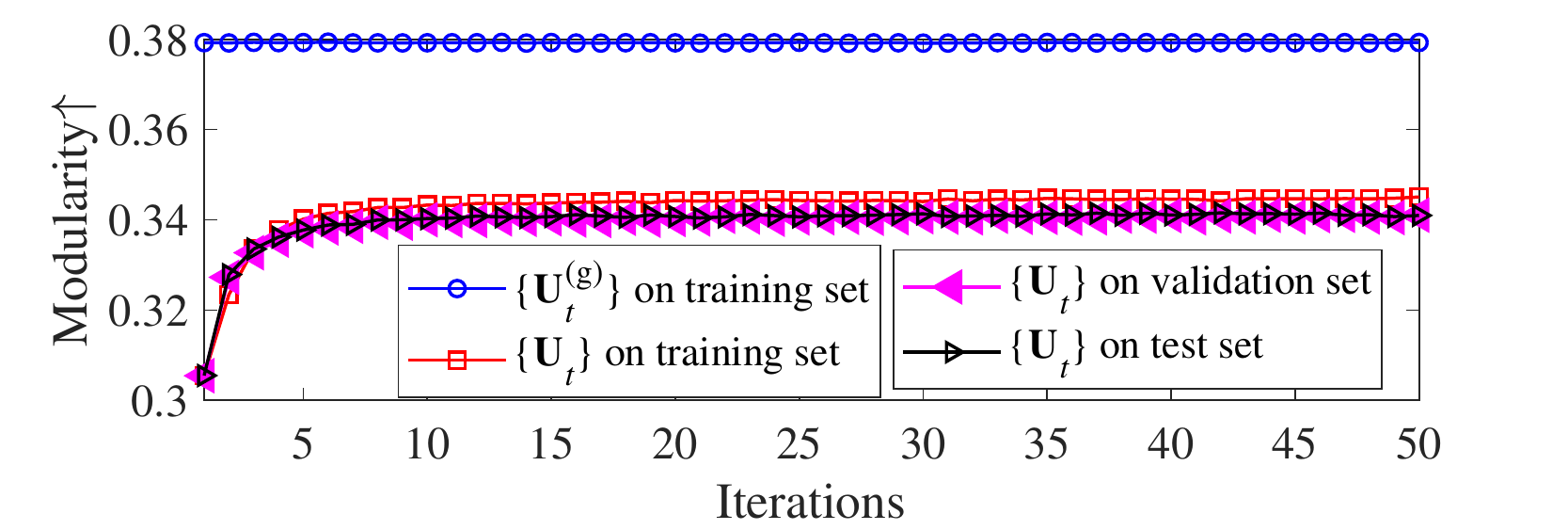}} %.pdf .eps
   }
\end{minipage}
\begin{minipage}{0.45\linewidth}
 \subfigure[ICD-C, modularity]{
   \frame{
   \includegraphics[width=\textwidth, trim=20 32 38 5, clip]{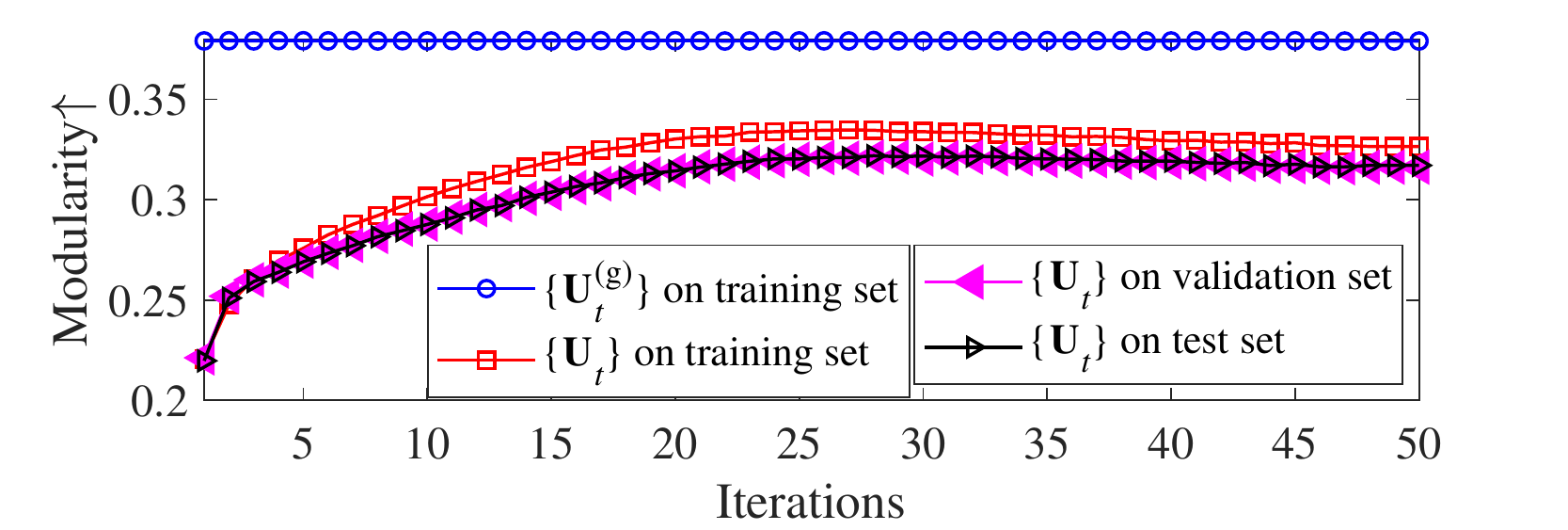}} %.pdf .eps
   }
\end{minipage}
\begin{minipage}{0.45\linewidth}
 \subfigure[ICD-M, NCut]{
   \frame{
   \includegraphics[width=\textwidth, trim=20 32 38 5, clip]{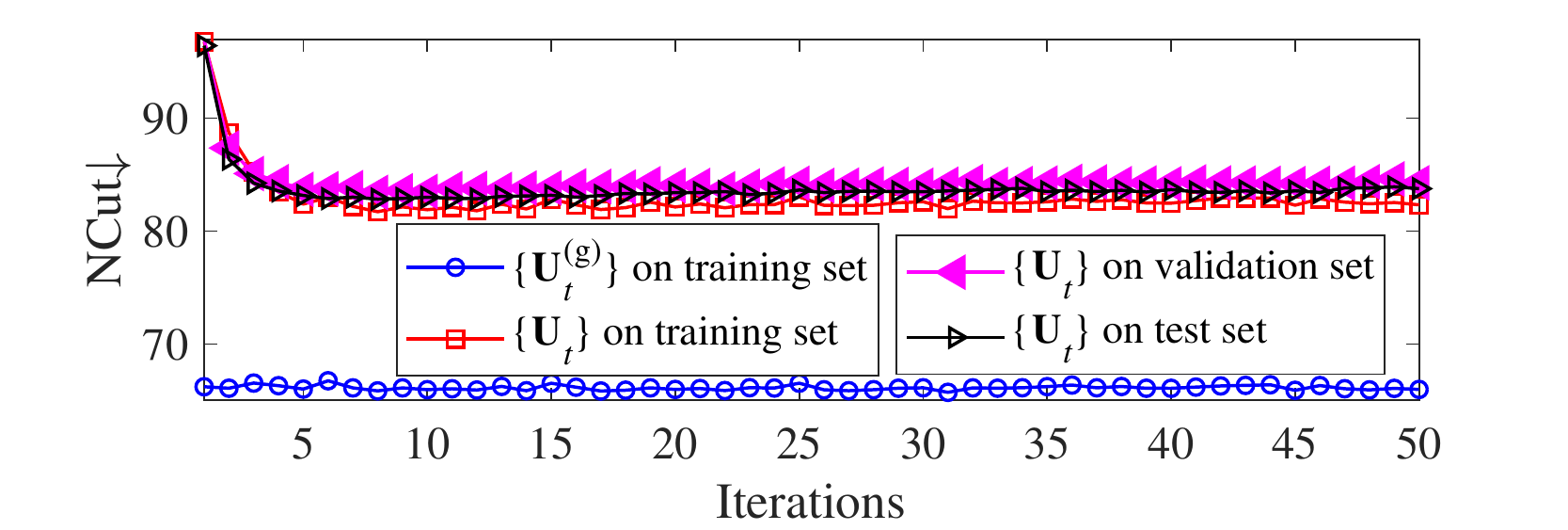}} %.pdf .eps
   }
\end{minipage}
\begin{minipage}{0.45\linewidth}
 \subfigure[ICD-C, NCut]{
   \frame{
   \includegraphics[width=\textwidth, trim=20 32 38 5, clip]{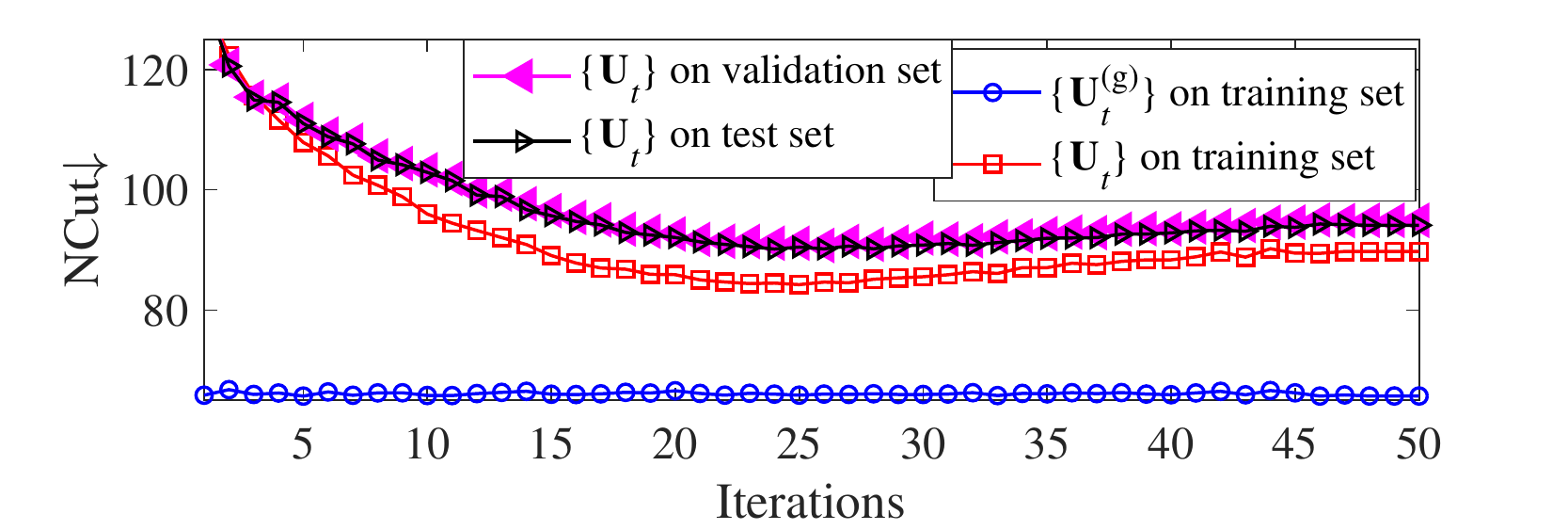}} %.pdf .eps
   }
\end{minipage}
\vspace{-0.3cm}
\caption{Convergence curves of ICD-M and ICD-C on \textit{L}(\textit{n},.6) w.r.t. \textbf{NMI}, \textbf{AC}, \textbf{modularity}, and \textbf{NCut} in the first $50$ iterations ($x$-axis).}
\label{Fig:Conv-ICD}
\vspace{-0.5cm}
\end{figure}

We also analyzed the convergence of the \textit{offline} training of ICD. In each epoch, we recorded the average CD quality (in terms of \textbf{NMI}, \textbf{AC}, \textbf{modularity}, and \textbf{NCut}) w.r.t. the learned embedding $\{ {\bf{U}}_t \}$ on the training set $\Gamma_{\rm{T}}$ and validation set $\Gamma_{\rm{V}}$ as well as the generalized embedding $\{ {\bf{U}}_t \}$ on the test set $\Gamma'$. To further validate the effectiveness of \textit{label-induced embedding} $\{ {\bf{U}}_t^{(g)} \}$, the average CD quality of $\{ {\bf{U}}_t^{(g)} \}$ on $\Gamma_{\rm{T}}$ was also recorded.
We use results on \textit{L}(\textit{n}, 0.6), whose community structures are difficult to identify, as an example. The convergence curves 
%within the first $50$ epochs 
of the average quality w.r.t. (\romannumeral1) $\{ {\bf{U}}_t^{(g)} \}$ on $\Gamma_{\rm{T}}$, (\romannumeral2) $\{ {\bf{U}}_t \}$ on $\Gamma_{\rm{T}}$, (\romannumeral3) $\{ {\bf{U}}_t \}$ on $\Gamma_{\rm{V}}$, and (\romannumeral4) $\{ {\bf{U}}_t \}$ on $\Gamma'$ for ICD-M and ICD-C are shown in Fig.~\ref{Fig:Conv-ICD}.

According to Fig.~\ref{Fig:Conv-ICD}, the CD quality of $\{ {\bf{U}}_t \}$ continuously improves in the first several epochs on $\Gamma_{\rm{T}}$, $\Gamma_{\rm{V}}$, and $\Gamma'$. In particular, the average \textbf{NMI} and \textbf{AC} of $\{ {\bf{U}}_t^{(g)} \}$ on $\Gamma_{\rm{T}}$ are always $1$ for ICD-M and ICD-C. It indicates that the CD results derived from the \textit{auxiliary label-induced embedding} $\{ {\bf{U}}_t^{(g)} \}$ have perfect mapping to the ground-truth.
Therefore, $\{ {\bf{U}}_t^{(g)} \}$ can capture key information of the permutation invariant training `ground-truth' of $\Gamma_{\rm{T}}$ and further enhance the learned embedding $\{ {\bf{U}}_t \}$ by incorporating the training labels via an adversarial process. In summary, Fig.~\ref{Fig:Conv-ICD} validates the effectiveness of the \textit{offline} training and adversarial dual GNN structure of ICD.

\section{Conclusion}\label{Sec:Cons}
In this paper, we proposed an ICD method to alleviate the NP-hard challenge of CD, obtaining a better trade-off between quality and efficiency. We first conduct the \textit{offline} training of an adversarial dual GNN on historical known graphs and generalize the model to newly generated graphs for fast high-quality \textit{online} CD. ICD is a generic method that can tackle the \textit{inductive} CD across graphs with non-fixed number of nodes and communities based on its \textit{inductive} graph embedding scheme and efficient feature extraction module. It can also incorporate the permutation invariant training labels to the \textit{offline} training by combining unsupervised CD objectives (e.g., modularity maximization and NCut minimization) with the adversarial dual GNN. Extensive experiments validate that ICD can achieve a significant trade-off between quality and efficiency over various baselines. In the rest of this section, we conclude this paper by highliting some possible future research directions.

In this study, we considered \textit{inductive} CD without attributes. Although ICD can be easily extended to include node attributes, prior research \cite{qin2018adaptive,qin2021dual} has demonstrated the complicated correlation between graph topology and attributes. On the one hand, \textit{attributes may provide complementary information beyond topology to further improve the CD quality}. On the other hand, \textit{attributes may also carry noise and mismatched characteristics, resulting in unexpected quality declines}. We intend to consider an \textit{adaptive incorporation scheme} between these two heterogeneous sources for ICD. When attributes carry consistent characteristics with topology, we can fully utilize attributes to improve CD quality. When topology `mismatches' with attributes, we need to control the contribution of attributes to avoid quality declines.

%\textcolor{red}{ICD has a similar motivation with existing pre-training methods for graphs \cite{hu2019strategies,qiu2020gcc}. We conduct the \textit{offline} (pre-)training of the adversarial dual GNN on historical graphs but sacrifice opportunities to fine-tune ICD on new graphs for a better trade-off between quality and efficiency. Additional fine-tuning on new graphs of a targeted scenario can avoid the negative transferring by using the validation information from new targeted graphs. Moreover, to select the proper pre-training datasets is still an open issue in recent research of the pre-training and fine-tuning scheme. We intend to conduct the \textit{offline} (pre-)training of ICD on synthetic graphs and explore an efficient strategy to fine-tune ICD on real new graphs of other real scenarios to tackle negative transferring for fast \textit{online} CD.}

Scaling GNNs up to ultra-large graphs (in terms of the number of nodes) is a significant direction in recent research, where some sampling strategies are used to split a large graph into multiple subgraphs \cite{chiang2019cluster,feng2022grand+,zhang2022pasca} (i.e., mini-batches with small number of nodes). Note that ICD is an \textit{inductive} framework across graphs, which can also be trained on and generalized to multiple subgraphs sampled from a single large graph. Existing GNNs that can be scaled up to large graphs usually focus on (semi-)supervised node-level tasks (e.g., node classification) with available node attributes, while this study considers \textit{inductive} unsupervised node-level tasks (i.e., CD) across graphs without attributes. In our future work, we intend to explore efficient (\romannumeral1) feature extraction and (\romannumeral2) subgraph sampling strategies for ICD on ultra-large graphs without attributes.
A possible challenge is to ensure the consistency of embeddings $\{ {\bf{U}}_t \}$ of multiple subgraphs sampled from a common large graph (e.g., embeddings of all the subgraphs are mapped into a common latent space) without the auxiliary information of node attributes.

%\begin{acks}
%To Robert, for the bagels and explaining CMYK and color spaces.
%\end{acks}


\bibliographystyle{ACM-Reference-Format}
%\bibliography{ICD-Ref}

\appendix

\section{HEM Graph Coarsening Algorithm}\label{Sec:HEM}
The HEM graph coarsening procedure is summarized in Algorithm~\ref{Alg:HEM}.

\begin{algorithm}[t]\small
\caption{HEM Graph Coarsening}
\label{Alg:HEM}
\LinesNumbered
\KwIn{input graph ${{\mathcal{G}}_t}$; neighbor-induced features ${\bf{X}}_t$; number of nodes ${N_t}$; reduced dimensionality $L$}
\KwOut{supernode set (i.e., merging membership) $\mathcal{V}_t^*$, coarsening matrix ${{\bf{C}}_t}$}
get node set ${{\mathcal{V}}_t}$ and reweighted edges ${\mathcal{E}}_t^{w}$ from $\{{{\mathcal{G}}_t}, {\bf{X}}_t\}$\\
initialize level index: $k \leftarrow 0$\\
initialize 1st level's node and edge sets: ${\mathcal{V}}_t^{(k)} \leftarrow {\mathcal{V}}_t$, ${\mathcal{E}}_t^{(k)} \leftarrow {\mathcal{E}}_t^{w}$\\
initialize number of (super)nodes: ${N'_t} \leftarrow |{\mathcal{V}}_t^{(0)}|$\\
\While{$N'_t > L$}
{
    initialize next level's node and edge set: ${\mathcal{V}}_t^{(k + 1)} \leftarrow \emptyset$,${\mathcal{E}}_t^{(k + 1)} \leftarrow \emptyset$\\
    %${\mathcal{E}}_t^{(k + 1)} \leftarrow \emptyset$ //Initialize edge set of next level\\
    sort ${{\mathcal{E}}^{(k)}_t}$ based on edge weights in descending order\\ 
    arrange the sorted result as a queue $q({\mathcal{E}}_t^{(k)})$\\
    \While{$q({\mathcal{E}}_t^{(k)}) \ne \emptyset$}
    {
        pop $(v_i^{(k)},v_j^{(k)})$ from $q({\mathcal{E}}_t^{(k)})$\\
        \If{$v_i^{(k)} \in {\mathcal{V}}_t^{(k)}$ {\bf{and}} $v_j^{(k)} \in {\mathcal{V}}_t^{(k)}$}
        {
            delete $v_i^{(k)}$ and $v_j^{(k)}$ from ${\mathcal{V}}_t^{(k)}$\\
            merge $v_i^{(k)}$ and $v_j^{(k)}$ into supernode $v_l^{(k+1)}$\\
            add $v_l^{(k+1)}$ to ${\mathcal{V}}_t^{(k + 1)}$\\
            adjust ${\mathcal{E}}_t^{(k)}$ and ${\mathcal{E}}_t^{(k+1)}$ w.r.t. ${\mathcal{V}}_t^{(k)}$ and ${\mathcal{V}}_t^{(k + 1)}$\\
            update number of (super)nodes: ${N'_t} \leftarrow {N'_t} - 1$\\
        }
        \If{$N'_t=L$}
        {
            Get supernode set output: ${\mathcal{V}}_t^{*} \leftarrow {\mathcal{V}}_t^{(k + 1)} \cup {\mathcal{V}}_t^{(k)}$\\
            derive coarsening matrix ${\bf{C}}_t$ based on ${\mathcal{V}}_t^{*}$\\
            {\bf{return}} ${\mathcal{V}}_t^{*}$ and ${{\bf{C}}_t}$
        }
    }
    ${\mathcal{V}}_t^{(k + 1)} \leftarrow {\mathcal{V}}_t^{(k + 1)} \cup {\mathcal{V}}_t^{(k)}$,
    ${\mathcal{E}}_t^{(k + 1)} \leftarrow {\mathcal{E}}_t^{(k + 1)} \cup {\mathcal{E}}_t^{(k)}$\\
    update level index: $k \leftarrow k+1$\\
}
get supernode set output: ${\mathcal{V}}_t^{*} \leftarrow {\mathcal{V}}_t^{(k)}$\\
derive coarsening matrix ${\bf{C}}_t$ based on ${\mathcal{V}}_t^{*}$\\
{\bf{return}} ${\mathcal{V}}_t^{*}$ and ${{\bf{C}}_t}$
\end{algorithm}

\section{Proof of Fact 1}\label{Sec:Proof}
By the disjoint constraint of CD that each node can only belong to one unique community (i.e., $\forall r \ne s$ s.t. $C_r^t \cap C_s^t = \emptyset$), only one entry in each row of ${\bf{R}}_t$ must be $1$ with rest entries in the same row set to $0$. 
For adjacency matrix ${\bf{A}}_t^{(g)} = {{\bf{R}}_t}{\bf{R}}_t^T$, consider each entry ${({\bf{A}}_t^{(g)})_{ij}} = \sum\nolimits_{r = 1}^{{K_t}} {[{{({{\bf{R}}_t})}_{ir}} {{({{\bf{R}}_t})}_{jr}}]}$.
\begin{itemize}
    \item If nodes $(v_i^t , v_j^t)$ are in the same community $C_s^t$, we have ${{\bf{R}}_{is}} = {{\bf{R}}_{js}} = 1$ and ${({\bf{A}}_t^{(g)})_{ij}} = {({{\bf{R}}_t})_{is}}{({{\bf{R}}_t})_{js}} = 1$. Hence, there is an edge between $(v_i^t, v_j^t)$ with weight $1$ in ${\mathcal{G}}_t^{(g)}$ when they are in the same community.
    \item If nodes $(v_i^t , v_j^t)$ are in different communities, i.e., $v_i^t \in C_r^t$ and $v_j^t \in C_s^t$ ($r \ne s$), we have ${{\bf{R}}_{ir}} = 1$ and ${{\bf{R}}_{js}} = 1$ but ${({\bf{A}}_t^{(g)})_{ij}} = 0$. Thus, there is no edge between $(v_i^t, v_j^t)$ in ${\mathcal{G}}_t^{(g)}$ when they are in different communities.
\end{itemize}
In summary, there is an edge between nodes $(v_i^t, v_j^t)$ in ${\mathcal{G}}_t^{(g)}$ only if they are partitioned into the same community according to the given CD result $C_t = \{ C_1^t, \cdots, C_{K_t}^t \}$.
In particular, each node $v_i^t \in C_r^t$ (\romannumeral1) must have edges connected to all the other nodes in the same community $C_r^t$ and (\romannumeral2) do not have edges connected to nodes in different communities $C_s^t$ ($r \ne s$), forming a fully connected component w.r.t. community $C_r^t$. Hence, for a graph ${{\mathcal{G}}_t}$ with ${K_t}$ communities, its \textit{auxiliary label-induced graph} ${\mathcal{G}}_t^{(g)}$ has ${K_t}$ fully connected components with each component corresponding to one unique community of ${{\mathcal{G}}_t}$, which completes the proof.

\section{Additional Experiment Details}

\subsection{Detailed Experiment Setups}\label{Sec:App-Setup}

\textbf{Experiment Environment}. We used PyTorch to implement ICD while we adopted the official or widely-used open-source implementations of other baselines. All the experiments were conducted on a server with Intel Xeon CPU (E5-2650v4@2.20GHz, 48 cores), 1 Tesla V100 GPU, 61GB memory, as well as the Ubuntu Linux OS. In this setting, methods implemented by PyTorch or TensorFlow were speeded up via GPUs.

\begin{table}[t]\scriptsize
\centering
\caption{Parameter settings and layer configurations of ICD on all the datasets.}
\label{Tab:Param-Layer}
\vspace{-0.3cm}
\begin{tabular}{l|ll|ll|ll}
\hline
\multirow{2}{*}{} & \multicolumn{2}{c|}{\textbf{ICD-M}} & \multicolumn{2}{c|}{\textbf{ICD-C}} & \multicolumn{2}{c}{\textbf{Layer Configurations}} \\ \cline{2-7} 
 & ($\alpha$, $\beta$, $m$, $p$) & ($\eta_G$, $\eta_D$) & ($\alpha$, $\beta$, $m$, $p$) & ($\eta_G$, $\eta_D$) & \multicolumn{1}{l|}{$G$} & $D$ \\ \hline
\textit{GN}-0.3 & (1,5,1,1000) & (5e-4,5e-4) & (1,1,1,1000) & (1e-4,1e-4) & \multicolumn{1}{l|}{2048-1024-512} & 512-128-64-16-1 \\
\textit{GN}-0.4 & (1,1,1,1000) & (5e-4,5e-4) & (1,1,1,1000) & (1e-4,1e-4) & \multicolumn{1}{l|}{2048-1024-512} & 512-128-64-16-1 \\ \hline
\textit{L}(\textit{f},0.3) & (1,1,1,1000) & (5e-4,5e-4) & (1,1e2,5,1000) & (1e-4,1e-4) & \multicolumn{1}{l|}{4096-2048-512-256} & 256-128-64-16-1 \\
\textit{L}(\textit{f},0.6) & (1,3,1,1000) & (1e-4,1e-4) & (1,1e3,2,1000) & (5e-5,5e-5) & \multicolumn{1}{l|}{4096-2048-512-256} & 256-128-64-16-1 \\ \hline
\textit{L}(\textit{n},0.3) & (1,1,1,1000) & (5e-4,5e-4) & (1,1e2,5,1000) & (1e-4,1e-4) & \multicolumn{1}{l|}{4096-2048-512-256} & 256-128-64-16-1 \\
\textit{L}(\textit{n},0.6) & (1,3,1,1000) & (1e-4,1e-4) & (1,1e3,2,1000) & (5e-5,5e-5) & \multicolumn{1}{l|}{4096-2048-512-256} & 256-128-64-16-1 \\ \hline
\textit{Taxi} & (1,0.1,1,2000) & (5e-5,5e-5) & (1,1e2,1,2000) & (5e-5,5e-5) & \multicolumn{1}{l|}{1024-512-256} & 256-64-32-16-1 \\ \hline
\textit{Reddit} & (1,0.1,1,1000) & (5e-5,5e-5) & (1,0.1,1,1000) & (5e-5,5e-5) & \multicolumn{1}{l|}{2048-1024-512-256-100} & 100-64-16-1 \\ \hline
\textit{AS} & (1,0.1,1,500) & (5e-5,5e-5) & (1,1e2,1,500) & (5e-5,5e-5) & \multicolumn{1}{l|}{6000-4096-2048-1024-512-256} & 256-128-64-16-1 \\ \hline
\textit{Enron} & (1,5,1,300) & (5e-5,5e-5) & (1,1e2,1,300) & (5e-5,5e-5) & \multicolumn{1}{l|}{1024-512-256} & 256-128-64-16-1 \\ \hline
\end{tabular}
\vspace{-0.3cm}
\end{table}

\textbf{Parameter Settings and Layer Configurations}. On validation set ${\Gamma_{\rm{V}}}$ of each dataset, we tuned parameters and determined layer configurations for ICD-M and ICD-C. Concretely, we adjusted $\alpha \in \{1, 5, 10\}$ and $m \in \{1,2, \cdots 5\}$ for both the variants, while we tuned $\beta \in \{0.1, 1,2, \cdots 5\}$ and $\beta \in \{0.1, 1, 100, 1000\}$ for ICD-M and ICD-C, respectively. The recommended parameter settings and layer configurations for each dataset are depicted in Table~\ref{Tab:Param-Layer}, where dimensionality of the first and last layer of $G$ is set to the dimensionality of the reduced feature and graph embedding. Note that we use the same layer configurations for both ICD-M and ICD-C on each dataset. Moreover, we set the number of epochs $n=100$ for all the datasets.

\subsection{Detailed Quantitative Evaluation Results}\label{Sec:App-Eva-Res}

\begin{table}[t]\scriptsize
\centering
\caption{Quantitative evaluation of CD quality in terms of \textbf{NMI}$\uparrow$ (\%) and \textbf{AC}$\uparrow$ (\%).}
\label{Tab:Eva-NMI-AC}
\vspace{-0.3cm}
\begin{tabular}{l|p{0.7cm}p{0.7cm}p{0.7cm}p{0.7cm}p{0.7cm}l|p{0.7cm}p{0.7cm}p{0.7cm}p{0.7cm}p{0.7cm}l}
\hline
\multirow{2}{*}{} & \multicolumn{6}{c|}{\textbf{NMI}$\uparrow$ (\%)} & \multicolumn{6}{c}{\textbf{AC}$\uparrow$ (\%)} \\ \cline{2-13} 
 & \textit{GN}.3 & \textit{GN}.4 & \textit{L}(\textit{f},.3) & \textit{L}(\textit{f},.6) & \textit{L}(\textit{n},.3) & \textit{L(n,0.6)} & \textit{GN}.3 & \textit{GN}.4 & \textit{L}(\textit{f},.3) & \textit{L}(\textit{f},.6) & \textit{L}(\textit{n},.3) & \textit{L}(\textit{n},.6) \\ \hline
SNMF & 94.37~(0.69) & 98.79~(0.36) & 83.53~(1.60) & 49.15~(2.56) & 83.59~(1.59) & 49.98~(2.19) & 90.25~(1.32) & 97.46~(0.78) & 73.37~(3.20) & 41.61~(2.56) & 72.96~(2.97) & 41.95~(2.40) \\
SC & 94.26~(0.41) & 98.73~(0.24) & 98.19~(0.39) & 35.78~(3.65) & 98.21~(0.35) & 24.66~(2.76) & 89.29~(0.85) & 96.67~(0.54) & 96.35~(1.07) & 26.01~(3.62) & 97.27~(0.96) & 12.26~(1.59) \\
GraClus & 90.72~(0.76) & 98.10~(0.30) & 90.58~(1.14) & 53.06~(3.00) & 90.60~(1.09) & 53.90~(2.75) & 83.16~(1.54) & 95.11~(0.94) & 72.00~(3.40) & 46.19~(3.02) & 71.81~(3.33) & 46.50~(2.78) \\
MC-SBM & 84.63~(1.84) & 95.33~(2.68) & 96.67~(0.56) & 36.49~(5.36) & 96.62~(0.56) & 38.44~(4.79) & 36.15~(7.32) & 67.00~(20.23) & 89.97~(2.13) & 24.89~(4.27) & 89.60~(2.05) & 25.88~(3.84) \\
Locale & 93.44~(0.41) & 99.60~(0.09) & 92.40~(1.26) & 50.83~(3.19) & 92.42~(1.31) & 51.36~(2.72) & 88.13~(0.95) & 98.33~(0.56) & 86.49~(2.38) & 44.50~(3.07) & 86.27~(2.51) & 44.51~(2.63) \\
N2V & 93.94~(0.52) & 98.41~(0.25) & 94.35~(0.74) & 57.35~(2.45) & 94.47~(0.73) & 58.13~(2.18) & 87.24~(1.01) & 94.63~(0.66) & 88.53~(2.12) & 56.22~(2.39) & 88.51~(2.08) & 56.35~(2.11) \\
M-NMF & 95.96~(0.41) & 99.12~(0.13) & 83.78~(1.51) & 43.17~(3.17) & 83.49~(1.56) & 43.95~(3.01) & 89.63~(0.96) & 95.03~(0.68) & 72.01~(2.79) & 33.48~(2.52) & 70.82~(3.01) & 33.77~(2.57) \\
DNR & 89.79~(0.72) & 98.70~(0.23) & 84.84~(1.31) & 46.51~(2.73) & 84.84~(1.47) & 47.48~(2.47) & 79.11~(1.44) & 95.18~(0.70) & 70.53~(2.88) & 36.31~(2.69) & 69.94~(3.12) & 36.81~(2.50) \\
NECS & 91.67~(0.74) & 98.58~(0.23) & 88.86~(1.39) & 50.63~(3.03) & 88.80~(1.43) & 51.39~(2.81) & 82.30~(1.41) & 93.56~(0.87) & 80.01~(2.35) & 44.81~(2.78) & 79.93~(2.43) & 45.09~(2.79) \\
RandNE & 79.18~(1.01) & 94.08~(0.47) & 57.63~(2.06) & 18.57~(2.16) & 58.08~(2.08) & 18.87~(1.95) & 63.77~(1.49) & 86.56~(1.03) & 40.67~(1.77) & 12.91~(1.25) & 40.64~(1.73) & 12.80~(1.17) \\
NRP & 93.06~(0.55) & 98.51~(0.15) & 80.04~(1.47) & 45.00~(2.94) & 80.35~(1.59) & 46.57~(2.63) & 83.27~(1.10) & 91.22~(0.86) & 65.74~(3.17) & 36.55~(2.71) & 65.28~(3.28) & 37.69~(2.59) \\
ProNE & 96.55~(0.30) & 99.54~(0.08) & 93.33~(0.81) & 56.17~(2.80) & 93.31~(0.89) & 56.53~(2.51) & 92.10~(0.66) & 97.66~(0.42) & 82.56~(2.55) & 53.98~(2.65) & 82.23~(2.52) & 53.83~(2.64) \\ \hline
GSAGE & 43.21~(0.21) & 45.98~(0.09) & 22.52~(2.21) & 15.79~(1.89) & 24.67~(2.03) & 16.66~(1.77) & 7.44~(0.12) & 7.66~(0.13) & 7.55~(0.25) & 5.79~(0.15) & 7.58~(0.34) & 5.66~(0.19) \\
GAT & 44.55~(0.20) & 45.03~(0.25) & 22.99~(1.88) & 15.49~(1.78) & 25.91~(1.94) & 15.83~(1.70) & 8.00~(0.13) & 8.66~(0.16) & 8.52~(0.36) & 6.22~(0.16) & 8.78~(0.40) & 5.69~(0.19) \\
GIN & 43.49~(0.17) & 43.26~(0.19) & 18.49~(1.83) & 13.32~(1.47) & 17.61~(1.71) & 13.43~(1.22) & 7.46~(0.11) & 7.51~(0.11) & 6.97~(0.23) & 5.83~(0.15) & 6.08~(0.30) & 5.37~(0.21) \\
GAP & 56.13~(0.31) & 64.37~(0.39) & 93.73~(1.76) & 47.30~(3.23) & 93.64~(1.65) & 48.01~(3.11) & 20.30~(0.57) & 29.71~(0.76) & 83.63~(4.16) & 40.06~(3.34) & 83.06~(3.92) & 40.08~(3.41) \\
ClusNet & 13.10~(1.44) & 17.65~(2.10) & 89.84~(1.31) & 55.49~(2.04) & 90.10~(1.27) & 56.04~(1.78) & 3.13~(0.23) & 2.94~(0.28) & 86.45~(2.47) & 58.20~(1.82) & 86.87~(2.36) & 58.25~(1.65) \\ \hline
\textbf{ICD-M} & 96.26~(0.32) & 99.38~(0.11) & 97.21~(0.49) & 56.07~(2.33) & 97.30~(0.44) & 56.48~(2.08) & 91.96~(0.73) & 97.36~(0.46) & 94.29~(1.41) & 53.57~(2.21) & 94.24~(1.38) & 53.42~(2.04) \\
\textbf{ICD-C} & 95.50~(0.43) & 99.46~(0.09) & 96.44~(0.57) & 52.11~(2.79) & 96.56~(0.59) & 52.71~(2.52) & 90.47~(0.85) & 97.48~(0.46) & 92.58~(1.49) & 47.38~(2.68) & 92.46~(1.46) & 47.47~(2.59) \\ \hline
\end{tabular}
%\vspace{-0.2cm}
\end{table}

\begin{table}[t]\scriptsize
\centering
\caption{Quantitative evaluation of CD quality in terms of \textbf{modularity}$\uparrow$ (\%).}
\label{Tab:Eva-Mod}
\vspace{-0.3cm}
\begin{tabular}{l|llllllllll}
\hline
 & \textit{GN}-0.3 & \textit{GN}-0.4 & \textit{L}(\textit{f},0.3) & \textit{L}(\textit{f},0.6) & \textit{L}(\textit{n},0.3) & \textit{L}(\textit{n},0.6) & \textit{Taxi} & \textit{Reddit} & \textit{AS} & \textit{Enron} \\ \hline
SNMF & 27.28~(0.29) & 37.79~(0.27) & 56.12~(1.58) & 29.22~(0.63) & 56.03~(1.50) & 29.43~(0.60) & 73.72~(0.41) & 68.29~(7.16) & 53.32~(2.55) & 73.88~(14.55) \\
SC & 27.07~(0.23) & 37.81~(0.23) & 67.32~(0.20) & 26.64~(1.06) & 67.50~(0.26) & 23.62~(2.83) & 73.80~(0.59) & 61.19~(7.68) & 43.14~(6.60) & 71.43~(19.05) \\
GraClus & 25.00~(0.48) & 37.34~(0.30) & 53.54~(1.98) & 31.61~(0.92) & 53.43~(1.87) & 31.71~(0.83) & 73.05~(0.64) & 45.27~(11.59) & 46.81~(4.83) & 62.02~(21.93) \\
MC-SBM & 28.36~(0.63) & 38.51~(0.27) & 65.99~(1.66) & 22.85~(2.74) & 66.26~(2.14) & 23.87~(2.36) & 42.02~(13.46) & 3.55~(13.69) & 55.36~(4.19) & 13.34~(19.12) \\
Locale & 26.87~(0.25) & 38.26~(0.20) & 63.11~(1.00) & 29.93~(0.84) & 63.19~(1.07) & 30.04~(0.73) & 71.56~(1.56) & 68.85~(5.32) & 51.82~(1.85) & 73.07~(12.04) \\
N2V & 27.08~(0.27) & 37.56~(0.22) & 62.90~(1.07) & 34.28~(0.52) & 63.02~(1.03) & 34.40~(0.47) & 73.64~(0.54) & 69.31~(7.90) & 58.19~(2.80) & 76.86~(13.31) \\
M-NMF & 27.67~(0.24) & 38.05~(0.20) & 55.85~(1.41) & 24.07~(0.92) & 55.35~(1.57) & 24.32~(0.94) & 73.32~(0.72) & 12.68~(12.98) & 14.55~(2.62) & 4.70~(12.19) \\
DNR & 24.72~(0.44) & 37.70~(0.25) & 51.34~(2.16) & 16.07~(1.76) & 51.22~(2.20) & 16.61~(1.69) & 70.04~(1.92) & 5.00~(4.47) & 13.06~(3.41) & 6.67~(10.21) \\
NECS & 26.02~(0.34) & 37.62~(0.26) & 60.46~(1.06) & 29.50~(0.80) & 60.60~(1.11) & 29.80~(0.79) & 73.46~(0.76) & 40.87~(19.41) & 24.53~(3.36) & 58.46~(22.54) \\
RandNE & 21.08~(0.47) & 35.61~(0.34) & 18.44~(1.20) & 3.26~(0.34) & 18.48~(1.21) & 3.38~(0.32) & 73.12~(0.72) & 5.52~(5.04) & 5.30~(0.54) & 21.53~(12.32) \\
NRP & 26.06~(0.36) & 37.47~(0.26) & 33.99~(2.51) & 12.82~(1.75) & 34.17~(2.65) & 14.23~(1.83) & 71.60~(1.01) & 7.61~(4.41) & 5.88~(3.92) & 1.87~(13.15) \\
ProNE & 28.10~(0.20) & 38.20~(0.19) & 60.96~(1.19) & 33.66~(0.69) & 60.98~(1.17) & 34.00~(0.61) & 73.61~(0.64) & 63.57~(4.87) & 49.10~(3.87) & 62.80~(9.98) \\ \hline
GSAGE & 0.01~(0.01) & 0.04~(0.03) & 0.28~(0.17) & 0.18~(0.11) & 0.56~(0.27) & 0.17~(0.16) & 21.85~(1.11) & 6.94~(3.72) & 5.69~(1.47) & 9.41~(6.55) \\
GAT & 0.08~(0.03) & 0.27~(0.05) & 0.99~(0.17) & 0.08~(0.12) & 1.71~(0.22) & 0.02~(0.11) & 1.11~(0.42) & 8.69~(11.82) & 4.60~(0.84) & 4.20~(4.89) \\
GIN & 0.01~(0.03) & 0.03~(0.03) & 0.50~(0.11) & 0.06~(0.08) & 4.73~(0.29) & 4.19~(0.18) & 8.09~(0.44) & 4.82~(7.30) & 5.62~(7.60) & 10.82~(3.61) \\
GAP & 5.24~(0.19) & 9.95~(0.36) & 66.45~(0.59) & 32.96~(0.73) & 66.65~(0.57) & 32.96~(0.77) & 71.79~(1.30) & 65.08~(6.24) & 58.31~(2.68) & 75.47~(12.57) \\
ClusNet & 0.01~(0.13) & 0.02~(0.12) & 65.52~(0.53) & 35.10~(0.51) & 65.77~(0.54) & 35.25~(0.44) & 72.54~(1.27) & 55.31~(10.37) & 55.87~(2.42) & 72.56~(18.75) \\ \hline
\textbf{ICD-M} & 27.92~(0.22) & 38.17~(0.20) & 66.26~(0.54) & 34.14~(0.53) & 66.45~(0.56) & 34.22~(0.47) & 74.01~(0.51) & 69.38~(5.49) & 56.82~(2.71) & 74.33~(13.51) \\
\textbf{ICD-C} & 27.68~(0.25) & 38.19~(0.19) & 65.72~(0.59) & 31.81~(0.65) & 65.83~(0.64) & 31.95~(0.62) & 74.00~(0.42) & 69.37~(5.47) & 55.40~(2.26) & 74.15~(11.78) \\ \hline
\end{tabular}
%\vspace{-0.2cm}
\end{table}

\begin{table}[t]\scriptsize
\centering
\caption{Quantitative evaluation of CD quality in terms of \textbf{NCut}$\downarrow$.}
\label{Tab:Eva-NCut}
\vspace{-0.3cm}
\begin{tabular}{l|p{0.95cm}p{0.95cm}p{0.95cm}p{0.95cm}p{0.95cm}p{0.95cm}p{0.95cm}p{0.95cm}p{0.95cm}l}
\hline
 & \textit{GN}.3 & \textit{GN}.4 & \textit{L}(\textit{f},.3) & \textit{L}(\textit{f},.6) & \textit{L}(\textit{n},.3) & \textit{L}(\textit{n},.6) & \textit{Taxi} & \textit{Reddit} & \textit{AS} & \textit{Enron} \\ \hline
SNMF & 352.04~(8.94) & 213.23~(14.28) & 38.80~(5.62) & 98.49~(10.95) & 43.22~(6.33) & 106.72~(11.89) & 0.79~(0.13) & 6.71~(8.44) & 10.05~(1.60) & 23.07~(134.44) \\
SC & 342.36~(34.29) & 286.04~(118.40) & 35.38~(35.34) & 115.48~(29.30) & 40.20~(45.35) & 88.11~(43.17) & 0.77~(0.13) & 5.85~(1.57) & 2.24~(0.29) & 2.79~(10.56) \\
GraClus & 396.54~(24.44) & 229.87~(74.71) & 39.55~(7.91) & 79.57~(9.25) & 44.04~(7.73) & 86.74~(10.02) & 0.85~(0.11) & 2.21e3~(1.75e3) & 17.86~(6.75) & 308.32~(782.49) \\
MC-SBM & 319.61~(673.35) & 157.73~(216.67) & 63.38~(209.66) & 116.96~(587.71) & 117.97~(404.02) & 132.02~(553.07) & 43.96~(214.21) & 1.61e3~(2.55e3) & 242.10~(325.87) & 2.00e3~(2.36e3) \\
Locale & 485.66~(158.64) & 291.16~(81.18) & 71.52~(72.48) & 91.94~(21.52) & 98.17~(131.05) & 106.27~(33.27) & 0.93~(0.09) & 4.80~(3.82) & 12.03~(4.52) & 12.15~(52.66) \\
N2V & 367.96~(8.05) & 237.47~(33.18) & 36.49~(48.37) & 74.62~(8.14) & 37.08~(33.68) & 81.56~(9.01) & 0.81~(0.12) & 4.45~(4.05) & 6.11~(1.05) & 14.91~(87.09) \\
M-NMF & 1.05e3~(513.45) & 650.96~(179.14) & 347.52~(337.84) & 152.24~(121.42) & 351.10~(324.69) & 150.62~(97.02) & 80.68~(12.86) & 1.04e3~(777.17) & 1.21e3~(770.07) & 1.20e3~(692.38) \\
DNR & 3.84e3~(1.14e3) & 951.12~(407.76) & 558.60~(683.21) & 3.36e3~(2.11e3) & 666.72~(738.18) & 3.73e3~(2.16e3) & 1.03~(0.13) & 5.11e3~(3.48e3) & 273.69~(233.73) & 2.57e3~(1.65e3) \\
NECS & 873.78~(389.12) & 480.63~(230.88) & 161.73~(226.52) & 133.21~(123.22) & 186.38~(244.50) & 133.89~(115.84) & 0.81~(0.14) & 246.91~(544.06) & 274.28~(297.94) & 106.10~(124.51) \\
RandNE & 5.07e3~(877.58) & 1.36e3~(372.75) & 2.24e3~(1.50e3) & 1.31e4~(2.30e3) & 2.83e3~(1.83e3) & 1.41e4~(2.56e3) & 0.81~(0.12) & 5.33e3~(3.50e3) & 6.62e3~(3.08e3) & 3.26e3~(2.35e3) \\
NRP & 3.21e3~(762.04) & 1.06e3~(251.08) & 450.30~(381.71) & 3.05e3~(1.34e3) & 561.89~(432.96) & 3.18e3~(1.32e3) & 0.97~(0.17) & 3.33e3~(1.64e3) & 172.04~(83.36) & 1.68e3~(708.02) \\
ProNE & 343.15~(36.02) & 295.37~(92.08) & 47.03~(125.66) & 74.99~(8.24) & 59.34~(168.78) & 81.52~(9.22) & 0.78~(0.13) & 5.65~(2.00) & 12.88~(4.02) & 35.16~(109.78) \\ \hline
GSAGE & 1.42e5~(1.17e4) & 2.05e5~(1.29e4) & 3.42e4~(4.58e3) & 3.71e4~(5.51e3) & 4.79e4~(6.36e3) & 5.50e4~(6.79e3) & 12.41~(2.18) & 6.59e3~(3.97e3) & 1.05e4~(3.85e3) & 4.59e3~(1.68e3) \\
GAT & 1.36e5~(1.27e4) & 9.49e4~(9.97e3) & 1.81e4~(5.16e3) & 3.17e4~(5.66e3) & 3.26e4~(5.31e3) & 3.40e4~(6.86e3) & 1.20e3~(4.26e2) & 5.21e3~(2.86e3) & 5.27e3~(1.84e3) & 2.17e3~(953.05) \\
GIN & 1.53e5~(1.11e4) & 1.45e5~(1.10e4) & 2.42e4~(6.69e3) & 2.77e4~(7.50e3) & 4.45e4~(8.52e3) & 4.05e4~(9.36e3) & 25.32~(4.64) & 6.16e3~(3.74e3) & 1.18e4~(3.01e3) & 6.34e3~(2.60e3) \\
GAP & 1.28e4~(2.70e3) & 7.99e3~(2.00e3) & 17.33~(2.74) & 69.55~(7.80) & 18.35~(2.51) & 76.24~(9.55) & 0.94~(0.15) & 4.15~(4.48) & 6.13~(1.24) & 1.35~(0.41) \\
ClusNet & 2.91e4~(5.12e3) & 2.80e4~(6.12e3) & 297.67~(166.97) & 1.11e3~(418.13) & 291.28~(156.83) & 1.19e3~(422.28) & 6.92~(42.51) & 28.30~(53.44) & 31.47~(49.07) & 23.61~(99.44) \\ \hline
\textbf{ICD-M} & 345.82~(19.15) & 317.82~(91.99) & 32.39~(38.31) & 75.92~(8.29) & 34.84~(35.44) & 82.76~(8.97) & 0.77~(0.12) & 4.74~(2.19) & 7.26~(1.27) & 4.53~(16.94) \\
\textbf{ICD-C} & 354.95~(29.97) & 289.53~(88.19) & 29.22~(21.72) & 85.00~(9.02) & 38.96~(47.41) & 92.40~(10.05) & 0.77~(0.13) & 4.71~(2.15) & 9.66~(3.46) & 4.36~(16.50) \\ \hline
\end{tabular}
%\vspace{-0.1cm}
\end{table}

\begin{table}[t]\scriptsize
\centering
\caption{Quantitative evaluation of CD quality in terms of \textbf{runtime} (sec).}
\label{Tab:Eva-Time}
\vspace{-0.3cm}
\begin{tabular}{l|p{0.95cm}p{1cm}p{0.95cm}p{0.95cm}p{0.95cm}p{0.95cm}p{0.95cm}p{0.95cm}p{0.95cm}l}
\hline
 & \textit{GN}.3 & \textit{GN}.4 & \textit{L}(\textit{f},.3) & \textit{L}(\textit{f},.6) & \textit{L}(\textit{n},.3) & \textit{L}(\textit{n},.6) & \textit{Taxi} & \textit{Reddit} & \textit{AS} & \textit{Enron} \\ \hline
SNMF & 326.90~(63.67) & 232.32~(51.10) & 129.15~(32.03) & 136.21~(0.64) & 155.84~(43.87) & 194.13~(54.34) & 2.13~(1.40) & 6.01~(8.97) & 112.31~(83.53) & 1.68~(1.25) \\
SC & 62.14~(1.08) & 46.73~(0.69) & 21.04~(1.28) & 42.54~(2.21) & 27.00~(3.85) & 53.26~(6.95) & 1.34~(0.08) & 0.35~(0.52) & 15.13~(9.66) & 0.28~(0.33) \\
GraClus & 4.39~(0.14) & 4.23~(0.17) & 2.89~(1.15) & 3.04~(1.35) & 2.83~(1.11) & 3.23~(1.37) & 2.90~(0.17) & 0.62~(0.19) & 1.51~(1.50) & 0.54~(0.17) \\
MC-SBM & 30.10~(23.77) & 25.66~(25.48) & 5.29~(1.61) & 11.00~(6.85) & 6.19~(3.42) & 11.62~(5.59) & 1.45~(0.38) & 0.20~(0.32) & 2.16~(1.11) & 0.17~(0.20) \\
Locale & 13.62~(0.65) & 11.81~(0.78) & 4.40~(0.38) & 5.64~(0.51) & 5.89~(1.01) & 6.57~(1.07) & 0.18~(0.02) & 0.32~(0.43) & 2.71~(1.83) & 0.18~(0.02) \\
N2V & 81.34~(1.14) & 79.07~(0.99) & 52.21~(0.81) & 53.21~(0.81) & 57.12~(3.40) & 58.16~(3.66) & 19.70~(0.31) & 25.37~(17.83) & 29.90~(14.21) & 4.27~(1.49) \\
M-NMF & 472.92~(53.45) & 455.71~(14.77) & 73.21~(8.35) & 84.30~(9.03) & 96.55~(14.63) & 102.42~(14.56) & 24.29~(2.30) & 13.36~(17.62) & 165.40~(105.72) & 4.36~(2.19) \\
DNR & 42.94~(0.30) & 40.92~(0.21) & 30.71~(0.23) & 31.19~(0.26) & 39.23~(3.87) & 40.02~(3.91) & 4.39~(0.80) & 4.70~(0.71) & 30.51~(16.07) & 3.42~(0.74) \\
NECS & 1670.29~(57.72) & 1470.47~(103.41) & 741.24~(32.72) & 822.94~(49.96) & 746.38~(72.79) & 929.72~(79.06) & 46.95~(1.22) & 3.64~(4.89) & 266.28~(166.50) & 1.60~(1.26) \\
RandNE & 14.73~(0.22) & 12.48~(0.14) & 2.12~(0.18) & 2.49~(0.22) & 2.54~(0.30) & 2.91~(0.33) & 0.18~(0.02) & 0.48~(0.27) & 0.74~(0.31) & 0.26~(0.07) \\
NRP & 29.24~(1.87) & 19.26~(0.85) & 6.72~(1.13) & 8.59~(0.39) & 7.41~(0.73) & 9.95~(0.91) & 1.60~(0.14) & 0.53~(0.31) & 3.68~(1.66) & 0.65~(0.25) \\
ProNE & 21.17~(0.33) & 18.85~(0.13) & 6.67~(0.23) & 7.48~(0.29) & 7.85~(0.70) & 8.77~(0.73) & 0.99~(0.03) & 0.48~(0.31) & 4.89~(2.65) & 0.42~(0.13) \\ \hline
GSAGE & 8.18~(0.30) & 8.24~(0.31) & 2.58~(0.17) & 2.50~(0.16) & 3.12~(0.33) & 3.38~(0.36) & 0.20~(0.02) & 0.35~(0.19) & 1.29~(0.6155) & 0.17~(0.04) \\
GAT & 5.37~(0.12) & 5.56~(0.15) & 2.03~(0.16) & 2.37~(0.19) & 2.77~(0.31) & 2.81~(0.29) & 0.21~(0.02) & 0.33~(0.16) & 0.69~(0.30) & 0.14~(0.03) \\
GIN & 4.56~(0.06) & 4.58~(0.10) & 1.86~(0.14) & 1.93~(0.15) & 1.83~(0.22) & 1.85~(0.23) & 0.16~(0.01) & 0.24~(0.10) & 0.67~(0.29) & 0.13~(0.03) \\
GAP & 19.59~(0.61) & 19.59~(0.59) & 66.45~(1.96) & 66.94~(1.75) & 76.79~(6.73) & 77.66~(6.57) & 8.66~(2.59) & 6.32~(4.48) & 70.53~(45.17) & 3.20~(0.89) \\
ClusNet & 40.16~(0.24) & 39.21~(0.21) & 73.28~(4.97) & 76.96~(4.27) & 88.50~(7.63) & 90.07~(8.14) & 19.10~(0.28) & 6.55~(3.92) & 49.93~(29.66) & 4.80~(1.15) \\ \hline
\textbf{ICD-M} & 13.63~(0.55) & 11.19~(0.32) & 2.48~(0.18) & 3.56~(0.26) & 2.93~(0.32) & 4.19~(0.46) & 0.43~(0.02) & 0.29~(0.14) & 2.37~(1.25) & 0.21~(0.06) \\
Feat & 0.42~(0.01) & 0.42~(0.01) & 0.20~(0.01) & 0.20~(0.01) & 0.23~(0.02) & 0.23~(0.02) & 0.24~(0.0036) & 0.007~(0.0077) & 1.12~(0.73) & 0.03~(0.01) \\
Prop & 0.02~(0.0005) & 0.02~(0.001) & 0.03~(0.0006) & 0.03~(0.0005) & 0.03~(0.0023) & 0.03~(0.0023) & 0.003~(0.0001) & 0.01~(0.0085) & 0.13~(0.05) & 0.003~(0.0001) \\
Clus & 13.18~(0.55) & 10.76~(0.32) & 2.25~(0.18) & 3.33~(0.26) & 2.67~(0.31) & 3.93~(0.45) & 0.19~(0.02) & 0.2741~(0.12) & 1.12~(0.48) & 0.18~(0.05) \\ \hline
\textbf{ICD-C} & 14.11~(0.50) & 11.42~(0.30) & 2.84~(0.20) & 3.55~(0.25) & 3.36~(0.37) & 4.24~(0.44) & 0.43~(0.02) & 0.30~(0.14) & 2.38~(1.28) & 0.22~(0.06) \\
Feat & 0.47~(0.01) & 0.46~(0.01) & 0.24~(0.009) & 0.25~(0.009) & 0.28~(0.03) & 0.28~(0.03) & 0.24~(0.005) & 0.008~(0.0093) & 1.16~(0.75) & 0.03~(0.01) \\
Prop & 0.02~(0.0003) & 0.01~(0.0003) & 0.03~(0.0004) & 0.03~(0.0004) & 0.03~(0.0024) & 0.03~(0.002) & 0.003~(0.0001) & 0.01~(0.0087) & 0.13~(0.05) & 0.003~(0.0011) \\
Clus & 13.62~(0.50) & 10.95~(0.30) & 2.57~(0.20) & 3.28~(0.25) & 3.05~(0.35) & 3.93~(0.42) & 0.19~(0.01) & 0.28~(0.12) & 1.10~(0.49) & 0.19~(0.05) \\ \hline
\end{tabular}
\vspace{-0.3cm}
\end{table}

In our evaluation, we recorded mean $\mu$ and standard deviation $\sigma$ of all the metrics on the test set $\Gamma'$ of each dataset. Number records in the format $\mu (\sigma)$ w.r.t. \textbf{NMI}, \textbf{AC}, \textbf{modularity}, \textbf{NCut} and \textbf{runtime} on each dataset are depicted in Table~\ref{Tab:Eva-NMI-AC}, \ref{Tab:Eva-Mod}, \ref{Tab:Eva-NCut}, and \ref{Tab:Eva-Time}. For ICD, we also recorded the \textbf{runtime} of (\romannumeral1) feature extraction of ${\bf{Z}}_t$, (\romannumeral2) one feedforward propagation through the GNN encoder, and (\romannumeral3) downstream $K$Means clustering (with $10$ independent runs), which are denoted as `Feat', `Prop', and `Clus' in Table~\ref{Tab:Eva-Time}.

For the total runtime of ICD on all the datasets (see Table~\ref{Tab:Eva-Time}), the downstream clustering is a major bottleneck, but it is also the key component that enables ICD to tackle \textit{online} CD with non-fixed $K$. In our future work, we intend to further reduce the runtime of ICD by replacing the downstream clustering module with a generic E2E module that can directly derive the CD result w.r.t. a specific (non-fixed) $K$ via an output layer.

\end{document}